\documentclass[aps,groupedaddress,showpacs,floatfix]{revtex4}
\usepackage{amssymb}
\usepackage{amsmath}
\usepackage{graphicx}
\DeclareMathOperator{\sgn}{sgn}
\DeclareMathOperator{\Ai}{Ai}
\begin{document}

\thispagestyle{empty}

\title{Statistical properties of one-dimensional directed polymers in a random potential}

\author{Victor Dotsenko$^{\, a,b}$ }

\affiliation{$^a$LPTMC, Universit\'e Paris VI, 75252 Paris, France}

\affiliation{$^b$L.D.\ Landau Institute for Theoretical Physics,
   119334 Moscow, Russia}

\date{\today}

\begin{abstract}
This review is devoted to the detailed consideration
of the universal statistical properties of
one-dimensional directed polymers in a random potential.
In terms of the replica Bethe ansatz technique we derive
several exact results for different types of the free energy
probability distribution functions. In the second part of the review
we discuss the problems which are still waiting for their solutions.
Several mathematical appendices in the ending part of the review contain 
various technical  details of the performed calculations.

\end{abstract}


\maketitle

\medskip

\begin{center}

{\bf \large Contents}

\end{center}

\begin{tabbing}
 AAA \= AAAA \= {\bf GUE Tracy-Widom distribution in the model with fixed boundary conditions} \= ....... 1 \kill

\> {\bf I.  } \> {\bf   Introduction                                                                } \> ........1  \\
\> {\bf II. } \> {\bf   Replica method                                                              } \> ........3  \\
\> {\bf III.} \> {\bf   Mapping to quantum bosons                                                   } \> ........6  \\
\\
\> \> {\bf  Part 1: Exact results                                                            }   \\
\\
\> {\bf IV. } \> {\bf   GUE Tracy-Widom distribution in the model with fixed boundary conditions    } \> ........10  \\
\>            \>  A. Generating function approach                                                     \> ........10  \\
\>            \>  B. Inverse Laplace transformation approach                                          \> ........12  \\
\> {\bf V.  } \> {\bf   GOE Tracy-Widom distribution in the model with free boundary conditions     } \> ........14  \\
\> {\bf VI. } \> {\bf   Multi-point free energy distribution functions                              } \> ........21  \\   
\>            \>  A. Two-point distribution                                                           \> ........21  \\  
\>            \>  A. $N$-point distribution                                                           \> ........26  \\
\> {\bf VII.} \> {\bf   Probability distribution function of the endpoint fluctuations              } \> ........32  \\ 
\\
\>  \> {\bf Part 2:  Unsolved problems                                                        }   \\
\\
\> {\bf VIII.  } \> {\bf   Zero temperature limit                                                   } \> ........38  \\
\> {\bf IX.   } \> {\bf   Random force Burgers turbulence                                           } \> ........43  \\          
\> {\bf X.  } \> {\bf   Joint distribution function of free energies at two different temperatures   } \> ........52  \\  
\\
\> {\bf XI  } \> {\bf Conclusions                                                                   } \> ........57  \\
\\
\> {\bf Appendix A: GUE Tracy-Widom distribution function}                           \>  \> ........58  \\
\> {\bf Appendix B: The Airy function integral relations}                            \>  \> ........59  \\
\> {\bf Appendix C: Fredholm determinant with the Airy kernel }                      \>  \> ........60  \\
\> {\bf Appendix D: Useful combinatorial identities}                                 \>  \> ........63  \\
\> {\bf Appendix E: Technical part of Chapter V}                                     \>  \> ........64 \\
\end{tabbing}




\newpage

\setcounter{page}{1}

\section{Introduction}

\newcounter{1}
\setcounter{equation}{0}
\renewcommand{\theequation}{1.\arabic{equation}}

It is well known that  macroscopic characteristics
of any random system defined in terms of macroscopic number of {\it independent} random parameters
(according to the central limit theorem)
are described by  the Gaussian distribution function.
In a sense, such type of the universal behavior is trivial, and not so much interesting.
On the other hand, any non-trivial system  usually
requires individual consideration, and although there are lot of
universal macroscopic properties among microscopically different systems
(e.g. scaling and critical phenomena at the phase transitions)
until very recently no one would expect to have a
{\it universal function} (different from the Gaussian one) which would describe
macroscopic statistical properties of a whole class of non-trivial random systems.

One of the main achievement of the last three decades in the scope of the disordered
systems is the discovery of the entire class of random systems
whose macroscopic properties  are described
by {\it the same} universal probability distribution function which is
called the Tracy-Widom (TW) distribution \cite{TW-GUE}. Traditionally it is called the
KPZ universality class by the name of the celebrated Kardar-Parisi-Zang equation \cite{KPZ}
which describes the time evolution of an interface (separating two homogeneous bulk phases)
in the disordered inhomogeneous media.

Originally the solution of Tracy and Widom \cite{TW-GUE} were devoted to
rather specific mathematical problem, namely the distribution function of the largest eigenvalue
of $N\times N$ Hermitian matrices (Gaussian Unitary Ensemble (GUE)) in the limit $N \to \infty$.
Nowadays we have got rather comprehensive list of various systems (both purely mathematical
and physical) whose macroscopic statistical properties are described by the same universal
TW distribution function. Among these systems are:
   the longest increasing subsequences (LIS) model \cite{LIS};
   zero-temperature lattice directed polymers with geometric disorder \cite{DP_johansson};
   the polynuclear growth (PNG) system \cite{PNG_Spohn};
   the oriented digital boiling model \cite{oriented_boiling};
   the ballistic decomposition model \cite{ballistic_decomposition1,ballistic_decomposition2};
   the longest common subsequences (LCS) \cite{LCS};
   the fixed-trace ensembles of random matrices \cite{fixed-trace-ensembles}
   the asymmetric simple exclusion process (ASEP)\cite{ASEP1,ASEP2,ASEP3}; etc.
Besides, there are lot of
physical and biological systems in which the KPZ universality class scaling behavior
has been convincingly demonstrated experimentally:
   bacterial colony growth \cite{bacteria};
   paper combustion \cite{paper};
   liquid-crystal electroconvection \cite{liquid-crystal};
   eucariotic cell colony growth \cite{eucariotic};
   particle decomposition in coffee rings \cite{coffe};
   chemical reaction front \cite{chemical-reaction}
(see also \cite{exper-rev} for a review on experiments).

\vspace{5mm}

In this review we are going to concentrate on the model of directed polymers defined in terms
of an elastic string $\phi(\tau)$
directed along the $\tau$-axes within an interval $[0,t]$ which passes through a random medium
described by a random potential $V(\phi,\tau)$. The energy of a given polymer's trajectory
$\phi(\tau)$ is
\begin{equation}
   \label{1.1}
   H[\phi(\tau); V] = \int_{0}^{t} d\tau
   \Bigl\{\frac{1}{2} \bigl[\partial_\tau \phi(\tau)\bigr]^2
   + V[\phi(\tau),\tau]\Bigr\};
\end{equation}
Here the disorder potential $V[\phi,\tau]$
is supposed to be Gaussian distributed with a zero mean $\overline{V(\phi,\tau)}=0$
and the $\delta$-correlations
\begin{equation}
\label{1.2}
{\overline{V(\phi,\tau)V(\phi',\tau')}} = u \delta(\tau-\tau') \delta(\phi-\phi')
\end{equation}
where $\overline{(...)}$ denotes the disorder average and the parameter $u$ describes the strength of the
random potential.
Diverse physical systems such as domain walls in magnetic films
\cite{lemerle_98}, vortices in superconductors \cite{blatter_94}, wetting
fronts on planar systems \cite{wilkinson_83}, or Burgers turbulence
\cite{burgers_74, turbulence} can be mapped to this model, which exhibits numerous
non-trivial features deriving from the interplay between elasticity and
disorder.

The system defined by the above Hamiltonian (\ref{1.1})
has been the subject of intense investigations during the past almost three
decades (see e.g.
\cite{hh_zhang_95,kardar_book,hhf_85,numer1,numer2,kardar_87,bouchaud-orland,Brunet-Derrida,
Johansson,Prahofer-Spohn,Ferrari-Spohn1}).
Historically, the problem of central interest was the scaling behavior of the
polymer mean squared displacement  which in the thermodynamic limit
($t \to \infty$)  reveals a universal scaling form
$\overline{\langle\phi^{2}\rangle}(t) \propto t^{2\zeta} $
(where $\langle \dots \rangle$  denotes the thermal average), 
with $\zeta=2/3$, the so-called wandering exponent.

More general and more interesting problem
is the statistical properties  of the  free energy of this system.
For a given realization of the random potential the partition function of the considered
system is defined in terms of the functional integral:
\begin{equation}
\label{1.3}
   Z(x,t) = \int_{\phi(0)=0}^{\phi(t)=x}
              {\cal D} [\phi(\tau)]  \;  \exp\Bigl\{-\beta H[\phi(\tau); V]\Bigr\} \; = \;
\exp\bigl\{-\beta F(x,t)\}
\end{equation}
where $\beta = 1/T$ is the inverse temperature and the integration goes over all trajectories
with fixed boundary conditions $\phi(0)=0$ and $\phi(t)=x$, and $F(x,t)$ is the (random) free energy.
According to the above definition one can easily show that the partition function $Z(x,t)$
satisfies the differential equation
\begin{equation}
 \label{1.4}
\partial_{\tau} Z(x,t) \; = \; \frac{1}{2\beta} \partial_{x}^{2} Z(x,t) \; - \; \beta V(x,t) Z(x,t)
\end{equation}
substituting here $Z(x,t) \, = \, \exp\{-\beta F(x,t)\}$ one obtains
\begin{equation}
 \label{1.5}
-\partial_{\tau} F(x,t) \; = \; \frac{1}{2} \Bigl(\partial_{x} F(x,t)\Bigr)^{2} \; - \;
\frac{1}{2\beta} \partial_{x}^{2} F(x,t) \; - \; \beta V(x,t)
\end{equation}
which is the KPZ equation \cite{KPZ} where the free energy $F(x,t)$ of the original directed polymer
problem, eqs.(\ref{1.1})-(\ref{1.3}), plays now the role of the interface front evolving in time in the presence
quenched random potential $V(x,t)$.

First of all it is evident that in the absence of the random potential the  partition function $Z(x,t)$
describes simple thermal diffusion. Indeed, according to eq.(\ref{1.3}) the probability
that at time $\tau = t$ the trajectory arrives to the point $\phi(t) = x$ is given by
\begin{equation}
\label{1.6}
   Z_{0}(x, t) = \int_{\phi(0)=0}^{\phi(t)=x}
              {\cal D} [\phi(\tau)]  \;
\exp\biggl\{-\frac{1}{2} \beta \int_{0}^{t} d\tau \bigl[\partial_\tau \phi(\tau)\bigr]^{2} \biggr\}
\end{equation}
Simple Gaussian integration (with the proper choice of the integration measure
of the functional integral) yields
\begin{equation}
\label{1.7}
   Z_{0}(x, t) = \sqrt{\frac{\beta}{2\pi t}} \exp\biggl\{-\frac{\beta x^{2}}{2t} \biggr\}
\end{equation}
In other words the typical deviation $\langle\phi(t)\rangle$ of the trajectory due to the {\it thermal
fluctuations} growth as $ t^{1/2}$ which is much smaller than  the typical value of
the trajectory deviations due to the action of the random potentials which scales
as $\phi(t) \sim t^{2/3}$
On the other hand, in the presence of the random potential
the two terms of the Hamiltonian (\ref{1.1}) must balance each other.
For a given value of the typical deviation $\phi \sim t^{2/3}$ the contribution of the elastic term
can be estimated  as $\phi^{2}/ t \sim t^{1/3}$. Thus, in the presence of disorder the free energy
fluctuations of this system must scale as $t^{1/3}$.
In other words, in the limit $t\to\infty$, besides the usual extensive (linear in $t$) self-averaging part 
and the elastic term, 
the total free energy $F$ of the considered systems must contain disorder dependent
fluctuating contribution $\sim t^{1/3}$:
\begin{equation}
\label{1.8}
F \; = \; f_{0} t \; + \frac{x^{2}}{2 \, t} \; + \;  c \, t^{1/3}\; f
\end{equation}
where $f_{0}$ is the (non-random) linear free energy density, $x^{2}/2t$ is the trivial 
elastic contribution,  $c$ is a non-universal parameter, which depends on the temperature 
and the strength of disorder, and  finally 
$f\sim 1$ is the random quantity which in the thermodynamic limit $t\to\infty$
is expected to be described by a non-trivial universal distribution function $P(f)$.


\vspace{5mm}

The breakthrough in the studies of the problem defined above took place in 2010
when the exact solution for the free energy
probability distribution function (PDF) $P(f)$ for the model with fixed boundary condition has been found
\cite{KPZ-TW1a,KPZ-TW1b,KPZ-TW1c,KPZ-TW2,BA-TW2,LeDoussal1,BA-TW3}.
It was shown that this PDF is given by the Tracy-Widom (TW) distribution of the largest eigenvalue of
the Gaussian Unitary Ensemble (GUE) \cite{TW-GUE} (see Appendix A).
Since that time important progress in understanding of the statistical properties of the KPZ-class
systems has been achieved (for the review see \cite{Corwin,Borodin}).
In particular, by this time it is shown that the free energy PDF of the directed polymer model
(\ref{1.1}) with {\it free} boundary conditions is given by the Gaussian Orthogonal Ensemble (GOE) TW distribution
\cite{LeDoussal2,goe}, while in the presence of a "wall" ($\phi(\tau) \geq 0, \; 0 \leq \tau \leq t$)
such PDF is given by the Gaussian Simplectic Ensemble (GSE) TW distribution \cite{LeDoussal3}.
The two-point as well as $N$-point free energy distribution function
which describes joint statistics of the free energies of the directed polymers
coming to different endpoints has been derived in
\cite{Prolhac-Spohn,2pointPDF,Imamura-Sasamoto-Spohn,N-point-1,N-point-2}.
Besides the free energy,
the explicit expression for the PDF for the end-point $\phi(t)$ fluctuations
has been also obtained \cite{math1, math2, math3, end-point,math4}.
In all the above studies, however, the problems were considered in the so called
"one-time" situation. The problem of the joint statistical properties
of the free energies at  two different times has been studied in
\cite{2time-1,2time-2,2time-J,Ferrari-Spohn_time-corr,Takeuchi_time-corr,2time-3,Nardis-LeDoussal-Takeuchi_time-corr}.
Unfortunately at present stage most of the theoretical results for the  two-time quantities remain 
analytically intractable and moreover the most recent studies indicate that the replica treatment of this problem contains
controversial technical issues (see \cite{Nardis-LeDoussal_time-tail}). For that reason we will not 
touch the two-time problem in the present review.

The consideration will be done in terms of the so called replica Bethe ansatz technique
which is the combination of the  replica method (used for systems with quenched
disorder) and the Bethe ansatz wave function solution for one-dimensional quantum boson
system (which can be shown to be equivalent to the replicated representation of the original 
directed polymer problem). It should be stressed that although in some cases this technique 
includes rather "heavy mathematical machinery", everything represented in this 
review is physics and not mathematics. Namely, although most of the results obtained
in terms of this technique are {\it exact}, their derivation is {\it not rigorous}.
First of all, the replica technique itself (which includes unjustified analytic continuations
as well as summations of formally divergent series) is mathematically ill grounded.
Besides, in some cases other unproved assumptions (such as multiple Gamma functions poles
cancellations, etc.) are used. On the other hand, until now in {\it all} the cases where
it was possible to give mathematically rigorous derivation the results
of the replica Bethe ansatz technique are confirmed. Thus, by "analytic continuation"
it is supposed that all the other (not confirmed yet)  results obtained in terms of this 
technique are also correct. We see very well that the method works. The problem is that
no one understands, why it works. 


The review is structured in the following way:

In Chapter II the replica technique  is formulated
and a brief discussion of its mathematically ill grounded tricks is given.
In Chapter III the original replicated directed polymer problem is redefined in terms of one-dimensional
quantum bosons with paired attractive interactions and the Bethe ansatz wave function of this system
is introduced.
Chapter IV is devoted to the formal derivation of the GUE Tracy-Widom probability distribution
for the directed polymer problem with fixed boundary conditions.
In Chapter V similar derivation of the GOE distribution is given for the same system with
free boundary conditions.
In Chapter VI we derive the joint two-point and the $N$-point probability distribution functions.
In Chapter VII the explicit expression for the PDF of the end-point $\phi(t)$ fluctuations
is derived.

In the second part of the review we discuss some of the unsolved problems:

In Chapter VIII we consider the problem of the zero temperature limit for the directed polymers
in random potentials with finite correlation length.
In Chapter IX in terms of a particular "toy" model we consider the application of the directed polymer
studies for the problem of random force Burgers turbulence.
Chapter X is devoted to the problem of the joint  statistical properties of two free energies
computed at two different temperatures in the same sample (for the same realization of the random
potential).
Future perspectives are discussed in Conclusions.

Finally, several technical appendices contain all necessary mathematical machinery which hopefully
makes the whole review self-contained.


\vspace{10mm}

\section{Replica method}

\newcounter{2}
\setcounter{equation}{0}
\renewcommand{\theequation}{2.\arabic{equation}}

For the calculation of thermodynamic quantities averaged over quenched disorder parameters
(e.g. average free energy) the replica method assumes, first, calculation of the averages
of an integer $N$-th power of the partition function $Z(N)$, and second, analytic continuation of 
this function in the replica parameter $N$ from integer to arbitrary non-integer values
(and in particular, taking the limit $N\to 0$)\cite{DeGennes,Edwards-Anderson} 
(see also  \cite{RSB-general,book}). Usually one is facing difficulties at both
stages of this program. First of all, in  realistic disordered systems
the calculations of the replica partition function $Z(N)$ can be done only using
some kind of approximations, and in this case the status of further analytic continuation 
in the replica parameter $N$ becomes rather indefinite since the terms neglected 
at integer $N$ could become  essential at non-integer $N$ (in particular the limit $N\to 0$)
\cite{zirnbauer1,zirnbauer2}. 

The typical example of such type of trouble is provided by the old Kardar's solution 
of $(1+1)$ directed polymers
in random potential where due to the approximation used at the first stage of calculations
(when the parameter $N$ is still integer)
the resulting free energy distribution function appears to be  not positively defined
\cite{kardar1,kardar2} (see also \cite{dirpoly,replicas})

On the other hand, even in rare  cases when 
the derivation of the replica partition function $Z(N)$  can be done exactly, 
further analytic continuation to non-integer $N$ appears  to be ambiguous.
The classical example of this situation is provided by the Derrida's Random Energy Model
(REM) in which the momenta $Z(N)$ growths as $\exp(N^2)$ at large $N$, and in this case
there are many different distributions yielding the same values of $Z(N)$, but 
providing {\it different} values for the average free energy of the system \cite{REM}.
Performing "direct" analytic continuation to non-integer $N$
(just assuming that the parameter $N$ in the obtained expression for $Z(N)$ can take arbitrary
real values), one finds the so called replica symmetric (RS) solution which turns out to be 
correct at high temperatures, but which is apparently wrong (it provides negative entropy)
in the low temperature (spin-glass) phase. In the case of REM the situation is sufficiently
simple because here one can check what is right and what is wrong comparing with the 
available exact solution (which can be derived without replicas). Unfortunately in other systems 
the status of the results obtained by the replica method is less clear. 

It should be noted that during last two decades remarkable progress has been 
achieved in mathematically rigorous derivations of various results previously obtained in
terms of the replica method. 
First of all, a number of rigorous results have been obtained 
\cite{Talagrend,Gamarnik1,Gamarnik2,Archlioptas,Bayati} 
which prove the validity of the cavity method 
\cite{RSB-general, cavity} 
for the entire class of the random satisfiability problems 
\cite{RSP1,RSP2,RSP3}, revealing the physical phenomena similar to
what happens in REM and which are described by the one-step RSB solution.
Rigorous proof has been found  
\cite{Aldous} for the result of the replica solution of the random bipartite matching problem
in the thermodynamic limit
\cite{matching}.
The results obtained in terms of the continuous RSB scheme developed for
mean-field spin glasses \cite{RSB-general} has been also confirmed by independent
mathematically rigorous calculations (see \cite{guerra} and references therein).
Notable progress in the studies of the subtleties of the replica 
method has been achieved in the context of the random matrix 
theory, where the remarkable exact relation between replica partition functions
and Painlev\'e transcendents has been proved \cite{kanzieper1,splittorff,kanzieper2,kanzieper3}.
Recently rigorous replica method has been developed for $q$-TASEP and ASEP 
models \cite{Replicas_ASEP}, as well as for the Povolotsky particle system and for the general
class of stochastic higher spin vertex models \cite{Replicas_corwin,Replicas_corwin-petrov}.

All these studies convincingly demonstrate that 
the replica method is robust and reliable, 
although of course, in many cases it does not explain
{\it why} the replica trick, the way it is used in the actual calculations by physicists,
does provide correct results.

\vspace{5mm}

In this Chapter we will consider the application of the replica technique for the directed polymer model
(\ref{1.1})-(\ref{1.2}). For simplicity let us consider the  situation with the zero boundary conditions:
$\phi(0) = \phi(t) = 0$. In this case the partition function of a given sample  is (c.f. eq.(\ref{1.3}))
\begin{equation}
\label{2.1}
   Z(t) = \int_{\phi(0)=0}^{\phi(t)=0}
              {\cal D} [\phi(\tau)]  \;  \exp\Bigl\{-\beta H[\phi,V]\Bigr\} 
\end{equation}
on the other hand, the partition function $Z(t)$ is related to the total free energy $F(t)$ via
\begin{equation}
 \label{2.2}
 Z(t) \; = \; \exp\{-\beta \, F(t)\}
\end{equation}
The free energy $F(t)$ is defined for a specific
realization of the random potential $V$ and thus represent a random variable.
The free energy probability distribution function $P_{t}(F)$ can be studied in terms of the integer
moments of the above partition function.
Taking the $N$-th power of both sides of eq.(\ref{2.2})
and performing the averaging over the random potential $V$ we obtain
\begin{equation}
\label{2.3}
\overline{Z^{N}(t)} \equiv Z(N; t) = \overline{\exp\bigl\{ -\beta N F\bigr\} }
\end{equation}
where the quantity $Z(N; t)$ is called the {\it replica partition function}.
As the averaging in the rhs of the above equation can be represented in terms of the
distribution function $P_{t}(F)$  we arrive to the following general relation
between the replica partition function $Z(N; t)$ and the free energy distribution function
$P_{t}(F)$:
\begin{equation}
\label{2.4}
   Z(N; t)\; =\;
           \int_{-\infty}^{+\infty} dF \, P_{t} (F) \;
           \exp\bigl\{ -\beta N F\bigr\}
\end{equation}
 The above equation is the bilateral Laplace transform of  the function $P_{t}(F)$,
and at least formally it allows to restore this function via inverse Laplace transform
 of the replica partition function $Z(N; t)$. In order to do so one has to compute
$Z(N; t)$ for an  {\it arbitrary }
integer $N$ and then perform  analytical continuation of this function
from integer to arbitrary complex values of $N$. This is the standard  strategy of the replica method
in disordered systems where it is well known that  very often the procedure of such analytic
continuation turns out to be rather controversial point \cite{zirnbauer1,replicas}. Even in rare  cases where
the replica partition function $Z(N; t)$  can be derived exactly, its
further analytic continuation to non-integer $N$ appears  to be ambiguous. Compared with 
the Derrida's Random Energy Model in which the momenta $Z(N; t)$ growths as fast as  $\exp(N^2)$ 
at large $N$, in the present system the situation is even worse because, as we will see later, the replica
partition function growth here as  $\exp(N^3)$ at large $N$, and in this situation its
analytic continuation from integer to non integer $N$ is ambiguous.

There are two (physical) approaches which allows to cope with this problem. 
Both of them are mathematically ill grounded, but it is interesting to note that although they 
look as completely independent ways of calculations, both approaches 
provide the same finial result (see Chapter IV).

\vspace{5mm}

In the first approach one performs the "analytic continuation" from integer to arbitrary 
complex values of $N$ in the explicit expression for the replica partition function $Z(N; t)$ 
just by claiming that from now on $N$ is a complex parameter. As the free energy of the system
under consideration besides the fluctuating part $\sim t^{1/3} f$ contains also non-random (self-averaging)
contribution $f_{0} t$, eq.(\ref{1.8}), to extract the probability distribution function $P_{t}(f)$
of the fluctuating part let us redefine
\begin{equation}
 \label{2.5}
 Z(t) \; = \; \tilde{Z}(t) \, \exp\bigl\{-\beta \, f_{0} \, t \bigr\}
\end{equation}
so that 
\begin{equation}
 \label{2.6}
\tilde{Z}(t) \; = \; \exp\{-\lambda(t) \, f\}
\end{equation}
where 
\begin{equation}
 \label{2.7}
\lambda(t) \; = \; \beta \, c \, t^{1/3} \; \propto \; t^{1/3} 
\end{equation}
Correspondingly, for the replica partition function we have
\begin{equation}
 \label{2.8}
Z(N; t) \; = \; \tilde{Z}(N; t) \, \exp\bigl\{-\beta N \, f_{0} \, t \bigr\}
\end{equation}
where
\begin{equation}
 \label{2.9}
\tilde{Z}(N; t) \; = \;  \overline{\tilde{Z}^{N}(t)}
\end{equation}
Substituting eqs.(\ref{2.8}) and (\ref{1.8}) into eq.(\ref{2.4}) we get
\begin{equation}
\label{2.10}
   \tilde{Z}(N; t)\; =\;
           \int_{-\infty}^{+\infty} df \, P_{t} (f) \;
           \exp\bigl\{ -\lambda(t) \, N \,  f \bigr\}
\end{equation}
Formally the above relation allows to restore the probability distribution function
$P_{t}(f)$ via the inverse Laplace transform. Redefining $N \, = \, s/\lambda(t)$,
where $s$ is a new {\it complex} parameter, we get
\begin{equation}
 \label{2.11}
P_{t}(f) \; = \; \int_{-i\infty}^{+i\infty} \frac{ds}{2\pi i} \; 
\tilde{Z}\Bigl(\frac{s}{\lambda(t)}; \, t\Bigr) \, \exp\{ s\, f \}
\end{equation}
In the present study we are mostly interested in the asymptotic (universal) shape of the 
probability distribution function $P_{t}(f)$ in the limit $t \to \infty$. 
Introducing
\begin{equation}
 \label{2.12}
P_{*}(f) \; \equiv \; \lim_{t\to\infty} \, P_{t}(f)
\end{equation}
and
\begin{equation}
 \label{2.13}
\tilde{Z}_{*}(s) \; \equiv \; \lim_{t\to\infty} \, \tilde{Z}\Bigl(\frac{s}{\lambda(t)}; \, t\Bigr) \, ,
\end{equation}
provided the above limits exist, according to eq.(\ref{2.11}) we get
\begin{equation}
 \label{2.14}
P_{*}(f) \; = \; \int_{-i\infty}^{+i\infty} \frac{ds}{2\pi i} \; 
\tilde{Z}(s) \, \exp\{ s\, f \}
\end{equation}
In Chapter IV it will be shown that following the above procedure one eventually
obtains the Tracy-Widom distribution result for the function $P_{*}(f)$.

\vspace{5mm}

In an alternative approach, to bypass the problem of the analytic continuation in the replica parameter 
$N$ to non-integer values, instead of the free energy 
distribution function $P_{*}(f)$ one introduces its integral
representation
\begin{equation}
 \label{2.15}
W(f) \; = \; \int_{f}^{\infty} \; df' \; P_{*}(f')
\end{equation}
which gives the probability to find the fluctuation  bigger that a given value $f$.
According to the above definition
\begin{equation}
 \label{2.16}
P_{*}(f) \; = \; -\frac{\partial}{\partial f} \, W(f)
\end{equation}
In terms of the above replica partition function the function $W(f)$ can be defined as follows:
\begin{equation}
 \label{2.17}
W(f) \; = \; \lim_{t\to\infty} \, \sum_{N=0}^{\infty} \, \frac{(-1)^{N}}{N!} \, 
\exp\Bigl\{\lambda(t) N f\Bigr\} \; \overline{\tilde{Z}^{N}}
\end{equation}
Indeed, substituting here eqs.(\ref{2.9}) and (\ref{2.10}), we have
\begin{eqnarray}
 \nonumber
W(f) &=& \lim_{t\to\infty} \sum_{N=0}^{\infty} \frac{(-1)^{N}}{N!}
\int_{-\infty}^{+\infty} \; df' \; P_{t}(f') \,
\exp\Bigl\{\lambda(t) N (f - f') \Bigr\}
\\
\nonumber
\\
\nonumber
&=& \lim_{t\to\infty}
\int_{-\infty}^{+\infty} \; df' \; P_{t}(f') \, 
\exp\Bigl\{-\exp\bigl[\lambda(t) (f - f') \bigr] \Bigr\}
\\
\nonumber
\\
&=&
\int_{-\infty}^{+\infty} \; df' \; P_{*}(f') \;
\theta\bigl(f'-f\bigr)
\label{2.18}
\end{eqnarray}
which coincides with the definition, eq.(\ref{2.15}).
Thus, according to eq.(\ref{2.17}) the probability function $W(f)$
can be computed in terms of the replica partition function
$\tilde{Z}(N; t)$ by summing over all replica {\it integers}
\begin{equation}
 \label{2.19}
W(f) \; = \; \lim_{t\to\infty} \sum_{N=0}^{\infty} \frac{(-1)^{N}}{N!}
\exp\Bigl\{\lambda(t)\, N \, f \Bigr\} \; \tilde{Z}(N; t)
\end{equation}
It should be noted however, that in accordance with the {\it troubles conservation law}
the calculations of the probability distribution function
in terms of the above series  contain a delicate point 
which makes them mathematically ill-posed. On one hand, if we first perform the summation 
of the  series as it is done in eq.(\ref{2.18})
(which is perfectly convergent)  and only
after that perform the disorder averaging and take the limit $t \to \infty$,
everything looks mathematically well grounded (of course, provided the limit 
$\lim_{t\to\infty} P_{t}(f)$ exist).
On the other hand,  if we perform the disorder averaging first (which is done in the present
approach) and only after that perform the summation of the series, taking into account that 
$\tilde{Z}(N\gg 1; t) \sim \exp\bigl\{ N^{3} t\bigr\}$
we are facing formally divergent (sign alternating) series which require proper regularization.
For the moment the proof that such regularization in the limit $t \to \infty$ can be done in unambiguous 
way does not exist. Nevertheless, it is interesting to note that all apparent examples of the regularizations
of such series which demonstrate ambiguity of the result of its summation at finite value of $t$,
{\it in the limit $t \to \infty$} fall into the same unique result. 
This result for the  function $W(f)$, derived in Chapter IV, reveal the universal GUE
Tracy-Widom probability distribution function and it coincides with the one obtained 
via the inverse Laplace transform of analytically continued replica partition function, eqs(\ref{2.5})-(\ref{2.14}).


 \vspace{10mm}

\section{Mapping to quantum bosons}

\newcounter{3}
\setcounter{equation}{0}
\renewcommand{\theequation}{3.\arabic{equation}}

Explicitly, for the zero boundary conditions ($\phi(0)=\phi(t)=0$) the replica partition function, Eq.(\ref{2.3}), 
of the system described by the Hamiltonian, Eq.(\ref{1.1}), is
\begin{equation}
\label{3.1}
   Z(N; t) \; = \; \prod_{a=1}^{N} \Biggl[\int_{\phi_{a}(0)=0}^{\phi_{a}(t)=0} 
   {\cal D} \phi_{a}(\tau)\Biggr] \;
   \overline{\exp\Biggl\{-\beta \int_{0}^{t} d\tau \sum_{a=1}^{N}
   \Bigl[\frac{1}{2} \bigl[\partial_\tau \phi_{a}(\tau)\bigr]^2 
   + V[\phi_{a}(\tau),\tau]\Bigr]\Biggr\} }
\end{equation}
Since  the random potential $V[\phi,\tau]$ has the Gaussian 
distribution the disorder averaging $\overline{(...)}$ in the above equation 
is very simple:
\begin{equation}
\label{3.2}
\overline{\exp\Biggl\{-\beta \int_{0}^{t} d\tau \sum_{a=1}^{N}
     V[\phi_{a}(\tau),\tau]\Biggr\} } \; = \; 
\exp\Biggl[\frac{1}{2} \beta^{2}\int\int_{0}^{t} d\tau d\tau' \sum_{a,b=1}^{N}
     \overline{V[\phi_{a}(\tau),\tau] V[\phi_{b}(\tau'),\tau']}\Biggr]
\end{equation}
Using Eq.(\ref{1.2}) we find:
\begin{equation}
\label{3.3}
   Z(N; t) = \prod_{a=1}^{N} \Biggl[\int_{\phi_{a}(0)=0}^{\phi_{a}(t)=0} 
   {\cal D} \phi_{a}(\tau) \Biggr] \;
   \exp\Biggl\{-\frac{1}{2}\beta \int_{0}^{t} d\tau 
   \Bigl[\sum_{a=1}^{N} \bigl[\partial_\tau \phi_{a}(\tau)\bigr]^2 
   -\beta u \sum_{a,b=1}^{N} \delta\bigl(\phi_{a}(\tau)-\phi_{b}(\tau)\bigr)\Bigr]\Biggr\} 
\end{equation}
It should be noted that the second term in the exponential 
of the above equation contain formally divergent contributions proportional
to $\delta(0)$ (due to the terms with $a=b$). In fact, this is just an indication
that the {\it continuous} model, Eqs.(\ref{1.1})-(\ref{1.2}) is ill defined
as short distances and requires proper lattice regularization. Of course, the 
corresponding lattice model would contain no divergences, and the terms with
$a=b$ in the exponential of the corresponding replica partition function would produce
irrelevant constant $\frac{1}{2} t \beta^2 u N \delta(0)$ (where the lattice
version of $\delta(0)$ has a finite value). Since the  lattice 
regularization has no impact on the continuous long distance properties 
of the considered system this term will be just omitted in our further study.

Introducing the $N$-component scalar field replica Hamiltonian
\begin{equation}
\label{3.4}
   H_{N}\bigl[{\boldsymbol \phi}\bigr] =  
   \frac{1}{2} \int_{0}^{t} d\tau \Biggl(
   \sum_{a=1}^{N} \bigl[\partial_\tau\phi_{a}(\tau)\bigr]^2 
   - \beta u \sum_{a\not= b}^{N} 
   \delta\bigl(\phi_{a}(\tau)-\phi_{b}(\tau)\bigr) \Biggr)
\end{equation}
for the replica partition function, Eq.(\ref{3.3}), 
we obtain the standard expression
\begin{equation}
   \label{3.5}
   Z(N; t) = \prod_{a=1}^{N} \Biggl[\int_{\phi_{a}(0)=0}^{\phi_{a}(t)=0} {\cal D} \phi_{a}(\tau) \Biggr] \;
   \exp\Bigl\{-\beta H_{N}\bigl[{\boldsymbol \phi}\bigr] \Bigr\}
\end{equation}
where ${\boldsymbol \phi} \equiv \{\phi_{1},\dots, \phi_{N}\}$.
According to the above definition this partition function describe the statistics
of $N$ trajectories $\phi_{a}(\tau)$ with attractive $\delta$-interactions  all starting 
 and ending at zero: $\phi_{a}(0) = \phi_{a}(t) = 0$

In order to map the problem to one-dimensional quantum bosons, 
 let us introduce more general object
\begin{equation}
   \label{3.6}
   \Psi({\bf x}; t) = 
\prod_{a=1}^{N} \Biggl[\int_{\phi_{a}(0)=0}^{\phi_{a}(t)=x_a} {\cal D} \phi_{a}(\tau) \Biggr]
  \;  \exp\Bigl\{-\beta H_{N}\bigl[{\boldsymbol \phi}\bigr] \Bigr\}
\end{equation}
which describes $N$ trajectories $\phi_{a}(\tau)$ all starting at zero ($\phi_{a}(0) = 0$),
but ending at $\tau = t$ in arbitrary given points $\{x_{1}, ..., x_{N}\}$.
One can easily show that instead of using the path integral, $\Psi({\bf x}; t)$
can be obtained as the solution of the  linear differential equation
\begin{equation}
   \label{3.7}
\partial_t \Psi({\bf x}; t) \; = \;
\frac{1}{2\beta}\sum_{a=1}^{N}\partial_{x_a}^2 \Psi({\bf x}; t)
  \; + \; \frac{1}{2}\beta^2 u \sum_{a\not=b}^{N} \delta(x_a-x_b) \Psi({\bf x}; t)
\end{equation}
with the initial condition 
\begin{equation}
   \label{3.8}
\Psi({\bf x}; 0) = \Pi_{a=1}^{N} \delta(x_a)
\end{equation}
One can easily see that Eq.(\ref{3.7}) is the imaginary-time
Schr\"odinger equation
\begin{equation}
   \label{3.9}
-\partial_t \Psi({\bf x}; t) = \hat{H} \Psi({\bf x}; t)
\end{equation}
with the Hamiltonian
\begin{equation}
   \label{3.10}
   \hat{H} = 
    -\frac{1}{2\beta}\sum_{a=1}^{N}\partial_{x_a}^2 
   -\frac{1}{2}\beta^2 u \sum_{a\not=b}^{N} \delta(x_a-x_b) 
\end{equation}
which describes  $N$ bose-particles of mass $\beta$ interacting via
the {\it attractive} two-body potential $-\beta^2 u \delta(x)$. 
The original replica partition function, Eq.(\ref{3.5}), then is obtained via a particular
choice of the final-point coordinates,
\begin{equation}
   \label{3.11}
   Z(N; t) = \Psi({\bf 0}; t).
\end{equation}
with the initial condition $\Psi({\bf x}; 0) = \Pi_{a=1}^{N} \delta(x_a)$.
According to the standard procedure,
the wave function  $\Psi({\bf x}; t)$ of the quantum problem, eq.(\ref{3.7})-(\ref{3.8}), 
can be represented in terms of the linear combination
of the solutions of the the corresponding eigenvalue equation
\begin{equation}
   \label{3.7a}
2\beta \, E \,  \Psi({\bf x}) \; = \;
\sum_{a=1}^{N}\partial_{x_a}^2 \Psi({\bf x}; t)
  \; + \; \kappa \sum_{a\not=b}^{N} \delta(x_a-x_b) \Psi({\bf x}; t)
\end{equation}
where $\kappa = \beta^{3} u$. 
A generic  eigenstate of such system is described  in terms of the so called
{\it Bethe ansatz} eigenfunctions $\Psi_{{\bf Q}}^{(M)}({\bf x})$ and it is 
characterized by $N$ momenta
$\{ Q_{a} \} \; (a=1,...,N)$ which split into
$M$  ($1 \leq M \leq N$) "clusters" each described by
{\it continuous real} momenta $q_{\alpha}$ $(\alpha = 1,...,M)$
and by $n_{\alpha}$ {\it discrete imaginary} "components" 
(for details see \cite{Lieb-Liniger,McGuire,Yang,gaudin,bogolubov,Takahashi,Mehta}) :
\begin{equation}
   \label{3.12}
Q_{a} \; \to \; q^{\alpha}_{r} \; = \;
q_{\alpha} - \frac{i\kappa}{2}  (n_{\alpha} + 1 - 2r)
\;\; ; \; \;\;\; \;\;\; \;\;\;
(r = 1, ..., n_{\alpha}\,; \; \; \alpha = 1, ..., M)
\end{equation}
with the global constraint
\begin{equation}
 \label{3.13}
 \sum_{\alpha=1}^{M} n_{\alpha} = N
\end{equation}
Explicitly,
\begin{equation}
\label{3.14}
\Psi_{{\bf Q}}^{(M)}({\bf x}) =
\sum_{{\cal P}\in S_{N}}  \;
\prod_{1\leq a<b}^{N}
\Biggl[
1 +i \kappa \frac{\sgn(x_{a}-x_{b})}{Q_{{\cal P}_a} - Q_{{\cal P}_b}}
\Biggr] \;
\exp\Bigl[i \sum_{a=1}^{N} Q_{{\cal P}_{a}} x_{a} \Bigr]
\end{equation}
where the vector ${\bf Q}$ denotes the set of all $N$ momenta eq.(\ref{3.12}) and
the summation goes over $N!$ permutations ${\cal P}$ of $N$ momenta $Q_{a}$,
over $N$ particles $x_{a}$.
Note that recently it has been rigorously proved that the set of such Bethe ansatz eigenfunctions
is orthogonal and complete on the full continuous line of the interaction parameter $u$
which includes both attractive ($u > 0$) and repulsive ($u < 0$) sectors 
\cite{Bosons_borodin-corwin1,Bosons_borodin-corwin2}.

In terms of the above eigenfunctions,eqs.(\ref{3.12})-(\ref{3.14}), 
the solution of eq.(\ref{3.7}) can be expressed as follows:
\begin{equation}
\label{3.15}
\Psi({\bf x}; t) = \frac{1}{N!} \sum_{M=1}^{N} \, \frac{1}{M!}
\int {\cal D}^{(M)}_{\bf Q} \; \big|C_{M}({\bf Q})\big|^{2} \; 
\Psi_{\bf Q}^{(M)}({\bf x}) {\Psi_{\bf Q}^{(M)}}^{*}(0) \;
\exp\bigl(-t \, E_{M}({\bf Q}) \bigr)
\end{equation}
where we have introduced the notation
\begin{equation}
   \label{3.16}
\int {\cal D}^{(M)}_{\bf Q}  \equiv
\prod_{\alpha=1}^{M} \Biggl[\int_{-\infty}^{+\infty} \frac{dq_{\alpha}}{2\pi} \sum_{n_{\alpha}=1}^{\infty}\Biggr]
{\boldsymbol \delta}\Bigl(\sum_{\alpha=1}^{M} n_{\alpha} \; , \;  N\Bigr)
\end{equation}
and ${\boldsymbol \delta}(k , m)$ is the Kronecker symbol 
(note that the presence of this Kronecker symbol in the above equation 
allows to extend the summations over $n_{\alpha}$'s to infinity).
$|C_{M}({\bf Q})|^{2}$ in eq.(\ref{3.15}) is the normalization factor,
\begin{equation}
  \label{3.17}
\big|C_{M}({\bf Q})\big|^{2} \; = \; \frac{\kappa^{N}}{\prod_{\alpha=1}^{M}\bigl(\kappa n_{\alpha}\bigr)}
\prod_{\alpha<\beta}^{M}
\frac{\big|q_{\alpha}-q_{\beta} -\frac{i\kappa}{2}(n_{\alpha}-n_{\beta})\big|^{2}}{
      \big|q_{\alpha}-q_{\beta} -\frac{i\kappa}{2}(n_{\alpha}+n_{\beta})\big|^{2}}
\end{equation}
and $E_{M}({\bf Q})$ is the eigenvalue (energy) of the
eigenstate $\Psi_{\bf Q}^{(M)}({\bf x})$,
\begin{equation}
\label{3.18}
E_{M}({\bf Q}) \; = \;
\frac{1}{2\beta} \sum_{\alpha=1}^{N} Q_{a}^{2} =
 \frac{1}{2\beta} \sum_{\alpha=1}^{M} \; n_{\alpha} q_{\alpha}^{2}
- \frac{\kappa^{2}}{24\beta}\sum_{\alpha=1}^{M} \bigl(n_{\alpha}^{3}-n_{\alpha}\bigr)
\end{equation}
Taking into account the global constraint, eq.(\ref{3.13}), the energy $E_{M}({\bf Q})$
can be represented as the sum of two contributions:
\begin{equation}
 \label{3.19}
 E_{M}({\bf Q}) \; = \; \tilde{E}_{M}({\bf Q}) \; + \; \beta \, N \, f_{0} \, 
\end{equation}
where 
\begin{equation}
 \label{3.20}
\tilde{E}_{M}({\bf Q}) \; = \; 
 \frac{1}{2\beta} \sum_{\alpha=1}^{M} \; n_{\alpha} q_{\alpha}^{2}
- \frac{\kappa^{2}}{24\beta}\sum_{\alpha=1}^{M} n_{\alpha}^{3}
\end{equation}
and the factor
\begin{equation}
 \label{3.21}
f_{0} \; = \; \frac{\kappa^{2}}{24\beta^{2}} \; = \; \frac{1}{24} \, \beta^{4} \, u^{2}
\end{equation}
provides the linear in $t$ non-random  contribution to the total free energy, eq.(\ref{1.8}).
Redefining the replica partition function $Z(N; t)$ according to eq.(\ref{2.8}) 
(with $f_{0}$ given in eq.(\ref{3.21})) and taking into account the relations (\ref{3.11}) and
(\ref{3.15}) for the reduced replica partition function $\tilde{Z}(N; t)$ we obtain the following 
sufficiently simple representation
\begin{equation}
 \label{3.22}
\tilde{Z}(N; t) \; = \; 
\frac{1}{N!} \sum_{M=1}^{N} \, \frac{1}{M!}\int {\cal D}_{\bf Q}^{(M)} \; \big|C_{M}({\bf Q})\big|^{2} \; 
\big|\Psi_{\bf Q}^{(M)}({\bf 0})\big|^{2} \;
\exp\bigl(-t \, \tilde{E}_{M}({\bf Q}) \bigr)
\end{equation}
where according to eq.(\ref{3.14})
\begin{equation}
\label{3.23}
\Psi_{\bf Q}^{(M)}({\bf 0}) \; = \; N!
\end{equation}
and the explicit expressions for the reduced
energy $\tilde{E}_{M}({\bf Q})$ and the normalization factor $|C_{M}({\bf Q})|^{2}$ 
are given in eqs.(\ref{3.20}) and (\ref{3.17}) correspondingly. 

Finally, using the standard Cauchy double alternant identity (see Appendix D)
\begin{equation}
 \label{3.24}
\frac{\prod_{\alpha<\beta}^{M} (a_{\alpha} - a_{\beta})(b_{\alpha} - b_{\beta})}{
     \prod_{\alpha,\beta=1}^{M} (a_{\alpha} + b_{\beta})} \; = \; 
 \det\Bigl[\frac{1}{a_{\alpha}+b_{\beta}}\Bigr]_{\alpha,\beta=1,...M}
\end{equation}
the normalization factor, eq.(\ref{3.17}) can be represented in the compact determinant form:
\begin{equation}
\label{3.25}
\big|C_{M}({\bf Q})\big|^{2} \; = \; \kappa^{N} \det\Biggl[
   \frac{1}{\frac{1}{2}\kappa n_{\alpha} - i q_{\alpha}
          + \frac{1}{2}\kappa n_{\beta} + iq_{\beta}}\Biggr]_{\alpha,\beta=1,...M}
\end{equation}
%


\newpage

\begin{center}
 {\bf \Large PART 1: Exact results}
\end{center}

\vspace{5mm}

\section{GUE Tracy-Widom distribution in the model with fixed boundary conditions}

\newcounter{4}
\setcounter{equation}{0}
\renewcommand{\theequation}{4.\arabic{equation}}

\subsection{Generating function approach}

Substituting eqs.(\ref{3.16}), (\ref{3.23}) and (\ref{3.25}) into eq.(\ref{3.22})  we get
\begin{eqnarray}
 \nonumber
 \tilde{Z}(N; t) &=& N!
\sum_{M=1}^{N} \, \frac{1}{M!}
\prod_{\alpha=1}^{M} \Biggl[
\sum_{n_{\alpha}=1}^{\infty} 
\int_{-\infty}^{+\infty} \frac{dq_{\alpha} \, \kappa^{n_{\alpha}}}{2\pi} 
\exp\Bigl\{
-\frac{t}{2\beta} \, n_{\alpha} q_{\alpha}^{2} \, + \, 
\frac{\kappa^{2}\, t}{24\beta} n_{\alpha}^{3}
\Bigr\}
\Biggr]
{\boldsymbol \delta}\Bigl(\sum_{\alpha=1}^{M} n_{\alpha} \; , \;  N\Bigr) \times
\\
 \label{4.1}
\\
\nonumber
&\times&
\det\Biggl[
   \frac{1}{\frac{1}{2}\kappa n_{\alpha} - i q_{\alpha}
          + \frac{1}{2}\kappa n_{\beta} + iq_{\beta}}\Biggr]_{\alpha,\beta=1,...M}
\end{eqnarray}
Substituting this into eq.(\ref{2.19}) and taking into account that $N \; = \; \sum_{\alpha=1}^{M} n_{\alpha}$
we obtain
\begin{eqnarray}
 \nonumber
W(f) &=& \lim_{t\to\infty} \sum_{M=0}^{\infty} \frac{(-1)^{M}}{M!}
\prod_{\alpha=1}^{M} \Biggl[
\sum_{n_{\alpha}=1}^{\infty} \, (-1)^{n_{\alpha}-1} 
\int_{-\infty}^{+\infty} \frac{dq_{\alpha} \, \kappa^{n_{\alpha}}}{2\pi} 
\exp\Bigl\{
\lambda(t) n_{\alpha} \, f 
-\frac{t}{2\beta} \, n_{\alpha} q_{\alpha}^{2} \, + \, 
\frac{\kappa^{2}\, t}{24\beta} n_{\alpha}^{3}
\Bigr\}
\Biggr] \times
\\
 \label{4.2}
\\
\nonumber
&\times&
\det\Biggl[
   \frac{1}{\frac{1}{2}\kappa n_{\alpha} - i q_{\alpha}
          + \frac{1}{2}\kappa n_{\beta} + iq_{\beta}}\Biggr]_{\alpha,\beta=1,...M}
\end{eqnarray}
Let us redefine the momenta
\begin{equation}
 \label{4.3}
 q_{\alpha} \; = \; \frac{\kappa}{2\lambda(t)} \, p_{\alpha}
\end{equation}
with
\begin{equation}
 \label{4.4}
\lambda(t) \; = \; \frac{1}{2} \,
\Bigl(\frac{\kappa^{2} t}{\beta}\Bigr)^{1/3} \; = \;
\frac{1}{2} \, \bigl(\beta^{5} u^{2} t\bigr)^{1/3}
\end{equation}
Then for the last two term in the exponential of eq.(\ref{4.2}) we find
\begin{equation}
\label{4.5}
-\frac{t}{2\beta} \, n_{\alpha} \, q_{\alpha}^{2} \; + \; \frac{\kappa^{2} \, t}{24 \beta} \, n_{\alpha}^{3}
\; = \;
-\lambda(t) \, n_{\alpha} \, p_{\alpha}^{2} \; + \; \frac{1}{3} \, \lambda^{3}(t) \, n_{\alpha}^{3}
\end{equation}
The determinant factor 
\begin{equation}
 \label{4.6}
\det\Biggl[
   \frac{1}{\frac{1}{2}\kappa n_{\alpha} - i q_{\alpha}
          + \frac{1}{2}\kappa n_{\beta} + iq_{\beta}}\Biggr]_{\alpha,\beta=1,...M}  \; = \; 
 \Biggl(\frac{2\lambda(t)}{\kappa} \Biggr)^{M} \, 
\det\Biggl[
   \frac{1}{\lambda(t) \, n_{\alpha} - i p_{\alpha}
          + \lambda(t) \, n_{\beta} + ip_{\beta}}\Biggr]_{\alpha,\beta=1,...M} 
\end{equation}
The cubic  term in the exponent can be linearized using the Airy function relation (see Appendix B)
\begin{equation}
   \label{4.7}
\exp\Bigl\{ \frac{1}{3} \, \lambda^{3}(t) \, n_{\alpha}^{3} \Bigr\} \; = \;
\int_{-\infty}^{+\infty} dy_{\alpha} \; \Ai(y_{\alpha}) \;
\exp\Bigl\{\lambda(t) \, n_{\alpha} \, y_{\alpha} \Bigr\}
\end{equation}
Substituting eqs.(\ref{4.3}), (\ref{4.5}), (\ref{4.6}) and (\ref{4.7}) into eq.(\ref{4.2}), and redefining
$y_{\alpha} \; \to \; y_{\alpha} + p_{\alpha}^{2}$, we get
\begin{eqnarray}
 \nonumber
W(f) &=& \lim_{t\to\infty}
\sum_{M=1}^{\infty} \; \frac{(-1)^{M}}{M!} \;
\prod_{\alpha=1}^{M}
\Biggl[
\int\int_{-\infty}^{+\infty} \frac{dy dp_{\alpha}}{2\pi}
\Ai\bigl(y + p_{\alpha}^{2}\bigr)
\sum_{n_{\alpha}=1}^{\infty} (-1)^{n_{\alpha}-1} \; \kappa^{n_{\alpha}}
\exp\bigl\{\lambda(t) \, n_{\alpha} \, (y + f) \bigr\} 
\Biggr] \times
\\
 \label{4.8}
\\
\nonumber
&\times&
\det\Biggl[
   \frac{1}{\lambda(t) \, n_{\alpha} - i p_{\alpha}
          + \lambda(t) \, n_{\beta} + ip_{\beta}} \Biggr]_{\alpha,\beta=1,...M}
\end{eqnarray}
The above expression in nothing else but the expansion of the Fredholm determinant $\det(1 - \hat{K})$
(see e.g. \cite{Mehta} and Appendix C) with the kernel
\begin{equation}
\label{4.9}
\hat{K} \equiv
K\bigl[(n,p); (n',p')\bigr] = 
\int_{-\infty}^{+\infty} dy \Ai(y+p^{2}) \; \frac{(-1)^{n-1} \kappa^{n} \exp\bigl\{\lambda(t) \, n \, (y+f)\bigr\} }{
\lambda(t) \, n - ip + \lambda(t) \, n' + ip'} 
\end{equation}
Using the exponential representation of this determinant we get
\begin{equation}
 \label{4.10}
W(f)  = 
\lim_{t\to\infty}
\exp\Biggl\{-\sum_{M=1}^{\infty} \frac{1}{M} \; \mbox{Tr} \; \hat{K}^{M} \Biggr\}
\end{equation}
where 
\begin{eqnarray}
\nonumber
\mbox{Tr} \; \hat{K}^{M} &=& 
\prod_{\alpha=1}^{M} \Biggl[
\int\int_{-\infty}^{+\infty}
\frac{dy dp_{\alpha}}{2\pi}  
\Ai(y+p^{2}_{\alpha})
\sum_{n_{\alpha}=1}^{\infty} (-1)^{n_{\alpha}-1} \kappa^{n_{\alpha}} 
\exp\bigl\{\lambda(t) \, n_{\alpha} \, (y+f)\bigr\}
\Biggr] \times
\\
\nonumber
\\
&\times&
\frac{1}{\bigl[\lambda(t) n_{1} -ip_{1} + \lambda(t) n_{2} + ip_{2}\bigr]
         \bigl[\lambda(t) n_{2} -ip_{2} + \lambda(t) n_{3} + ip_{3}\bigr] ...
\bigl[\lambda(t) n_{M} -ip_{M} + \lambda(t) n_{1} + ip_{1}\bigr]}
 \label{4.11}
\end{eqnarray}
Substituting
\begin{equation}
 \label{4.12}
\frac{1}{\lambda(t) \, n_{\alpha} -ip_{\alpha} + \lambda(t) \, n_{\alpha+1} + ip_{\alpha+1}}  \; = \; 
\int_{0}^{\infty} d\omega_{\alpha} 
\exp\Bigl\{
-\bigl[ \lambda(t) \, n_{\alpha} - ip_{\alpha} + \lambda(t) \, n_{\alpha+1} + ip_{\alpha+1}\bigr]\omega_{\alpha}\Bigr\}
\end{equation}
into eq.(\ref{4.11}) we get
\begin{eqnarray}
\nonumber
\mbox{Tr} \; \hat{K}^{M} &=& 
\prod_{\alpha=1}^{M} \Biggl[
\int_{\infty}^{\infty} dy_{\alpha} 
\int_{-\infty}^{+\infty} \frac{dp_{\alpha}}{2\pi}  
\int_{0}^{\infty} d\omega_{\alpha} 
\Ai(y_{\alpha}+p^{2}_{\alpha}) \; 
\exp\bigl\{ip_{\alpha} (\omega_{\alpha}-\omega_{\alpha-1})\bigr\} \times
\\
\nonumber
\\
&\times&
\sum_{n=1}^{\infty} (-1)^{n-1} \kappa^{n} 
\exp\bigl\{\lambda(t) \, \bigl[y_{\alpha} +\omega_{\alpha} + \omega_{\alpha-1} + f\bigr] \, n \bigr\}
\Biggr]
\label{4.13}
\end{eqnarray}
where by definition it is assumed that $\omega_{0} \equiv \omega_{M}$. The series 
\begin{equation}
 \label{4.14}
S(z) \; = \; \sum_{n=1}^{\infty} (-1)^{n-1} \; z^{n} \; = \; \frac{z}{1 + z}
\end{equation}
is converging only for $|z| < 1$, but the function $S(z)$ can be unambiguously analytically continued 
for a whole complex plain (except the pole at $z = -1$). 
Thus, in the limit $t\to\infty$ (where $\lambda(t) \to \infty$) we have
\begin{eqnarray}
 \nonumber
\lim_{t\to\infty} 
\sum_{n=1}^{\infty} (-1)^{n-1} \kappa^{n} 
\exp\bigl\{\lambda(t) \,\bigl[y_{\alpha} +\omega_{\alpha} + \omega_{\alpha-1} + f\bigr] \, n\bigr\} &=&
\lim_{t\to\infty} \, 
\frac{\kappa \, \exp\bigl\{\lambda(t) \,\bigl[y_{\alpha} +\omega_{\alpha} + \omega_{\alpha-1} + f\bigr]\bigr\}}{
1 \, + \, \kappa \, \exp\bigl\{\lambda(t) \,\bigl[y_{\alpha} +\omega_{\alpha} + \omega_{\alpha-1} + f\bigr]\bigr\}}
\, = \, 
\\
\nonumber
\\
&=&
\theta\bigl(y_{\alpha} +\omega_{\alpha} + \omega_{\alpha-1} + f\bigr)
\label{4.15}
\end{eqnarray}
Substituting this into eq.(\ref{4.13}) 
and shifting the integration parameters, $y_{\alpha} \to y_{\alpha} - f + \omega_{\alpha} + \omega_{\alpha-1}$
and $\omega_{\alpha} \to  \omega_{\alpha} + f/2$, we obtain
\begin{equation}
\lim_{t\to\infty} \mbox{Tr} \, \hat{K}^{M} \equiv  \mbox{Tr} \, \hat{K}_{A}^{M} = 
\prod_{\alpha=1}^{M}
\Biggl[
\int_{-f/2}^{\infty} d\omega_{\alpha} 
\int_{0}^{\infty} dy
\int_{-\infty}^{+\infty} \frac{dp}{2\pi}  
\Ai(y + p^{2} + \omega_{\alpha} + \omega_{\alpha-1}) \; 
\exp\bigl\{ip (\omega_{\alpha} - \omega_{\alpha-1} )\bigr\}
\Biggr]
 \label{4.16}
\end{equation}
Using the standard Airy function integral relations (see Appendix B)
\begin{equation}
\label{4.17}
\int_{-\infty}^{+\infty} dp \;
\Ai\bigl(p^{2} + \omega_{1} + \omega_{2}\bigr) \; 
\exp\bigl[i p ( \omega_{1} - \omega_{2}) \bigr]
\; = \; 
2^{2/3} \pi \Ai\bigl(2^{1/3} \omega_{1}\bigr)  \Ai\bigl(2^{1/3} \omega_{2}\bigr) 
\end{equation}
and 
\begin{equation}
\label{4.18}
\int_{0}^{\infty} dy \; 
\Ai\bigl(y + \omega_{1}\bigr) \Ai\bigl(y + \omega_{2}\bigr) 
\; = \; 
\frac{\Ai\bigl(\omega_{1}\bigr) \Ai'\bigl(\omega_{2}\bigr) \; - \; 
\Ai'\bigl(\omega_{1}\bigr) \Ai\bigl(\omega_{2}\bigr)}{\omega_{1} - \omega_{2}}
\end{equation}
and 
redefining $\omega_{\alpha} \to \omega_{\alpha} 2^{-1/3}$ we eventually find
\begin{equation}
\mbox{Tr} \, \hat{K}_{A}^{M} = 
\int\int ...\int_{-f/2^{2/3}}^{\infty} d\omega_{1} d\omega_{2} ... d\omega_{M}
K_{A}(\omega_{1},\omega_{2}) K_{A}(\omega_{2},\omega_{3}) ... K_{A}(\omega_{M},\omega_{1})
\label{4.19}
\end{equation}
where
\begin{equation}
 \label{4.20}
K_{A}(\omega,\omega') \; = \; 
\frac{\Ai(\omega) \Ai'(\omega') - \Ai'(\omega) \Ai(\omega')}{
\omega - \omega'}
\end{equation}
is the Airy kernel. This proves that in the  limit $t \to \infty$
the probability function $W(f)$, eqs.(\ref{2.15}), (\ref{4.10}), is defined by the Fredholm determinant,
\begin{equation}
 \label{4.21}
W_{GUE}(f)  \; = \; \exp\Biggl\{-\sum_{M=1}^{\infty} \frac{1}{M} \; \mbox{Tr} \; \hat{K}_{A}^{M} \Biggr\}
\, = \, \det[1 - \hat{K}_{A}] \; \equiv \; F_{2}(-f/2^{2/3})
\end{equation}
where $\hat{K}_{A}$ is the integral operator on $[-f/2^{2/3}, \infty)$ with the 
Airy kernel, eq.(\ref{4.20}). The function $F_{2}(s)$ is the GUE Tracy-Widom distribution \cite{TW-GUE}
(see Appendix A). It can be shown to admit the following explicit representation (see Appendix C)
\begin{equation}
 \label{4.22}
F_{2}(s) \; = \; \exp\Bigl(-\int_{s}^{\infty} dt \; (t-s) \; q^{2}(t)\Bigr)
\end{equation}
where the function $q(t)$ is the solution of the Panlev\'e II equation, 
$q'' = t q + 2 q^{3}$
with the boundary condition, $q(t\to +\infty) \sim Ai(t)$ \cite{Panleve,Clarkson}.

\subsection{Inverse Laplace transformation approach}

The Kronecker symbol in eq.(\ref{4.1}) can be represented in the integral form
\begin{equation}
 \label{4.23}
 {\boldsymbol \delta}\Bigl(\sum_{\alpha=1}^{M} n_{\alpha} \; , \;  N\Bigr) \; = \; 
 \oint_{{\cal C}}\frac{dz}{2\pi i \, z} \, z^{N} \, \prod_{\alpha=1}^{M} \, z^{-n_{\alpha}}
\end{equation}
where the integration over $z$ in the complex plane is goes over the closed contour around zero.
Substituting representation (\ref{4.23}) as well as eqs.(\ref{4.3}) and  (\ref{4.5})-(\ref{4.7}) 
into eq.(\ref{4.1}) we get (s.f. eq.(\ref{4.8}))
\begin{eqnarray}
 \nonumber
\tilde{Z}(N; t) &=& N! \, \oint_{{\cal C}}\frac{dz}{2\pi i \, z} \, z^{N} 
\sum_{M=1}^{\infty} \; \frac{(-1)^{M}}{M!} \;
\prod_{\alpha=1}^{M}
\Biggl[
\int\int_{-\infty}^{+\infty} \frac{dy_{\alpha} dp_{\alpha}}{2\pi}
\Ai\bigl(y_{\alpha} + p_{\alpha}^{2}\bigr)
\sum_{n_{\alpha}=1}^{\infty} (-1)^{n_{\alpha}-1} \; \kappa^{n_{\alpha}} \, z^{-n_{\alpha}}
\exp\bigl\{\lambda(t) \, y_{\alpha} \, n_{\alpha}\bigr\} 
\Biggr] \times
\\
 \label{4.24}
\\
\nonumber
&\times&
\det\Biggl[
   \frac{1}{\lambda(t) \, n_{\alpha} - i p_{\alpha}
          + \lambda(t) \, n_{\beta} + ip_{\beta}} \Biggr]_{\alpha,\beta=1,...M}
\end{eqnarray}
In this way the replica partition function $\tilde{Z}(N; t)$ can be represented in the form 
of the integral of the Fredholm determinant (s.f. (\ref{4.10})-(\ref{4.14})):
\begin{equation}
 \label{4.25}
\tilde{Z}(N; t) \; = \;  N! \, \oint_{{\cal C}}\frac{dz}{2\pi i \, z} \, z^{N} \,
\exp\bigl\{-G_{t}(z) \bigr\}
\end{equation}
where
\begin{equation}
 \label{4.26}
G_{t}(z) \; = \;  \sum_{M=1}^{\infty} \frac{1}{M} \; \mbox{Tr} \; \hat{K}_{t}^{M}(z)
\end{equation}
with
\begin{equation}
\mbox{Tr} \, \hat{K}_{t}^{M}(z) = 
\prod_{\alpha=1}^{M}
\Biggl[
\int_{0}^{\infty} d\omega_{\alpha} 
\int_{-\infty}^{\infty} dy
\int_{-\infty}^{+\infty} \frac{dp}{2\pi}  
\Ai(y + p^{2} + \omega_{\alpha} + \omega_{\alpha-1}) \; 
\exp\bigl\{ip (\omega_{\alpha} - \omega_{\alpha-1} )\bigr\} \; 
S_{t}(z, y) 
\Biggr]
 \label{4.27}
\end{equation}
and
\begin{equation}
 \label{4.28}
S_{t}(z, y) \; = \; 
\sum_{n=1}^{\infty} (-1)^{n-1} \kappa^{n} \, z^{-n_{\alpha}}
\exp\bigl\{\lambda(t) \, y  \, n \bigr\} 
\; = \; 
\frac{\kappa \, z^{-1} \, \exp\bigl\{\lambda(t) \, y \bigr\}}{
      1 \; + \; \kappa \, z^{-1} \, \exp\bigl\{\lambda(t) \, y \bigr\} }
\end{equation}
Now, the above expression for the replica partition function $\tilde{Z}(N; t)$, eqs.(\ref{4.25})-(\ref{4.28}),
we analytically continue to arbitrary non-integer values of the replica parameter $N$
and redefine $N \; = \; s/\lambda(t)$ (see the discussion in Chapter II). Besides, let us deform the 
contour of integration ${\cal C}$ to the configuration shown in Fig.1 so that
$z \, = \, \rho \, \exp\{ i \theta\}$, where $0 < \rho < +\infty$, while
$\theta = 0$ in the upper branch of the new contour
and $\theta = 2\pi$ in the lower branch. In this way we get:
$N! = \Gamma(1+N) \, \to \, \Gamma\bigl(1 + s/\lambda\bigr)$; 
the function $G_{t}(z) \, \to G_{t}(\rho)$ and it takes the same values at the upper 
and at the lower branch of the contour;
on the other hand, $z^{N} \, \to \, \rho^{s/\lambda}$ at the upper branch while 
$z^{N} \, \to \rho^{s/\lambda} \exp\{2\pi i s/\lambda\}$ at the lower branch. Thus instead of eq.(\ref{4.25})
we obtain
\begin{figure}[h]
\begin{center}
   \includegraphics[width=6.0cm]{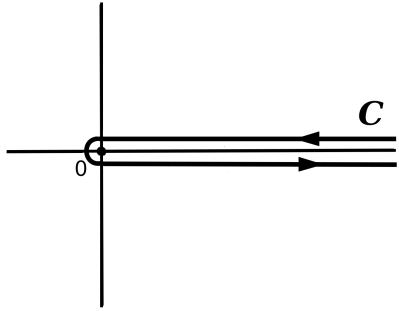}
\caption[]{The contour of integration which result in eq.(\ref{4.29})}
\end{center}
\label{figure1}
\end{figure}
\begin{equation}
 \label{4.29}
\tilde{Z}\Bigl(\frac{s}{\lambda(t)}; t\Bigr) \; = \; 
\frac{\Gamma\bigl(1+s/\lambda(t)\bigr)}{2\pi i} \,
\Biggl[
\int_{+\infty}^{0} d\rho \, \rho^{s/\lambda(t)} \, \exp\bigl\{-G_{t}(\rho) \bigr\} \; + \; 
\int_{0}^{+\infty} d\rho \, \rho^{s/\lambda(t)} \exp\{2\pi i s/\lambda(t)\} \, \exp\bigl\{-G_{t}(\rho) \bigr\}
\Biggr]
\end{equation}
Substituting $\rho \, = \, \exp\{\lambda(t) \, \xi\}$ and reorganizing the integration terms we gets
\begin{equation}
 \label{4.30}
\tilde{Z}\Bigl(\frac{s}{\lambda(t)}; t\Bigr) \; = \; 
- \Gamma\bigl(1+s/\lambda(t)\bigr) \; \frac{\lambda(t)}{2\pi i} \,
\Bigl[1 - \exp\bigl\{2\pi i s/\lambda(t)\bigr\} \Bigr] \, 
\int_{-\infty}^{+\infty} \, d\xi \, \exp\bigl\{ s \, \xi \, - \, G_{t}(\xi) \bigr\}
\end{equation}
where the function $G_{t}(\xi)$ is defined in eqs.(\ref{4.26})-(\ref{4.27}) with
\begin{equation}
 \label{4.31}
S_{t}(z, y) \; \to \; S_{t}(\xi, y) \; = \; 
\frac{\kappa \, \exp\bigl\{\lambda(t) \, (y-\xi) \bigr\}}{
      1 \; + \; \kappa \,  \exp\bigl\{\lambda(t) \, (y-\xi) \bigr\} }
\end{equation}
Performing the inverse Laplace transformation, eq.(\ref{2.11}), for the free energy probability distribution function
we get 
\begin{equation}
 \label{4.32}
P_{t}(f) \; = \; - \Gamma\bigl(1+s/\lambda(t)\bigr) \; \frac{\lambda(t)}{2\pi i} \,
\int_{-\infty}^{+\infty} d\xi \int_{-i\infty}^{+i\infty} \frac{ds}{2\pi i} \, 
\Bigl[1 - \exp\bigl\{2\pi i s/\lambda(t)\bigr\} \Bigr] \, 
\exp\bigl\{ s \, (\xi+f) \, - \, G_{t}(\xi) \bigr\}
\end{equation}
Finally, taking the limit $t \to \infty$ (so that $\lambda(t) \to \infty$) we obtain
\begin{eqnarray}
 \nonumber
\lim_{t\to\infty} P_{t}(f) \, \equiv \, P_{GUE}(f) &=& 
\int_{-\infty}^{+\infty} d\xi \int_{-i\infty}^{+i\infty} \frac{ds}{2\pi i} \, 
s \,  \exp\bigl\{ s \, (\xi+f) \bigr\} \; 
\exp\bigl\{ - \, G_{*}(\xi) \bigr\}
\\
\nonumber
\\
\nonumber
&=&
\int_{-\infty}^{+\infty} d\xi \; \delta'(\xi + f) \exp\bigl\{ - \, G_{*}(\xi) \bigr\}
\
\nonumber
\\
&=&
-\frac{\partial}{\partial f} \; \exp\bigl\{ - \, G_{*}(-f) \bigr\}
\label{4.33}
\end{eqnarray}
where, according to eq.(\ref{4.26}) and (\ref{4.27}),
\begin{equation}
 \label{4.34}
G_{*}(-f) \; = \;  \sum_{M=1}^{\infty} \frac{1}{M} \; \mbox{Tr} \; \hat{K}_{A}^{M}
\end{equation}
with
\begin{equation}
\mbox{Tr} \, \hat{K}_{A}^{M} \equiv \lim_{t\to\infty}
\prod_{\alpha=1}^{M}
\Biggl[
\int_{0}^{\infty} d\omega_{\alpha} 
\int_{-\infty}^{\infty} dy
\int_{-\infty}^{+\infty} \frac{dp}{2\pi}  
\Ai(y + p^{2} + \omega_{\alpha} + \omega_{\alpha-1}) \; 
\exp\bigl\{ip (\omega_{\alpha} - \omega_{\alpha-1} )\bigr\} \; 
S_{t}(z, y) 
\Biggr]
 \label{4.35}
\end{equation}
Using  eq.(\ref{4.31}) we find
\begin{equation}
 \label{4.36}
\lim_{t\to\infty} S_{t}(-f, y) \; = \; \theta(y + f)
\end{equation}
Thus, according eqs.(\ref{4.33})-(\ref{4.36})  in the limit $t\to\infty$ 
the free energy probability density function is
\begin{equation}
 \label{4.37}
P_{GUE}(f) \; = \; -\frac{\partial}{\partial f} \; 
\exp\Biggl\{- \sum_{M=1}^{\infty} \frac{1}{M} \; \mbox{Tr} \; \hat{K}_{A}^{M} \Biggr\}
\end{equation}
where
\begin{equation}
 \label{4.38}
 \mbox{Tr} \; \hat{K}_{A}^{M} \; = \; 
\prod_{\alpha=1}^{M}
\Biggl[
\int_{0}^{\infty} d\omega_{\alpha} 
\int_{-\infty}^{\infty} dy
\int_{-\infty}^{+\infty} \frac{dp}{2\pi}  
\Ai(y + p^{2} + \omega_{\alpha} + \omega_{\alpha-1}) \; 
\exp\bigl\{ip (\omega_{\alpha} - \omega_{\alpha-1} )\bigr\} \; 
\theta(y + f) 
\Biggr]
\end{equation}
One can easily see that the above expression coincides with the one in (\ref{4.16}) and (\ref{4.19})
which defines the Fredholm determinant, eq.(\ref{4.21}).


\vspace{10mm}

\section{GOE Tracy-Widom distribution in the model with free boundary conditions}

\newcounter{5}
\setcounter{equation}{0}
\renewcommand{\theequation}{5.\arabic{equation}}

In this Chapter we consider the system in which the polymer is fixed
at the origin, $\phi(0)=0$ and it is free at $\tau = t$.
In other words, for a given realization of the random potential
$V$ the partition function of this system is:
\begin{equation}
\label{5.1}
   Z(t) = \int_{-\infty}^{+\infty} dx \; Z(x; t) \; = \; \exp\{-\beta F(t)\}
\end{equation}
where
\begin{equation}
\label{5.2}
   Z(x; t) = \int_{\phi(0)=0}^{\phi(t)=x}
              {\cal D} \phi(\tau)  \;  \exp\bigl\{-\beta H[\phi; V]\bigr\}
\end{equation}
is the partition function of the system with the fixed boundary conditions,
$\phi(0)=0$ and $\phi(t)=x$. Here $F = f_{0} t \, + \, c \, t^{1/3}\, f$ 
 is the total free energy (see discussion in Chapter I and eq.(\ref{1.8})
and the Hamiltonian $H[\phi; V]$ is given in eq.(\ref{1.1}). 
In this Chapter it will be shown that unlike the system with fixed 
boundary conditions (considered in Chapter IV) the probability distribution 
of the fluctuating part of the free energy $W(f)$, eq.(\ref{2.5}), in the present system
is the Gaussian orthogonal ensemble (GOE) Tracy-Widom distribution 
\cite{TW2,LeDoussal2,goe}. Namely, in the limit limit, $t \to \infty$,
this function is equal to the Fredholm
determinant
\begin{equation}
\label{5.3}
W_{GOE}(f) \; = \; \det(1 - \hat{K}) \; \equiv \; F_{1}(-f)
\end{equation}
where $\hat{K}$ is the integral operator on $[-f, \, +\infty)$ with the kernel \cite{Ferrari-Spohn}
\begin{equation}
 \label{5.4}
K(\omega, \omega') \; = \; \Ai(\omega + \omega' - f) \; ; \; \; \;  \; \; \;  \; \; \;
(\omega, \omega' \; > 0)
\end{equation}
Explicitly \cite{TW2},
\begin{equation}
 \label{5.5}
F_{1}(s) \; = \;
\exp\Biggl[-\frac{1}{2}\int_{s}^{+\infty} d\xi \; (\xi-s) \; q^{2}(\xi) \;
-\frac{1}{2}\int_{s}^{+\infty} d\xi \; q(\xi)\Biggr]
\end{equation}
where $q(\xi)$ is the solution of the Panlev\'e II differential equation,
$q''(\xi)=\xi q(\xi)+2 q^{3}(\xi)$,
with the boundary condition  $q(\xi\to+\infty)=\Ai(\xi)$.

It should be noted that derivation given below is rather technical and  
the purpose of this Chapter is not only the final result (which is well known anyway)
but the demonstration of the method and  new technical tricks used in the derivation.
The most cumbersome technical parts of the calculations are moved to the Appendix E.

\vspace{5mm}

Later on we will see that the integration over $x$ in the definition of the
partition function, eq.(\ref{5.1}), requires proper regularization
at both limits $\pm\infty$. For that reason it is convenient to represent it
in the form of two contributions:
\begin{equation}
\label{5.6}
Z (t)\; = \;
\int_{-\infty}^{0} dx \; Z(x; t) \; + \;
\int_{0}^{+\infty} dx \; Z(x; t) \; \equiv \;
Z_{(-)}(t) \; + \; Z_{(+)}(t)
\end{equation}
Correspondingly, following the same route as in Chapter III, for the replica partition function 
$\overline{Z^{N}(t)} = Z(N; t)$ instead of eq.(\ref{3.11}) we get
\begin{equation}
 \label{5.7}
Z(N; t) \; = \;  \sum_{K,L=0}^{N} \frac{N!}{K! \, L!} \, \delta_{K+L, \, N}
\int_{-\infty}^{0} dx_{1}...dx_{K}
\int_{0}^{+\infty} dy_{1}...dy_{L}
\Psi(x_{1},...,x_{K},y_{L},...,y_{1} ; t)
\end{equation}
where
\begin{equation}
\label{5.8}
\Psi(x_{1},...,x_{K},y_{L},...,y_{1} ; t) \; = \;
\overline{Z(x_{1}; t) \, Z(x_{2}; t) \, ... \, Z(x_{K}; t) \, Z(y_{L}; t) \, Z(y_{L-1}; t) \, ... \, Z(y_{1}; t)}
\end{equation}
Explicitly, in terms of the Bethe ansatz solution the above wave function is given in 
eqs.(\ref{3.15}) and (\ref{3.14}). 
Next, repeating the calculations of Chapter III, instead of eq.(\ref{4.1}) we eventually get
\begin{eqnarray}
 \nonumber
 \tilde{Z}(N; t) &=& N! \sum_{K,L=0}^{N}  \, \delta_{K+L, \, N} 
\sum_{M=1}^{N} \, \frac{1}{M!}
\prod_{\alpha=1}^{M} \Biggl[
\sum_{n_{\alpha}=1}^{\infty} 
\int_{-\infty}^{+\infty} \frac{dq_{\alpha} \, \kappa^{n_{\alpha}}}{2\pi} 
\exp\Bigl\{
-\frac{t}{2\beta} \, n_{\alpha} q_{\alpha}^{2} \, + \, 
\frac{\kappa^{2}\, t}{24\beta} n_{\alpha}^{3}
\Bigr\}
\Biggr]
{\boldsymbol \delta}\Bigl(\sum_{\alpha=1}^{M} n_{\alpha} \; , \;  N\Bigr) \times
\\
 \label{5.9}
\\
\nonumber
&\times&
\det\Biggl[
   \frac{1}{\frac{1}{2}\kappa n_{\alpha} - i q_{\alpha}
          + \frac{1}{2}\kappa n_{\beta} + iq_{\beta}}\Biggr]_{\alpha,\beta=1,...M}
I_{K,L} ({\bf q}, {\bf n})
\end{eqnarray}
where
\begin{eqnarray}
\nonumber
I_{K,L} ({\bf q}, {\bf n}) &=& \frac{1}{K! \, L!}
\sum_{{\cal P}\in S_{N}}   \;
\int_{-\infty}^{0} dx_{1}...dx_{K}
\int_{0}^{+\infty} dy_{1}...dy_{L} \; 
\exp\Biggl[
i \sum_{a=1}^{K} \Bigl(Q_{{\cal P}_{a}} - i\epsilon\Bigr) x_{a} \; + \; 
i \sum_{c=1}^{L} \Bigl(Q_{{\cal P}_{c}} - i\epsilon\Bigr) y_{c}
\Biggr]
\times
\
\nonumber
\\
&\times&
\prod_{a<b}^{K}
\Biggl[1 +i \kappa \frac{\sgn(x_{a}-x_{b})}{Q_{{\cal P}_a} - Q_{{\cal P}_b}} \Biggr] \;
\prod_{c<d}^{L}
\Biggl[1 +i \kappa \frac{\sgn(y_{c}-y_{d})}{Q_{{\cal P}_c} - Q_{{\cal P}_d}} \Biggr] \;
\prod_{a=1}^{K} \prod_{c=1}^{L}
\Biggl[1 +i \kappa \frac{\sgn(x_{a}-y_{c})}{Q_{{\cal P}_a} - Q_{{\cal P}_c}} \Biggr] 
\label{5.10}
\end{eqnarray}
Here the summation goes over all permutations $P$ of $N$ momenta $\{Q_{1}, ..., Q_{N}\}$
over $K+L = N$ particles $\{x_{1}, ..., x_{K}, y_{L}, ..., y_{1}\}$.
Note also that the integrations both over $x_{a}$'s and over $y_{c}$'s
in eq.(\ref{5.10}) require proper regularization at $-\infty$ and $+\infty$ correspondingly.
This is done in the standard way by introducing a supplementary parameter $\epsilon$
which will be set to zero in final results.
Due to the symmetry of the above expression with respect to  permutations among $x_{a}$'s and 
$y_{c}$'s it can be represented as follows
\begin{eqnarray}
\nonumber
I_{K,L} ({\bf q}, {\bf n}) &=&
\sum_{{\cal P}^{(K,L)}}  \sum_{{\cal P}^{(K)}} \sum_{{\cal P}^{(L)}} \;
\prod_{a=1}^{K} \prod_{c=1}^{L}
\Biggl[
\frac{Q_{{\cal P}_a^{(K)}} - Q_{{\cal P}_c^{(L)}} - i \kappa}{Q_{{\cal P}_a^{(K)}} - Q_{{\cal P}_c^{(L)}}}
\Biggr]
\times
\prod_{a<b}^{K}\Biggl[\frac{Q_{{\cal P}_a^{(K)}} - Q_{{\cal P}_b^{(K)}}  - i \kappa }{Q_{{\cal P}_a^{(K)}} - Q_{{\cal P}_b^{(K)}}}\Biggr]
\times
\prod_{c<d}^{L}\Biggl[\frac{Q_{{\cal P}_c^{(L)}} - Q_{{\cal P}_d^{(L)}}  + i \kappa }{Q_{{\cal P}_c^{(L)}} - Q_{{\cal P}_d^{(L)}}}\Biggr]
\times
\\
\nonumber
\\
\nonumber
&\times&
\int_{-\infty < x_{1} \leq ... \leq x_{K}\leq 0} dx_{1} ... dx_{K} \;
\exp\Bigl[i \sum_{a=1}^{K} (Q_{{\cal P}_{a}^{(K)}}-i\epsilon) x_{a} \Bigr]
\\
\nonumber
\\
&\times&
\int_{0 \leq y_{L} \leq ... \leq y_{1} < +\infty} dy_{L} ... dy_{1} \;
\exp\Bigl[i \sum_{c=1}^{L} (Q_{{\cal P}_{c}^{(L)}}+i\epsilon) y_{c} \Bigr]
\label{5.11}
\end{eqnarray}
Here the summation over all permutations ${\cal P}$ 
is split  into three parts: the permutations ${\cal P}^{(K)}$
of $K$ momenta (taken at random out of the total list $\{Q_{1}, ..., Q_{K+L}\}$)
over $K$ "negative" particles $\{x_{1}, ..., x_{K}\}$, the permutations ${\cal P}^{(L)}$
of the remaining $L$ momenta over $L$ "positive" particles $\{y_{L}, ..., y_{1}\}$, and
finally the permutations ${\cal P}^{(K,L)}$ (or the exchange) of the
momenta between the group $"K"$ and the group $"L"$.

Performing the integrations in eq.(\ref{5.11}) we get:
\begin{eqnarray}
\nonumber
I_{K,L} ({\bf q}, {\bf n}) &=& i^{-(K+L)}
\sum_{{\cal P}^{(K,L)}}  \; \;
\prod_{a=1}^{K} \prod_{c=1}^{L}
\Biggl[
\frac{Q_{{\cal P}_a^{(K)}} - Q_{{\cal P}_c^{(L)}} - i \kappa}{Q_{{\cal P}_a^{(K)}} - Q_{{\cal P}_c^{(L)}}}
\Biggr]
\times
\\
\nonumber
\\
\nonumber
&\times&
\sum_{{\cal P}^{(K)}} \; \;
\frac{1}{Q^{(-)}_{{\cal P}_{1}^{(K)}} \bigl(Q^{(-)}_{{\cal P}_{1}^{(K)}} + Q^{(-)}_{{\cal P}_{2}^{(K)}}\bigr)... \bigl(Q^{(-)}_{{\cal P}_{1}^{(K)}} + ... + Q^{(-)}_{{\cal P}_{K}^{(K)}}\bigr)}
\prod_{a<b}^{K}\Biggl[\frac{Q^{(-)}_{{\cal P}_a^{(K)}} - Q^{(-)}_{{\cal P}_b^{(K)}}  - i \kappa }{Q^{(-)}_{{\cal P}_a^{(K)}} - Q^{(-)}_{{\cal P}_b^{(K)}}}\Biggr]
\times
\\
\nonumber
\\
&\times&
\sum_{{\cal P}^{(L)}} \; \;
\frac{(-1)^{L}}{Q^{(+)}_{{\cal P}_{1}^{(L)}} \bigl(Q^{(+)}_{{\cal P}_{1}^{(L)}} + Q^{(+)}_{{\cal P}_{2}^{(L)}}\bigr)... \bigl(Q^{(+)}_{{\cal P}_{1}^{(L)}} + ... + Q^{(+)}_{{\cal P}_{L}^{(L)}}\bigr)}
\prod_{c<d}^{L}\Biggl[\frac{Q^{(+)}_{{\cal P}_c^{(L)}} - Q^{(+)}_{{\cal P}_d^{(L)}}  + i \kappa }{Q^{(+)}_{{\cal P}_c^{(L)}} - Q^{(+)}_{{\cal P}_d^{(L)}}}\Biggr]
\label{5.12}
\end{eqnarray}
where
\begin{equation}
\label{5.13}
Q^{(\pm)}_a \; \equiv \; Q_{a} \pm i\epsilon
\end{equation}
Using the  Bethe ansatz combinatorial identity \cite{LeDoussal2} (Appendix D),
\begin{equation}
 \label{5.14}
\sum_{P\in S_{N}} \frac{1}{Q_{P_{1}} (Q_{P_{1}} + Q_{P_{2}})... (Q_{P_{1}} + ... + Q _{P_{N}})}
\prod_{a<b}^{N}\Biggl[\frac{Q_{P_a} - Q_{P_b}  - i \kappa }{Q_{P_a} - Q_{P_b}}\Biggr] \; = \;
\frac{1}{\prod_{a=1}^{N} Q_{a}} \;
\prod_{a<b}^{N}\Biggl[\frac{Q_{a} + Q_{b}  + i \kappa }{Q_{a} + Q_{b}}\Biggr]
\end{equation}
(where the summation goes over all permutations of $N$ momenta $\{Q_{1}, ..., Q_{N}\}$) we get:
\begin{eqnarray}
\nonumber
I_{K,L} ({\bf q}, {\bf n}) &=& i^{-(K+L)}
\sum_{{\cal P}^{(K,L)}}  \; \;
\prod_{a=1}^{K} \prod_{c=1}^{L}
\Biggl[
\frac{Q_{{\cal P}_a^{(K)}} - Q_{{\cal P}_c^{(L)}} - i \kappa}{Q_{{\cal P}_a^{(K)}} - Q_{{\cal P}_c^{(L)}}}
\Biggr]
\times
\\
\nonumber
\\
&\times&
\frac{1}{\prod_{a=1}^{K} Q^{(-)}_{{\cal P}_{a}^{(K)}} }
\prod_{a<b}^{K}\Biggl[\frac{Q^{(-)}_{{\cal P}_a^{(K)}} + Q^{(-)}_{{\cal P}_b^{(K)}}  + i \kappa }{Q^{(-)}_{{\cal P}_a^{(K)}} + Q^{(-)}_{{\cal P}_b^{(K)}}}\Biggr]
\times
\frac{(-1)^{L}}{\prod_{c=1}^{L} Q^{(+)}_{{\cal P}_{c}^{(L)}} }
\prod_{c<d}^{L}\Biggl[\frac{Q^{(+)}_{{\cal P}_c^{(L)}} + Q^{(+)}_{{\cal P}_d^{(L)}}  - i \kappa }{Q^{(+)}_{{\cal P}_c^{(L)}} + Q^{(+)}_{{\cal P}_d^{(L)}}}\Biggr]
\label{5.15}
\end{eqnarray}

Substituting now eq.(\ref{5.9}) with $I_{K,L} ({\bf q}, {\bf n})$ given in eq.(\ref{5.15}) into the expression for the probability distribution 
function, eq.(\ref{2.19}), and taking into account that $N \; = \; K + L$
we obtain
\begin{eqnarray}
 \nonumber
W(f) &=& \lim_{t\to\infty} \sum_{K,L=0}^{\infty} (-1)^{K+L} \sum_{M=0}^{\infty} \frac{(-1)^{M}}{M!}
\prod_{\alpha=1}^{M} \Biggl[
\sum_{n_{\alpha}=1}^{\infty} \, (-1)^{n_{\alpha}-1} 
\int_{-\infty}^{+\infty} \frac{dq_{\alpha} \, \kappa^{n_{\alpha}}}{2\pi} 
\exp\Bigl\{
\lambda(t) n_{\alpha} \, f 
-\frac{t}{2\beta} \, n_{\alpha} q_{\alpha}^{2} \, + \, 
\frac{\kappa^{2}\, t}{24\beta} n_{\alpha}^{3}
\Bigr\}
\Biggr] \times
\\
 \label{5.16}
\\
\nonumber
&\times&
\det\Biggl[
  \frac{1}{\frac{1}{2}\kappa n_{\alpha} - i q_{\alpha}
          + \frac{1}{2}\kappa n_{\beta} + iq_{\beta}}\Biggr]_{\alpha,\beta=1,...M} 
 I_{K,L} ({\bf q}, {\bf n})  \; \, 
{\boldsymbol \delta}\Bigl(\sum_{\alpha=1}^{M} n_{\alpha} \; , \;  K+L\Bigr)
\end{eqnarray}

Further simplification comes due to the special 
Bethe ansatz product structure of the factor $I_{K,L}({\bf q}, {\bf n})$, eq.(\ref{5.15}). 
One can easily see that in these products, according to the definition (\ref{3.12}),
the momenta $Q_{a}$ belonging
to the same cluster must be {\it ordered}. In other words,
if we consider the momenta, eq.(\ref{3.12}), of a cluster $\alpha$, 
$\{q_{1}^{\alpha}, q_{2}^{\alpha}, ..., q_{n_{\alpha}}^{\alpha}\}$, 
the permutation of any two momenta $q_{r}^{\alpha}$
and $q_{r'}^{\alpha}$ of this {\it ordered} set gives zero contribution to the factor $I_{K,L}({\bf q}, {\bf n})$.
Thus, in order to perform the summation over the permutations ${\cal P}^{(K,L)}$
in eq.(\ref{5.15}) it is sufficient to split the momenta of each cluster into two parts:
$\{q_{1}^{\alpha}, ...,  q_{m_{\alpha}}^{\alpha} ||
q_{m_{\alpha}+1}^{\alpha}..., q_{n_{\alpha}}^{\alpha}\}$, where $m_{\alpha} = 0, 1, ..., n_{\alpha}$ and 
the momenta $q_{1}^{\alpha}, ...,  q_{m_{\alpha}}^{\alpha}$ belong to the 
sector $"K"$, while the momenta $q_{m_{\alpha}+1}^{\alpha}..., q_{n_{\alpha}}^{\alpha}$ 
belong to the sector $"L"$. 

It is convenient to  introduce the numbering of the momenta
of the sector $"L"$ in the reversed order:
\begin{eqnarray}
\nonumber
q_{n_{\alpha}}^{\alpha} &\to&  {q^{*}}_{1}^{\alpha}
\\
\nonumber
q_{n_{\alpha}-1}^{\alpha} &\to&  {q^{*}}_{2}^{\alpha}
\\
\nonumber
&........&
\\
q_{m_{\alpha}+1}^{\alpha} &\to&  {q^{*}}_{s_{\alpha}}^{\alpha}
\label{5.17}
\end{eqnarray}
where $m_{\alpha} + s_{\alpha} = n_{\alpha}$ and (s.f. eq.(\ref{3.12}))
\begin{equation}
\label{5.18}
{q^{*}}_{r}^{\alpha} \; = \; q_{\alpha} + \frac{i \kappa}{2} (n_{\alpha} + 1 - 2r)
\; = \; q_{\alpha} + \frac{i \kappa}{2} (m_{\alpha} + s_{\alpha} + 1 - 2r)
\end{equation}
By definition, the integer parameters $\{m_{\alpha}\}$ and $\{s_{\alpha}\}$
fulfill the global constrains
\begin{eqnarray}
\label{5.19}
\sum_{\alpha=1}^{M} m_{\alpha} &=& K
\\
\nonumber
\\
\sum_{\alpha=1}^{M} s_{\alpha} &=& L
\label{5.20}
\end{eqnarray}
In this way the summation over permutations ${\cal P}^{(K,L)}$
in eq.(\ref{5.15}) is changed by the summations over the integer parameters
$\{m_{\alpha}\}$ and $\{s_{\alpha}\}$:
\begin{equation}
\label{5.21}
\sum_{{\cal P}^{(K,L)}} \; \bigl( ... \bigr) \; \to \;
\prod_{\alpha=1}^{M}
\Biggl[
\sum_{m_{\alpha}+s_{\alpha} \geq 1}^{\infty} \;
{\boldsymbol \delta}\Bigl(m_{\alpha}+s_{\alpha} \; , \;  n_{\alpha}\Bigr)
\Biggr] \;
{\boldsymbol \delta}\Bigl(\sum_{\alpha=1}^{M} m_{\alpha}\; , \; K\Bigr) \;
{\boldsymbol \delta}\Bigl(\sum_{\alpha=1}^{M} s_{\alpha}\; , \; L\Bigr)
\;
\bigl( ... \bigr)
\end{equation}
which allows to lift the summations over $K$, $L$, and $\{n_{\alpha}\}$
in eq.(\ref{5.16}):
\begin{eqnarray}
 \nonumber
W(f) &=& \lim_{t\to\infty} \sum_{M=0}^{\infty} \frac{(-1)^{M}}{M!}
\prod_{\alpha=1}^{M} \Biggl[
\sum_{m_{\alpha}+s_{\alpha}\geq 1}^{\infty} \, (-1)^{m_{\alpha}+s_{\alpha}-1} \, \kappa^{m_{\alpha}+s_{\alpha}}
\int_{-\infty}^{+\infty} \frac{dq_{\alpha} }{2\pi} 
\\
\nonumber
\\
\nonumber
&\times&
\exp\Bigl\{
\lambda(t) (m_{\alpha}+s_{\alpha}) \, f 
-\frac{t}{2\beta} \, (m_{\alpha}+s_{\alpha}) q_{\alpha}^{2} \, + \, 
\frac{\kappa^{2}\, t}{24\beta} (m_{\alpha}+s_{\alpha})^{3}
\Bigr\}
\Biggr] 
\\
 \label{5.22}
\\
\nonumber
&\times&
\det\Biggl[
  \frac{1}{\frac{1}{2}\kappa (m_{\alpha}+s_{\alpha}) - i q_{\alpha}
          + \frac{1}{2}\kappa (m_{\beta}+s_{\beta}) + iq_{\beta}}\Biggr]_{\alpha,\beta=1,...M} 
 I_{K,L} ({\bf q}; {\bf m}, {\bf s}) 
\end{eqnarray}
Next, after somewhat painful algebra 
the factor $I_{K,L} ({\bf q}; {\bf m}, {\bf s})$, eq.(\ref{5.15}), can be represented in terms of the product 
of the Gamma functions (see Appendix E). Then, redefining the momenta, $q_{\alpha} = \kappa \, p_{\alpha} /(2\lambda)$ 
and performing the same transformations as in eqs.(\ref{4.5})-(\ref{4.7})
we get (s.f. eq.(\ref{4.8}))
\begin{eqnarray}
 \nonumber
W(f) &=& \lim_{t\to\infty}
\sum_{M=1}^{\infty} \; \frac{(-1)^{M}}{M!} \;
\prod_{\alpha=1}^{M}
\Biggl[
\int\int_{-\infty}^{+\infty} \frac{dy_{\alpha} dp_{\alpha}}{2\pi}
\Ai\bigl(y_{\alpha} + p_{\alpha}^{2} - f \bigr)
\\
\nonumber
\\
\nonumber
&\times&
\sum_{m_{\alpha}+s_{\alpha}\geq 1}^{\infty} (-1)^{m_{\alpha}+s_{\alpha}-1}
\exp\bigl\{\lambda(t) (m_{\alpha}+s_{\alpha})y_{\alpha}\bigr\} \;
{\cal G} \Bigl(\frac{p_{\alpha}}{\lambda(t)}, \; m_{\alpha}, \; s_{\alpha}\Bigr) \;
\Biggr]
\\
\nonumber
\\
&\times&
\det\Biggl[
  \frac{1}{\lambda(t)(m_{\alpha}+s_{\alpha}) - i p_{\alpha}
         + \lambda(t)(m_{\beta}+s_{\beta}) + ip_{\beta}}\Biggr]_{\alpha,\beta=1,...M} 
\times \prod_{1\leq\alpha<\beta}^{M} \;
{\cal G}_{\alpha\beta}\Bigl(\frac{{\bf p}}{\lambda(t)}, \; {\bf m}, \; {\bf s}\Bigr)
\label{5.23}
\end{eqnarray}
where (see eq.(\ref{E.17})
\begin{equation}
\label{5.24}
{\cal G}\bigl(\frac{p_{\alpha}}{\lambda}, m_{\alpha}, s_{\alpha}\bigr) \; = \;
\frac{
\Gamma\Bigl(
s_{\alpha} +  i {p_{\alpha}}^{(-)}/\lambda 
\Bigr) \,
\Gamma\Bigl(
m_{\alpha} -  i {p_{\alpha}}^{(+)}/\lambda 
\Bigr) \,
\Gamma\bigl(1 + m_{\alpha} + s_{\alpha}\bigr)}{
2^{(m_{\alpha} + s_{\alpha})}
\Gamma\Bigl(
m_{\alpha} + s_{\alpha} +  i {p_{\alpha}}^{(-)}/\lambda 
\Bigr) \,
\Gamma\Bigl(
m_{\alpha} + s_{\alpha} -  i {p_{\alpha}}^{(+)}/\lambda 
\Bigr) \,
\Gamma\bigl(1 + m_{\alpha}\bigr) \Gamma\bigl(1 + s_{\alpha}\bigr)}
\end{equation}
and the explicit expression for the factor ${\cal G}_{\alpha\beta}\bigl({\bf q}, \; {\bf m}, \; {\bf s}\bigr)$
is given in eq.(\ref{E.18}).

\vspace{5mm}

The crucial point of the further calculations is 
taking the  limit $t \to \infty$. 
Reorganizing the terms, the expression in eq.(\ref{5.23}) can be represented as follows
\begin{equation}
 \label{5.25}
W(f) \; = \;
\sum_{M=0}^{\infty} \; \frac{(-1)^{M}}{M!} \;
\prod_{\alpha=1}^{M}
\Biggl[
\int\int_{-\infty}^{+\infty} \frac{dy_{\alpha} dp_{\alpha}}{2\pi}
\Ai\bigl(y_{\alpha} + p_{\alpha}^{2} - f \bigr)
\Biggr]
\; {\cal S}_{M}\bigl({\bf p}, {\bf y}\bigr)
\end{equation}
where
\begin{eqnarray}
\nonumber
 {\cal S}_{M}\bigl({\bf p}, {\bf y}\bigr) &=&
\lim_{t\to\infty}
\prod_{\alpha=1}^{M}
\Biggl[
\sum_{m_{\alpha}+s_{\alpha}\geq 1}^{\infty} (-1)^{m_{\alpha}+s_{\alpha}-1}
\exp\bigl\{
\lambda(t) m_{\alpha} y_{\alpha} +
\lambda(t) s_{\alpha} y_{\alpha} 
\bigr\}
\Biggr]
\times
\prod_{\alpha=1}^{M}
\Biggl[
{\cal G} \Bigl(\frac{p_{\alpha}}{\lambda(t)}, \; m_{\alpha}, \; s_{\alpha}\Bigr) \;
\Biggr] 
\times
\\
\nonumber
\\
&\times&
\prod_{1\leq\alpha<\beta}^{M}
\Biggl[
{\cal G}_{\alpha\beta}\Bigl(\frac{{\bf p}}{\lambda(t)}, \; {\bf m}, \; {\bf s}\Bigr)
\Biggr] \;
\det\Biggl[
  \frac{1}{\lambda(t)(m_{\alpha}+s_{\alpha}) - i p_{\alpha}
         + \lambda(t)(m_{\beta}+s_{\beta}) + ip_{\beta}}\Biggr]_{\alpha,\beta=1,...M} 
\label{5.26}
\end{eqnarray}
\begin{figure}[h]
\begin{center}
   \includegraphics[width=6.0cm]{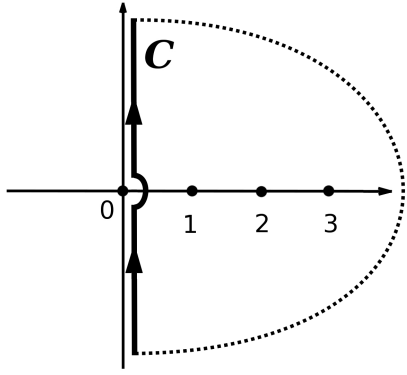}
\caption[]{The contour of integration in eq.(\ref{5.28})}
\end{center}
\label{figure2}
\end{figure}
In the limit $t\to\infty$ the summations over $\{m_{\alpha}\}$
and $\{s_{\alpha}\}$ in the above expression can performed according to the following algorithm.
Let us consider the sum of a general type:
\begin{equation}
\label{5.27}
R({\bf y}, {\bf p}) \; = \; \lim_{\lambda\to\infty} \prod_{\alpha=1}^{M}
\Biggl[
\sum_{n_{\alpha}=1}^{\infty} \; (-1)^{n_{\alpha} - 1}
\exp\{\lambda n_{\alpha} y_{\alpha}\}
\Biggr]\;
\Phi\Bigl({\bf p}, \; \frac{{\bf p}}{\lambda}, \; \lambda {\bf n}, \; {\bf n} \Bigr)
\end{equation}
where $\Phi$ is a "good" function which depend on the factors 
$p_{\alpha}$, $p_{\alpha}/\lambda$, $\lambda n_{\alpha}$ and $n_{\alpha}$.
One can easily see that the summations in eq.(\ref{5.27}) can be changed by 
of the integrations in the complex plane:
\begin{equation}
\label{5.28}
R({\bf y}, {\bf p}) \; = \; \lim_{\lambda\to\infty} \prod_{\alpha=1}^{M}
\Biggl[
\frac{1}{2i} \int_{{\cal C}} \frac{dz_{\alpha}}{\sin(\pi z_{\alpha})}
\exp\{\lambda z_{\alpha} y_{\alpha}\}
\Biggr]\;
\Phi\Bigl({\bf p}, \; \frac{{\bf p}}{\lambda}, \; \lambda {\bf z}, \; {\bf z} \Bigr)
\end{equation}
where the integration goes over the contour ${\cal C}$ shown in Fig.2, and it is assumed that 
the function $\Phi$ is such that there is no contribution from $\infty$. Indeed,
due to the sign alternating contributions of simple poles at integer $z = 1, 2, ...$, eq.(\ref{5.28}) 
reduces to eq.(\ref{5.27}). Then, redefining $z \to z/\lambda$, in the
limit $\lambda \to \infty$ we get:
\begin{eqnarray}
\nonumber
R({\bf y}, {\bf p}) &=&  \prod_{\alpha=1}^{M}
\Biggl[
\frac{1}{2\pi i} \int_{{\cal C}} \frac{dz_{\alpha}}{z_{\alpha}}
\exp\{z_{\alpha} y_{\alpha}\}
\Biggr]\;
\lim_{\lambda\to\infty}
\Phi\Bigl({\bf p}, \; \frac{{\bf p}}{\lambda}, \; {\bf z}, \; \frac{{\bf z}}{\lambda} \Bigr)
\\
\nonumber
\\
&=&
\prod_{\alpha=1}^{M}
\Biggl[
\frac{1}{2\pi i} \int_{{\cal C}} \frac{dz_{\alpha}}{z_{\alpha}}
\exp\{z_{\alpha} y_{\alpha}\}
\Biggr]\;
\Phi\Bigl({\bf p}, \; {\bf 0}, \; {\bf z}, \; {\bf 0} \Bigr)
\label{5.29}
\end{eqnarray}
Coming back to eq.(\ref{5.26}), the summations over $m_{\alpha}$ and $s_{\alpha}$ can
be represented as follows
\begin{equation}
\label{5.30}
\sum_{m_{\alpha}+s_{\alpha}\geq 1}^{\infty} (-1)^{m_{\alpha}+s_{\alpha}-1}
 =
\sum_{m_{\alpha}=1}^{\infty} (-1)^{m_{\alpha}-1} \delta(s_{\alpha}, 0) +
\sum_{s_{\alpha}=1}^{\infty} (-1)^{s_{\alpha}-1} \delta(m_{\alpha}, 0) -
\sum_{m_{\alpha}=1}^{\infty} (-1)^{m_{\alpha}-1}
\sum_{s_{\alpha}=1}^{\infty} (-1)^{s_{\alpha}-1}
\end{equation}
Thus in the integral representation, following the algorithm eqs.(\ref{5.27})-(\ref{5.29}),  we get
\begin{eqnarray}
\nonumber
{\cal S}_{M}\bigl({\bf p}, {\bf y}\bigr) &=&
\prod_{\alpha=1}^{M}
\Biggl[
\int\int_{{\cal C}}
\frac{d{z_{1}}_{\alpha}d{z_{2}}_{\alpha}}{(2\pi i)^{2}}
\Bigl(
\frac{2\pi i}{{z_{1}}_{\alpha}} \delta({z_{2}}_{\alpha}) +
\frac{2\pi i}{{z_{2}}_{\alpha}} \delta({z_{1}}_{\alpha}) -
\frac{1}{{z_{1}}_{\alpha}{z_{2}}_{\alpha}}
\Bigr)
\exp\bigl\{
{z_{1}}_{\alpha} y_{\alpha}  +
{z_{2}}_{\alpha} y_{\alpha}
\bigr\}
\Biggr]
\times
\\
\nonumber
\\
\nonumber
&\times&
\det 
\Bigl[
\frac{1}{{z_{1}}_{\alpha} + {z_{2}}_{\alpha} - i p_{\alpha} +
         {z_{1}}_{\beta} + {z_{2}}_{\beta} + i p_{\beta}}
\Bigr]_{(\alpha,\beta)=1,2,...,M}
\times
\\
\nonumber
\\
&\times&
\lim_{t\to\infty}
\Biggl\{
\prod_{\alpha=1}^{M}
\Biggl[
{\cal G} \Bigl(\frac{p_{\alpha}}{\lambda(t)}, \;\frac{{z_{1}}_{\alpha}}{\lambda(t)} , \;\frac{{z_{2}}_{\alpha}}{\lambda(t)} \Bigr) \;
\Biggr] \;
\prod_{1\leq\alpha<\beta}^{M}
\Biggl[
{\cal G}_{\alpha\beta}\Bigl(\frac{{\bf p}}{\lambda(t)}, \; \frac{{\bf z_{1}}}{\lambda(t)}, \; \frac{{\bf z_{2}}}{\lambda(t)}\Bigr)
\Biggr]
\Biggr\}
\label{5.31}
\end{eqnarray}
Taking into account the Gamma function properties,
$\Gamma(z)|_{|z|\to 0} = 1/z$ and
$\Gamma(1+z)|_{|z|\to 0} = 1$, for the factors ${\cal G}$, eq.(\ref{5.24}), and
${\cal G}_{\alpha\beta}$, eq.(\ref{E.18}), we obtain
\begin{equation}
\label{5.32}
\lim_{t\to\infty}
{\cal G}\Bigl(
\frac{p_{\alpha}}{\lambda(t)}, \;
\frac{{z_{1}}_{\alpha}}{\lambda(t)} , \;
\frac{{z_{2}}_{\alpha}}{\lambda(t)}
\Bigr)
\; = \;
\frac{
\bigl({z_{1}}_{\alpha}+{z_{2}}_{\alpha} + i p_{\alpha}^{(-)}\bigr)
\bigl({z_{1}}_{\alpha}+{z_{2}}_{\alpha} - i p_{\alpha}^{(+)}\bigr)}{
\bigl({z_{2}}_{\alpha} + i p_{\alpha}^{(-)}\bigr)
\bigl({z_{1}}_{\alpha} - i p_{\alpha}^{(+)}\bigr)}
\end{equation}
and
\begin{equation}
\label{5.33}
\lim_{t\to\infty}
{\bf G}
\Bigl(
\frac{{\bf p}}{\lambda(t)}, \; \frac{{\bf z}_{1}}{\lambda(t)}, \; \frac{{\bf z}_{2}}{\lambda(t)}\Bigr)
\; = \; 1
\end{equation}
Substituting eqs.(\ref{5.32}) and (\ref{5.33}) into eq.(\ref{5.31}) and then eq.(\ref{5.31}) into eq.(\ref{5.25})
we see that in the limit $t \to \infty$ the expression for the probability distribution function, eq.(\ref{5.25}),
takes the form of the Fredholm determinant
\begin{eqnarray}
 \nonumber
W(f) &=&
\sum_{M=0}^{\infty} \; \frac{(-1)^{M}}{M!} \;
\prod_{\alpha=1}^{M}
\Biggl[
\int\int_{-\infty}^{+\infty} \frac{dy_{\alpha} dp_{\alpha}}{2\pi}
\Ai\bigl(y_{\alpha} + p_{\alpha}^{2} - f \bigr)
\\
\nonumber
\\
\nonumber
&\times&
\int\int_{{\cal C}}
\frac{d{z_{1}}_{\alpha}d{z_{2}}_{\alpha}}{(2\pi i)^{2}}
\Bigl(
\frac{2\pi i}{{z_{1}}_{\alpha}} \delta({z_{2}}_{\alpha}) +
\frac{2\pi i}{{z_{2}}_{\alpha}} \delta({z_{1}}_{\alpha}) -
\frac{1}{{z_{1}}_{\alpha}{z_{2}}_{\alpha}}
\Bigr)
\Bigl(
1 + \frac{{z_{1}}_{\alpha}}{{z_{2}}_{\alpha} + i p_{\alpha}^{(-)}}
\Bigr)
\Bigl(
1 + \frac{{z_{2}}_{\alpha}}{{z_{1}}_{\alpha} - i p_{\alpha}^{(+)}}
\Bigr)
\times
\\
\nonumber
\\
\nonumber
&\times&
\exp\bigl\{ 
{z_{1}}_{\alpha} y_{\alpha}  + {z_{2}}_{\alpha} y_{\alpha} 
\bigr\}
\Biggr]
\det \Bigl[
\frac{1}{{z_{1}}_{\alpha} + {z_{2}}_{\alpha} - i p_{\alpha} +
{z_{1}}_{\beta} + {z_{2}}_{\beta} + i p_{\beta}}
\Bigr]_{(\alpha,\beta)=1,2,...,M}
\\
\nonumber
\\
&=&
\det\bigl[1 \, - \, \hat{K} \bigr]
\label{5.34}
\end{eqnarray}
with the kernel
\begin{eqnarray}
\nonumber
\hat{K}\bigl[({z_{1}}, \, {z_{2}}, \, p); ({z_{1}}', \, {z_{2}}', \, p')\bigr]
&=&
\int_{-\infty}^{+\infty} \frac{dy}{2\pi}
\Ai\bigl(y + p^{2} - f \bigr)
\Bigl(
\frac{2\pi i}{z_{1}} \delta(z_{2}) +
\frac{2\pi i}{z_{2}} \delta(z_{1}) -
\frac{1}{z_{1}{z_{2}}}
\Bigr)
\times
\\
\nonumber
\\
&\times&
\Bigl(1 + \frac{z_{1}}{z_{2} + i p^{(-)}} \Bigr)
\Bigl(1 + \frac{z_{2}}{z_{1} - i p^{(+)}} \Bigr)
\frac{\exp\bigl\{ z_{1} y  + z_{2} y \bigr\} }{
{z_{1}} + {z_{2}} - ip + {z_{1}}' + {z_{2}}' + ip'}
\label{5.35}
\end{eqnarray}
In the exponential representation of this determinant we get
\begin{equation}
 \label{5.36}
W(f) \; = \;
\exp\Bigl[-\sum_{M=1}^{\infty} \frac{1}{M} \; \mbox{Tr} \, \hat{K}^{M} \Bigr]
\end{equation}
where
\begin{eqnarray}
 \nonumber
\mbox{Tr} \, \hat{K}^{M} &=&
\prod_{\alpha=1}^{M}
\Biggl[
\int\int_{-\infty}^{+\infty} \frac{dy_{\alpha} dp_{\alpha}}{2\pi}
\Ai\bigl(y_{\alpha} + p_{\alpha}^{2} - f \bigr)
\times
\\
\nonumber
\\
\nonumber
&\times&
\int\int_{{\cal C}}
\frac{d{z_{1}}_{\alpha}d{z_{2}}_{\alpha}}{(2\pi i)^{2}}
\Bigl(
\frac{2\pi i}{{z_{1}}_{\alpha}} \delta({z_{2}}_{\alpha}) +
\frac{2\pi i}{{z_{2}}_{\alpha}} \delta({z_{1}}_{\alpha}) -
\frac{1}{{z_{1}}_{\alpha}{z_{2}}_{\alpha}}
\Bigr)
\Bigl(
1 + \frac{{z_{1}}_{\alpha}}{{z_{2}}_{\alpha} + i p_{\alpha}^{(-)}}
\Bigr)
\Bigl(
1 + \frac{{z_{2}}_{\alpha}}{{z_{1}}_{\alpha} - i p_{\alpha}^{(+)}}
\Bigr)
\times
\\
\nonumber
\\
&\times&
\exp\bigl\{ {z_{1}}_{\alpha} y_{\alpha} + {z_{2}}_{\alpha} y_{\alpha}
\bigr\}
\Biggr]
\;
\prod_{\alpha=1}^{M}
\Biggl[
\frac{1}{
{z_{1}}_{\alpha} + {z_{2}}_{\alpha} - i p_{\alpha} +
{z_{1}}_{\alpha +1}  + {z_{2}}_{\alpha +1}  + i p_{\alpha +1}}
\Biggr]
\label{5.37}
\end{eqnarray}
Here, by definition, it is assumed that ${z_{i_{M +1}}} \equiv {z_{i_{1}}}$ ($i=1,2$)
and $p_{M +1} \equiv p_{1}$.
Substituting
\begin{equation}
\label{5.38}
\frac{1}{
{z_{1}}_{\alpha} + {z_{2}}_{\alpha} - i p_{\alpha} +
{z_{1}}_{\alpha +1}  + {z_{2}}_{\alpha +1}  + i p_{\alpha +1}}
\; = \;
\int_{0}^{\infty} d\omega_{\alpha}
\exp\Bigl[-\bigl(
{z_{1}}_{\alpha} + {z_{2}}_{\alpha} - i p_{\alpha} +
{z_{1}}_{\alpha +1}  + {z_{2}}_{\alpha +1}  + i p_{\alpha +1}
\bigr) \, \omega_{\alpha}
\Bigr]
\end{equation}
into eq.(\ref{5.37}), we obtain
\begin{equation}
 \label{5.39}
\mbox{Tr} \, \hat{K}^{M} \; = \;
\int_{0}^{\infty} d\omega_{1} \, ... \, d\omega_{M} \,
\prod_{\alpha=1}^{M}
\Biggl[
\int\int_{-\infty}^{+\infty} \frac{dy dp}{2\pi}
\Ai\bigl(y + p^{2} + \omega_{\alpha} + \omega_{\alpha -1} - f \bigr)
\exp\{i p \bigl(\omega_{\alpha} - \omega_{\alpha -1}\bigr)\}
\;
S\bigl(p, y \bigr)
\Biggr]
\end{equation}
where, by definition, $\omega_{0} \equiv \omega_{M}$, and
\begin{eqnarray}
 \nonumber
S(p, y) &=&
\int\int_{{\cal C}}
\frac{dz_{1}dz_{2}}{(2\pi i)^{2}}
\Bigl(
\frac{2\pi i}{z_{1}} \delta(z_{2}) +
\frac{2\pi i}{z_{2}} \delta(z_{1}) -
\frac{1}{z_{1}z_{2}}
\Bigr)
\Bigl(
1 + \frac{z_{1}}{z_{2} + i p^{(-)}}
\Bigr)
\Bigl(
1 + \frac{z_{2}}{z_{1} - i p^{(+)}}
\Bigr)
\times
\\
\nonumber
\\
&\times&
\exp\bigl\{ z_{1} y + z_{2} y \bigr\}
\label{5.40}
\end{eqnarray}
Simple calculations yield:
\begin{eqnarray}
\nonumber
S\bigl(p, \, y \bigr) &=&
\frac{1}{2\pi i}
\int_{{\cal C}}
\frac{dz_{1}}{z_{1}} \,
\Bigl(
1 + \frac{z_{1}}{ i (p-i\epsilon)}
\Bigr) \,
\exp\{ z_{1}  y\}
\; + \;
\frac{1}{2\pi i}
\int_{{\cal C}}
\frac{d z_{2}}{z_{2}} \,
\Bigl(
1 - \frac{z_{2}}{ i (p+i\epsilon)}
\Bigr) \,
\exp\{ z_{2}  y\} \; -
\\
\nonumber
\\
\nonumber
&-&
\frac{1}{(2\pi i)^{2}}
\int\int_{{\cal C}}
\frac{dz_{1}}{z_{1}} \,
\frac{dz_{2}}{z_{2}}
\Bigl(
1 + \frac{z_{1}}{z_{2} + i (p-i\epsilon)}
\Bigr)
\Bigl(
1 + \frac{z_{2}}{z_{1} - i (p+i\epsilon)}
\Bigr)
\exp\{ (z_{1} + z_{2}) y\}
\\
\nonumber
\\
\nonumber
\\
\label{5.41}
&=&
\Bigl[\frac{1}{ i (p-i\epsilon)} - \frac{1}{ i (p+i\epsilon)}\Bigr] \;
\delta(y)
\end{eqnarray}
Taking the limit $\epsilon \to 0$ we find:
\begin{equation}
\label{5.42}
S\bigl(p, \, y \bigr) \; = \; \delta(y) \delta(p)
\end{equation}
Substituting this result into eq.(\ref{5.39}) we obtain
\begin{equation}
 \label{5.43}
\mbox{Tr} \, \hat{K}^{M} \; = \;
\int_{0}^{\infty} d\omega_{1} \, ... \, d\omega_{M} \,
\prod_{\alpha=1}^{M}
\Bigl[
\Ai\bigl( \omega_{\alpha} + \omega_{\alpha -1} - f \bigr)
\Bigr]
\end{equation}
In other words, the free energy distribution function of our problem
is given by the Fredholm determinant,
\begin{equation}
\label{5.44}
W(f) \; = \; \det\Bigl[1 \; - \; \hat{K}\Bigr] \; = \; F_{1}(-f)
\end{equation}
with the kernel
\begin{equation}
\label{5.45}
K(\omega, \omega') \; =\;
\Ai\bigl( \omega \; + \; \omega' \; - \; f \bigr) \; ,
\; \; \; \; \; \; \; \; \; \; \; \; \; \; \; \; (\omega , \; \omega' \; > \; 0)
\end{equation}
which is the GOE Tracy-Widom distribution \cite{TW2,Ferrari-Spohn}.

\vspace{5mm}

In conclusion, the key technical tricks of the calculations presented in this Chapter 
includes the following points:

1. First of all, to make the integration over particle coordinates of the Bethe ansatz
propagator well defined one has to introduce proper
regularization at $\pm\infty$ which requires formal
splitting the partition function into two parts:
the one in the positive particles coordinates
sector (up to $+\infty$) and another one in
the negative particles coordinates sector (down to $-\infty$),
eqs.(\ref{5.6})-(\ref{5.7}).

2. Next is the "magic" Bethe ansatz combinatorial identity, eq.(\ref{5.14}),
which allows to perform the summation over the momenta permutations and
"disentangle" sophisticated  products contained in the Bethe ansatz
propagator. 

3. One more trick is the representation of the summation over
permutations of the momenta  in terms of the series summations, eq.(\ref{5.21}) and (\ref{5.22}).

4. Finally, the crucial point of the considered derivation is taking the 
limit $t \to \infty$.  In this limit, due to the integral representation
of the series, eqs.(\ref{5.27})-(\ref{5.29}), one obtains dramatic simplifications
of  the expression for the probability distribution function $W(f)$ 
which eventually makes possible to represent it in the form of the Fredholm determinant,
eqs.(\ref{5.34}) and (\ref{5.44})-(\ref{5.45}).


\vspace{10mm}

\section{Multi-point free energy distribution functions}

\newcounter{6}
\setcounter{equation}{0}
\renewcommand{\theequation}{6.\arabic{equation}}

Let us consider consider the system in which the polymer is fixed
at the origin, $\phi(0)=0$ and it  arrives at a given point $x$ at $\tau = t$.
In other words, for a given realization of the random potential
$V$ the partition function of this system is:
\begin{equation}
\label{6.1}
   Z(x; t) \; = \; \int_{\phi(0)=0}^{\phi(t)=x}
              {\cal D} \phi(\tau)  \;  \exp\bigl\{-\beta H[\phi; V]\bigr\} 
              \; = \; \exp\{-\beta F(x; t)\}
\end{equation}
where (in the limit $t\to \infty$) the total free energy  $\beta  F(x; t) = \beta f_{0} t \, + \, \lambda(t) \, f(x)$ 
with $\lambda(t) = \frac{1}{2} (\beta^{5} u^{2} t)^{1/3}$ (see discussion in Chapter I and eq.(\ref{1.8})
and the Hamiltonian $H[\phi; V]$ is given in eq.(\ref{1.1}). 
In this Chapter we will the  study the $N$-point free energy
probability distribution function
\begin{equation}
\label{6.2}
W_{N}(f_{1}, \, f_{2}, \, ...,  f_{N}; \; x_{1}, \, x_{2}, \, ..., \, x_{N}) \; = \; \lim_{t\to\infty} \;
\mbox{Prob}\bigl[f(x_{1}) >  f_{1}; \; f(x_{2}) >  f_{2}; \; ... ; \; f(x_{N})  >  f_{N}\bigr]
\end{equation}
which describes joint statistics of the free energies of $N$ directed polymers
coming to $N$ different endpoints 
\cite{Prolhac-Spohn,2pointPDF,Imamura-Sasamoto-Spohn,N-point-1,N-point-2}

\subsection{Two-point distribution}

The simplest case is of course $N = 2$. It is clear that in this case the probability
distribution of the type (\ref{6.1}) must depend only on the distance
between the two points $x \equiv |x_{2} - x_{1}|$, and therefore, to simplify formulas we will 
consider the particular case:
$x_{1} = -\frac{1}{2} x$ and  $x_{2} = +\frac{1}{2} x$.
In other words, we are going to calculate the probability distribution function
\begin{equation}
\label{6.3}
W(f_{1}, f_{2}; x) \; = \; \lim_{t\to\infty} \;
\mbox{Prob}\bigl[f(-x/2)  >  f_{1}; \; f(x/2)  > f_{2}\bigr]
\end{equation}
To remove the irrelevant linear contribution $f_{0} t$ to the total free energy we redefine the partition 
function as it is shown in eqs.(\ref{2.5})-(\ref{2.8}). Then, following the usual procedure of the 
generating function approach, eqs.(\ref{2.15})-(\ref{2.19}), for the probability function (\ref{6.3}) we get
\begin{eqnarray}
\nonumber
W(f_{1}, f_{2}; x) &=& \lim_{t\to\infty}
\sum_{L=0}^{\infty} \sum_{R=0}^{\infty}
\frac{(-1)^{L}}{L!} \frac{(-1)^{R}}{R!}
\exp\bigl\{\lambda(t) L f_{1} + \lambda(t) R f_{2}\bigr\} \;
\overline{\tilde{Z}^{L}(-x/2 ; t) \, \tilde{Z}^{R}(x/2 ; t)}
\\
\nonumber
\\
&=& 
\lim_{t\to\infty}
\sum_{L=0}^{\infty} \sum_{R=0}^{\infty}
\frac{(-1)^{L}}{L!} \frac{(-1)^{R}}{R!}
\exp\bigl\{\lambda(t) L f_{1} + \lambda(t) R f_{2} \bigr\} \;
\Psi\bigl(\underbrace{-x/2, ...,-x/2}_{L}, \underbrace{x/2, ..., x/2}_{R} ; \; t\bigr)
\label{6.4}
\end{eqnarray}
where the explicit form of the wave function $\Psi\bigl( {\bf x}; t\bigr)$ is given in eqs.(\ref{3.15}) and (\ref{3.14}). 
Then, repeating the calculations of Chapter III, instead of eq.(\ref{4.2}) we get
\begin{eqnarray}
\nonumber
W(f_{1}, f_{2}; x) &=& \lim_{t\to\infty}
\sum_{L,R=0}^{\infty} \;\frac{(-1)^{L+R}}{L! \, R!} \;
\exp\bigl\{\lambda(t) L f_{1} + \lambda(t) R f_{2}\bigr\}
\times
\\
\nonumber
\\
\nonumber
&\times&
\sum_{M=0}^{\infty} \frac{1}{M!}
\prod_{\alpha=1}^{M}
\Biggl[
\sum_{n_{\alpha}=1}^{\infty}
\int_{-\infty}^{+\infty} \frac{dq_{\alpha}}{2\pi} \, \kappa^{n_{\alpha}}
\exp\Bigl\{-\frac{t}{2\beta} n_{\alpha} q_{\alpha}^{2}
+ \frac{\kappa^{2} t}{24 \beta} n_{\alpha}^{3} \Bigr\}
\Biggr]
\; {\boldsymbol \delta}\Bigl(\sum_{\alpha=1}^{M} n_{\alpha} \; , \;  L+R\Bigr) 
\\
\nonumber
\\
\nonumber
&\times&
   \det\Biggl[
   \frac{1}{\frac{1}{2}\kappa n_{\alpha} - i q_{\alpha}
          + \frac{1}{2}\kappa n_{\beta} + iq_{\beta}}\Biggr]_{\alpha,\beta=1,...M}
\\
\nonumber
\\
&\times&
\sum_{{\cal P}^{(L,R)}}  \sum_{{\cal P}^{(L)}} \sum_{{\cal P}^{(R)}} \;
\prod_{a=1}^{L} \prod_{c=1}^{R}
\Biggl[
\frac{Q_{{\cal P}_a^{(L)}} - Q_{{\cal P}_c^{(R)}} - i \kappa}{Q_{{\cal P}_a^{(L)}} - Q_{{\cal P}_c^{(R)}}}
\Biggr]
\exp\Biggl\{-\frac{i}{2} x \sum_{a=1}^{L} Q_{{\cal P}_{a}^{(L)}}
          +\frac{i}{2} x \sum_{c=1}^{R} Q_{{\cal P}_{c}^{(R)}} \Biggr\}
\label{6.5}
\end{eqnarray}

In eq.(\ref{6.5}) the summation over all permutations ${\cal P}$ of $(L+R)$ momenta
$\{Q_{1}, ..., Q_{L+R}\}$  over $L$ "left" particles
$\{x_{1}= x_{2} = ... = x_{L} = -x/2\}$
and $R$ "right" particles $\{x_{L+1} = x_{L+2} = ... = x_{L+R} = x/2\}$
split   into three parts: the permutations ${\cal P}^{(L)}$
of $L$ momenta (taken at random out of the total list $\{Q_{1}, ..., Q_{L+R}\}$)
over $L$ "left" particles, the permutations ${\cal P}^{(R)}$
of the remaining $R$ momenta over $R$ "right" particles, and
finally the permutations ${\cal P}^{(L,R)}$ (or the exchange) of the
momenta between the group $"L"$ and the group $"R"$. It is evident that due to the
symmetry of the expression in eq.(\ref{6.5}) with respect to the permutations
${\cal P}^{(L)}$ and ${\cal P}^{(R)}$ the summations over these permutations give
just the factor $L! \, R!$.

Further simplification comes due to the special 
Bethe ansatz product structure of the factor in eq.(\ref{6.5}). 
One can easily see that in this product, according to the definition (\ref{3.12}),
the momenta $Q_{a}$ belonging
to the same cluster must be {\it ordered}. In other words,
if we consider the momenta, eq.(\ref{3.12}), of a cluster $\alpha$, 
$\{q_{1}^{\alpha}, q_{2}^{\alpha}, ..., q_{n_{\alpha}}^{\alpha}\}$, 
the permutation of any two momenta $q_{r}^{\alpha}$
and $q_{r'}^{\alpha}$ of this {\it ordered} set in the product in eq.(\ref{6.5}) gives zero contribution.
Thus, in order to perform the summation over the permutations ${\cal P}^{(L,R)}$
in eq.(\ref{6.5}) it is sufficient to split the momenta of each cluster into two parts:
$\{q_{1}^{\alpha}, ...,  q_{m_{\alpha}}^{\alpha} ||
q_{m_{\alpha}+1}^{\alpha}..., q_{n_{\alpha}}^{\alpha}\}$, where $m_{\alpha} = 0, 1, ..., n_{\alpha}$ and
where the momenta $q_{1}^{\alpha}, ...,  q_{m_{\alpha}}^{\alpha}$ belong to the particles
of the sector $"L"$ (whose coordinates are all equal to $-x/2$),
while the momenta $q_{m_{\alpha}+1}^{\alpha}..., q_{n_{\alpha}}^{\alpha}$
belong to the particles of the sector $"R"$ (whose coordinates are all equal to $+x/2$).

It is convenient to introduce the numbering of the momenta
of the sector $"R"$ in the reversed order:
\begin{eqnarray}
\nonumber
q_{n_{\alpha}}^{\alpha} &\to&  {q^{*}}_{1}^{\alpha}
\\
\nonumber
q_{n_{\alpha}-1}^{\alpha} &\to&  {q^{*}}_{2}^{\alpha}
\\
\nonumber
&........&
\\
q_{m_{\alpha}+1}^{\alpha} &\to&  {q^{*}}_{s_{\alpha}}^{\alpha}
\label{6.6}
\end{eqnarray}
where $m_{\alpha} + s_{\alpha} = n_{\alpha}$ and (s.f. eq.(\ref{3.12}))

\begin{equation}
\label{6.7}
{q^{*}}_{r}^{\alpha} \; = \; q_{\alpha} + \frac{i \kappa}{2} (n_{\alpha} + 1 - 2r)
\; = \; q_{\alpha} + \frac{i \kappa}{2} (m_{\alpha} + s_{\alpha} + 1 - 2r)
\end{equation}
By definition, the integer parameters $\{m_{\alpha}\}$ and $\{s_{\alpha}\}$
fulfill the global constrains
\begin{eqnarray}
\label{6.8}
\sum_{\alpha=1}^{M} m_{\alpha} &=& L
\\
\nonumber
\\
\sum_{\alpha=1}^{M} s_{\alpha} &=& R
\label{6.9}
\end{eqnarray}
In this way the summation over permutations ${\cal P}^{(L,R)}$
in eq.(\ref{6.5}) is changed by the summations over the integer parameters
$\{m_{\alpha}\}$ and $\{s_{\alpha}\}$,
which allows to lift the summations over $L$, $R$, and $\{n_{\alpha}\}$.
Straightforward  calculations result in the following
expression:
\begin{eqnarray}
 \nonumber
W(f_{1}, f_{2}; x) &=& \lim_{t\to\infty}
\sum_{M=0}^{\infty} \; \frac{(-1)^{M}}{M!} \;
\prod_{\alpha=1}^{M}
\Biggl[
\sum_{m_{\alpha}+s_{\alpha}\geq 1}^{\infty}
(-1)^{m_{\alpha}+s_{\alpha}-1} \, 
\int_{-\infty}^{+\infty} \;
\frac{dq_{\alpha}}{2\pi}
\times
\\
\nonumber
\\
\nonumber
&\times&
\exp\Bigl\{
\lambda m_{\alpha} f_{1} + \lambda s_{\alpha} f_{2}
-\frac{i}{2} x m_{\alpha} q_{\alpha} + \frac{i}{2} x s_{\alpha} q_{\alpha} -\frac{1}{2}\kappa x m_{\alpha} s_{\alpha}
-\frac{t}{2\beta} (m_{\alpha}+s_{\alpha}) q_{\alpha}^{2} +
\frac{\kappa^{2} t}{24 \beta} (m_{\alpha}+s_{\alpha})^{3}
\Bigr\}
\Biggr]
\times
\\
\nonumber
\\
&\times&
 \det\Biggl[
   \frac{1}{\frac{1}{2}\kappa (m_{\alpha}+s_{\alpha}) - i q_{\alpha}
          + \frac{1}{2}\kappa (m_{\beta}+s_{\beta}) + iq_{\beta}}\Biggr]_{\alpha,\beta=1,...M} \;
{\bf G}_{M} \bigl({\bf q}, {\bf m}, {\bf s}\bigr)
\label{6.10}
\end{eqnarray}
where
\begin{eqnarray}
\nonumber
{\bf G}_{M} \bigl({\bf q}, {\bf m}, {\bf s}\bigr)
&=&
\prod_{\alpha=1}^{M} \prod_{r=1}^{m_{\alpha}} \prod_{r'=1}^{s_{\alpha}}
\Biggl(
\frac{q^{\alpha}_{r} - {q^{*}}_{r'}^{\alpha} - i \kappa}{q^{\alpha}_{r} - {q^{*}}_{r'}^{\alpha}}
\Biggr)
\times
\prod_{\alpha<\beta}^{M} \prod_{r=1}^{m_{\alpha}} \prod_{r'=1}^{s_{\alpha}}
\Biggl(
\frac{q^{\alpha}_{r} - {q^{*}}_{r'}^{\beta} - i \kappa}{q^{\alpha}_{r} - {q^{*}}_{r'}^{\beta}}
\Biggr)
\\
\nonumber
\\
\label{6.11}
&=&
\prod_{\alpha=1}^{M}
\frac{
\Gamma\bigl(1 + m_{\alpha} + s_{\alpha}\bigr)}{
\Gamma\bigl(1 + m_{\alpha}\bigr) \Gamma\bigl(1 + s_{\alpha}\bigr)}
\times
\\
\nonumber
\\
\nonumber
&\times&
\prod_{\alpha\not= \beta}^{M}
\frac{
\Gamma
\Bigl[
1 + \frac{m_{\alpha} + m_{\beta} + s_{\alpha} + s_{\beta}}{2}
+\frac{i}{\kappa}\bigl(q_{\alpha} - q_{\beta}\bigr)
\Bigr] \,
\Gamma
\Bigl[
1 + \frac{-m_{\alpha} + m_{\beta} + s_{\alpha} - s_{\beta}}{2}
+\frac{i}{\kappa}\bigl(q_{\alpha} - q_{\beta}\bigr)
\Bigr]}{
\Gamma
\Bigl[
1 + \frac{-m_{\alpha} + m_{\beta} + s_{\alpha} + s_{\beta}}{2}
+\frac{i}{\kappa}\bigl(q_{\alpha} - q_{\beta}\bigr)
\Bigr] \,
\Gamma
\Bigl[
1 + \frac{m_{\alpha} + m_{\beta} + s_{\alpha} - s_{\beta}}{2}
+\frac{i}{\kappa}\bigl(q_{\alpha} - q_{\beta}\bigr)
\Bigr]}
\end{eqnarray}
After rescaling (s.f. eqs.(\ref{4.3})-(\ref{4.6}))
\begin{eqnarray}
\label{6.12}
q_{\alpha} &\to& \frac{\kappa}{2\lambda} \, q_{\alpha}
\\
\nonumber
\\
\label{6.13}
x &\to&  \frac{2 \lambda^{2}}{\kappa} \, x
\end{eqnarray}
with
\begin{equation}
\label{6.14}
\lambda \; = \; \frac{1}{2} \, \bigl(\beta^{5} u^{2} t\bigr)^{1/3}
\end{equation}
and using the Airy function relation
\begin{equation}
   \label{6.15}
\exp\Bigl\{ \frac{1}{3} \lambda^{3} (m_{\alpha}+s_{\alpha})^{3} \Bigr\} \; = \;
\int_{-\infty}^{+\infty} dy \; \Ai(y) \;
\exp\Bigl\{\lambda (m_{\alpha}+s_{\alpha}) \, y \Bigr\}
\end{equation}
we get 
\begin{eqnarray}
 \nonumber
W(f_{1},f_{2};x) &=& \lim_{t\to\infty}
\sum_{M=0}^{\infty} \; \frac{(-1)^{M}}{M!} \;
\prod_{\alpha=1}^{M}
\Biggl[
\int\int_{-\infty}^{+\infty} \frac{dy_{\alpha} dq_{\alpha}}{2\pi}
\Ai\bigl(y_{\alpha} + q_{\alpha}^{2}\bigr)
\sum_{m_{\alpha}+s_{\alpha}\geq 1}^{\infty} (-1)^{m_{\alpha}+s_{\alpha}-1}
\times
\\
\nonumber
\\
\nonumber
&\times&
\exp\Bigl\{
\lambda(t) m_{\alpha} (y_{\alpha} + f_{1} - \frac{1}{2} i q_{\alpha} x) +
\lambda(t) s_{\alpha} (y_{\alpha} + f_{2} + \frac{1}{2} i q_{\alpha} x)
- \lambda^{2} m_{\alpha} s_{\alpha} x
\Bigr\} \;
\Biggr]
\times
\\
\nonumber
\\
&\times&
 \det\Biggl[
   \frac{1}{\lambda(t)(m_{\alpha}+s_{\alpha}) - i q_{\alpha}
          + \lambda(t)(m_{\beta}+s_{\beta}) + iq_{\beta}}\Biggr]_{\alpha,\beta=1,...M} \;
\;
{\bf G}_{M} \Bigl(\frac{{\kappa \bf q}}{2\lambda(t)}, \; {\bf m}, \; {\bf s}\Bigr)
\label{6.16}
\end{eqnarray}
Using the relation
\begin{equation}
 \label{6.17}
\exp\{- \lambda^{2} \, m \, s \, x \} \; = \;
\int_{-\infty}^{+\infty}
\frac{d\xi_{1}d\xi_{2}d\xi_{3}}{(2\pi)^{3/2}}
\exp\Bigl\{
-\frac{1}{2} \xi_{1}^{2} -\frac{1}{2} \xi_{2}^{2}-\frac{1}{2} \xi_{3}^{2}
+\lambda m \sqrt{x}\xi_{1} + \lambda s \sqrt{x} \xi_{2} + i\lambda (m+s) \sqrt{x} \xi_{3}
\Bigr\}
\end{equation}
the expression in eq.(\ref{6.16}) can be represented as follows:
\begin{eqnarray}
 \nonumber
W(f_{1},f_{2};x) &=&
\sum_{M=0}^{\infty} \; \frac{(-1)^{M}}{M!} \;
\prod_{\alpha=1}^{M}
\Biggl[
\int_{-\infty}^{+\infty} \frac{dy_{\alpha} dq_{\alpha}}{2\pi}
\frac{d{\xi_{1}}_{\alpha} d{\xi_{2}}_{\alpha}d{\xi_{3}}_{\alpha}}{(2\pi)^{3/2}} \;
\Ai\bigl(y_{\alpha} + q_{\alpha}^{2} - i{\xi_{3}}_{\alpha} \sqrt{x}\bigr)
\times
\\
\nonumber
\\
\label{6.18}
&\times&
\exp\Bigl\{
-\frac{1}{2} \bigl({\xi_{1}}_{\alpha} + \frac{1}{2}iq_{\alpha}\sqrt{x}\bigr)^{2}
-\frac{1}{2} \bigl({\xi_{2}}_{\alpha} - \frac{1}{2}iq_{\alpha}\sqrt{x}\bigr)^{2}
-\frac{1}{2} {\xi_{3}}_{\alpha}^{2}
\Bigr\}
\Biggr] \;
{\cal S}_{M}\bigl({\bf q}, {\bf y}, {\boldsymbol \xi}_{1}, {\boldsymbol \xi}_{1},
 f_{1}, f_{2}, x\bigr)
\end{eqnarray}
where
\begin{eqnarray}
\nonumber
{\cal S}_{M}\bigl({\bf q}, {\bf y}, {\boldsymbol \xi}_{1}, {\boldsymbol \xi}_{1},
 f_{1}, f_{2}, x\bigr)
&=&
\lim_{t\to\infty}
\prod_{\alpha=1}^{M}
\Biggl[
\sum_{m_{\alpha}+s_{\alpha}\geq 1}^{\infty} (-1)^{m_{\alpha}+s_{\alpha}-1}
\exp\Bigl\{
\lambda(t) m_{\alpha} (y_{\alpha} + f_{1} + {\xi_{1}}_{\alpha}\sqrt{x}) +
\lambda(t) s_{\alpha} (y_{\alpha} + f_{2} + {\xi_{2}}_{\alpha}\sqrt{x})
\Bigr\} \;
\Biggr]
\\
\nonumber
\\
&\times&
 \det\Biggl[
   \frac{1}{\lambda(t)(m_{\alpha}+s_{\alpha}) - i q_{\alpha}
          + \lambda(t)(m_{\beta}+s_{\beta}) + iq_{\beta}}\Biggr]_{\alpha,\beta=1,...M} \;
{\bf G}_{M} \Bigl(\frac{{\kappa \bf q}}{2\lambda(t)}, \; {\bf m}, \; {\bf s}\Bigr)
\label{6.19}
\end{eqnarray}
The summations over $\{m_{\alpha}\}$ and $\{s_{\alpha}\}$ in the limit $t\to \infty$ can be performed 
according to the algorithm described in Chapter V, eqs.(\ref{5.27})-(\ref{5.30}).
In this way for the function in eq.(\ref{6.19}), we get (s.f. eq.(\ref{5.31}))
\begin{eqnarray}
\nonumber
{\cal S}_{M}\bigl({\bf q}, {\bf y}, {\boldsymbol \xi}_{1}, {\boldsymbol \xi}_{1},
 f_{1}, f_{2}, x\bigr)
 &=&
\prod_{\alpha=1}^{M}
\Biggl[
\int\int_{{\cal C}}
\frac{d{z_{1}}_{\alpha}d{z_{2}}_{\alpha}}{(2\pi i)^{2}}
\Bigl(
\frac{2\pi i}{{z_{1}}_{\alpha}} \delta({z_{2}}_{\alpha}) +
\frac{2\pi i}{{z_{2}}_{\alpha}} \delta({z_{1}}_{\alpha}) -
\frac{1}{{z_{1}}_{\alpha}{z_{2}}_{\alpha}}
\Bigr)
\times
\\
\nonumber
\\
\nonumber
&\times&
\exp\Bigl\{
{z_{1}}_{\alpha}\bigl(y_{\alpha} + f_{1} + {\xi_{1}}_{\alpha}\sqrt{x}\bigr) +
{z_{2}}_{\alpha}\bigl(y_{\alpha} + f_{2} + {\xi_{2}}_{\alpha}\sqrt{x}\bigr)
\Bigr\}
\Biggr]
\times
\\
\nonumber
\\
\nonumber
&\times&
\det 
\Bigl[
\frac{1}{{z_{1}}_{\alpha} + {z_{2}}_{\alpha} - i p_{\alpha} +
         {z_{1}}_{\beta} + {z_{2}}_{\beta} + i p_{\beta}}
\Bigr]_{(\alpha,\beta)=1,2,...,M}
\\
\nonumber
\\
&\times&
\lim_{t\to\infty}
{\bf G}_{M}\Bigl(
\frac{\kappa {\bf q}}{2\lambda(t)}, \;
\frac{{\bf z_{1}}}{\lambda(t)}, \;
\frac{{\bf z_{2}}}{\lambda(t)}\Bigr)
\label{6.20}
\end{eqnarray}
where the integrations over $\{ {z_{1}}_{\alpha}\}$ and  $\{ {z_{2}}_{\alpha}\}$ goes over the contour 
${\cal C}$ shown in Fig. 2.
Using the explicit form of the factor ${\bf G}_{M}$, eq.(\ref{6.11}), and taking into account
that $\Gamma(1+z)\big|_{|z|\to 0} = 1$, we find
\begin{equation}
\label{6.21}
\lim_{t\to\infty}
{\bf G}_{M}
\Bigl(\frac{\kappa {\bf q}}{2\lambda(t)}, \;
\frac{{\bf z_{1}}}{\lambda(t)}, \;
\frac{{\bf z_{2}}}{\lambda(t)}\Bigr)
\; = \;
1
\end{equation}
Thus, in the limit $t\to \infty$ the expression for the probability distribution function, eq.(\ref{6.10}),
takes the form of the Fredholm determinant
\begin{eqnarray}
 \nonumber
W(f_{1},f_{2};x) &=&
\sum_{M=0}^{\infty} \; \frac{(-1)^{M}}{M!} \;
\prod_{\alpha=1}^{M}
\Biggl[
\int_{-\infty}^{+\infty} \frac{dy_{\alpha} dq_{\alpha}}{2\pi}
\frac{d{\xi_{1}}_{\alpha} d{\xi_{2}}_{\alpha}d{\xi_{3}}_{\alpha}}{(2\pi)^{3/2}} \;
\Ai\bigl(y_{\alpha} + q_{\alpha}^{2} - i{\xi_{3}}_{\alpha} \sqrt{x}\bigr)
\times
\\
\nonumber
\\
\nonumber
&\times&
\exp\Bigl\{
-\frac{1}{2} \bigl({\xi_{1}}_{\alpha} + \frac{1}{2}iq_{\alpha}\sqrt{x}\bigr)^{2}
-\frac{1}{2} \bigl({\xi_{2}}_{\alpha} - \frac{1}{2}iq_{\alpha}\sqrt{x}\bigr)^{2}
-\frac{1}{2} {\xi_{3}}_{\alpha}^{2}
\Bigr\}
\times
\\
\nonumber
\\
\nonumber
&\times&
\int\int_{{\cal C}}
\frac{d{z_{1}}_{\alpha}d{z_{2}}_{\alpha}}{(2\pi i)^{2}}
\Bigl(
\frac{2\pi i}{{z_{1}}_{\alpha}} \delta({z_{2}}_{\alpha}) +
\frac{2\pi i}{{z_{2}}_{\alpha}} \delta({z_{1}}_{\alpha}) -
\frac{1}{{z_{1}}_{\alpha}{z_{2}}_{\alpha}}
\Bigr)
\times
\\
\nonumber
\\
\nonumber
&\times&
\exp\Bigl\{
{z_{1}}_{\alpha}\bigl(y_{\alpha} + f_{1} + {\xi_{1}}_{\alpha}\sqrt{x}\bigr) +
{z_{2}}_{\alpha}\bigl(y_{\alpha} + f_{2} + {\xi_{2}}_{\alpha}\sqrt{x}\bigr)
\Bigr\}
\Biggr] \;
\times
\\
\nonumber
\\
\label{6.22}
&\times&
\det \Bigl[
\frac{1}{{z_{1}}_{\alpha} + {z_{2}}_{\alpha} - i p_{\alpha} +
{z_{1}}_{\beta} + {z_{2}}_{\beta} + i p_{\beta}}
\Bigr]_{(\alpha,\beta)=1,2,...,M}
\\
\nonumber
\\
\nonumber
&=&
\det\bigl[\hat{1} \, - \, \hat{A} \bigr]
\end{eqnarray}
with the kernel
\begin{eqnarray}
 \nonumber
\hat{A}\bigl[({z_{1}}, \, {z_{2}}, \, q); ({z_{1}}', \, {z_{2}}', \, q')\bigr]
&=&
\int_{-\infty}^{+\infty} \frac{dy}{2\pi}
\frac{d{\xi_{1}} d{\xi_{2}}d{\xi_{3}}}{(2\pi)^{3/2}} \;
\Ai\bigl(y + q^{2} - i{\xi_{3}} \sqrt{x}\bigr)
\times
\\
\nonumber
\\
\nonumber
&\times&
\exp\Bigl\{
-\frac{1}{2} \bigl({\xi_{1}} + \frac{1}{2}iq\sqrt{x}\bigr)^{2}
-\frac{1}{2} \bigl({\xi_{2}} - \frac{1}{2}iq\sqrt{x}\bigr)^{2}
-\frac{1}{2} {\xi_{3}}^{2}
\Bigr\}
\times
\\
\nonumber
\\
\nonumber
&\times&
\Bigl(
\frac{2\pi i}{{z_{1}}} \delta({z_{2}}) +
\frac{2\pi i}{{z_{2}}} \delta({z_{1}}) -
\frac{1}{{z_{1}}{z_{2}}}
\Bigr)
\times
\\
\nonumber
\\
\nonumber
&\times&
\exp\Bigl\{
{z_{1}}\bigl(y + f_{1} + {\xi_{1}}\sqrt{x}\bigr) +
{z_{2}}\bigl(y + f_{2} + {\xi_{2}}\sqrt{x}\bigr)
\Bigr\}
\times
\\
\nonumber
\\
\label{6.23}
&\times&
\frac{1}{z_{1}+z_{2}-iq + z_{1}'+z_{2}'+iq'}
\end{eqnarray}
In the exponential representation of this determinant we get
\begin{equation}
 \label{6.24}
W(f_{1},f_{2},x) \; = \;
\exp\Bigl[-\sum_{M=1}^{\infty} \frac{1}{M} \; \mbox{Tr} \, \hat{A}^{M} \Bigr]
\end{equation}
where
\begin{eqnarray}
 \nonumber
\mbox{Tr} \, \hat{A}^{M} &=&
\prod_{\alpha=1}^{M}
\Biggl[
\int_{-\infty}^{+\infty} \frac{dy dq_{\alpha}}{2\pi}
\frac{d{\xi_{1}} d{\xi_{2}}d{\xi_{3}}}{(2\pi)^{3/2}} \;
\Ai\bigl(y + q_{\alpha}^{2} - i{\xi_{3}} \sqrt{x}\bigr)
\times
\\
\nonumber
\\
\nonumber
&\times&
\exp\Bigl\{
-\frac{1}{2} \bigl({\xi_{1}} + \frac{1}{2}iq_{\alpha}\sqrt{x}\bigr)^{2}
-\frac{1}{2} \bigl({\xi_{2}} - \frac{1}{2}iq_{\alpha}\sqrt{x}\bigr)^{2}
-\frac{1}{2} {\xi_{3}}^{2}
\Bigr\}
\times
\\
\nonumber
\\
\nonumber
&\times&
\int\int_{{\cal C}}
\frac{d{z_{1}}_{\alpha}d{z_{2}}_{\alpha}}{(2\pi i)^{2}}
\Bigl(
\frac{2\pi i}{{z_{1}}_{\alpha}} \delta({z_{2}}_{\alpha}) +
\frac{2\pi i}{{z_{2}}_{\alpha}} \delta({z_{1}}_{\alpha}) -
\frac{1}{{z_{1}}_{\alpha}{z_{2}}_{\alpha}}
\Bigr)
\times
\\
\nonumber
\\
\nonumber
&\times&
\exp\Bigl\{
{z_{1}}_{\alpha}\bigl(y + f_{1} + {\xi_{1}}\sqrt{x}\bigr) +
{z_{2}}_{\alpha}\bigl(y + f_{2} + {\xi_{2}}\sqrt{x}\bigr)
\Bigr\}
\Biggr]
\times
\\
\nonumber
\\
\label{6.25}
&\times&
\prod_{\alpha=1}^{M}
\Biggl[
\frac{1}{{z_{1}}_{\alpha} + {z_{2}}_{\alpha} - iq_{\alpha}
       + {z_{1}}_{\alpha+1} + {z_{2}}_{\alpha+1} + iq_{\alpha+1}}
\Biggr]
\end{eqnarray}
Here, by definition, it is assumed that ${z_{i_{M +1}}} \equiv {z_{i_{1}}}$ ($i=1,2$)
and $q_{M +1} \equiv q_{1}$.
Substituting
\begin{equation}
\label{6.26}
\frac{1}{
{z_{1}}_{\alpha} + {z_{2}}_{\alpha} - i q_{\alpha} +
{z_{1}}_{\alpha +1}  + {z_{2}}_{\alpha +1}  + i q_{\alpha +1}}
\; = \;
\int_{0}^{\infty} d\omega_{\alpha}
\exp\Bigl[-\bigl(
{z_{1}}_{\alpha} + {z_{2}}_{\alpha} - i q_{\alpha} +
{z_{1}}_{\alpha +1}  + {z_{2}}_{\alpha +1}  + i q_{\alpha +1}
\bigr) \, \omega_{\alpha}
\Bigr]
\end{equation}
into eq.(\ref{6.25}), we obtain
\begin{equation}
\label{6.27}
\mbox{Tr} \, \hat{A}^{M} \; =  \;
\int_{0}^{\infty} d\omega_{1} \, ... \, d\omega_{M} \,
\prod_{\alpha=1}^{M}
\Bigl[
A\bigl(\omega_{\alpha}, \omega_{\alpha +1}\bigr)
\Bigr]
\end{equation}
where
\begin{eqnarray}
 \nonumber
A\bigl(\omega, \omega'\bigr) & = &
\int_{-\infty}^{+\infty} \frac{dy dq}{2\pi}
\frac{d{\xi_{1}} d{\xi_{2}}d{\xi_{3}}}{(2\pi)^{3/2}} \;
\Ai\bigl(y + q^{2} + \omega + \omega' - i{\xi_{3}} \sqrt{x}\bigr)
\times
\\
\nonumber
\\
\nonumber
&\times&
\exp\Bigl\{
-\frac{1}{2} \bigl({\xi_{1}} + \frac{1}{2}iq\sqrt{x}\bigr)^{2}
-\frac{1}{2} \bigl({\xi_{2}} - \frac{1}{2}iq\sqrt{x}\bigr)^{2}
-\frac{1}{2} {\xi_{3}}^{2}
- i q (\omega - \omega')
\Bigr\}
\times
\\
\nonumber
\\
\nonumber
&\times&
\int\int_{{\cal C}}
\frac{d{z_{1}}d{z_{2}}}{(2\pi i)^{2}}
\Bigl(
\frac{2\pi i}{{z_{1}}} \delta({z_{2}}) +
\frac{2\pi i}{{z_{2}}} \delta({z_{1}}) -
\frac{1}{{z_{1}}{z_{2}}}
\Bigr)
\times
\\
\nonumber
\\
&\times&
\exp\Bigl\{
{z_{1}}\bigl(y + f_{1} + {\xi_{1}}\sqrt{x}\bigr) +
{z_{2}}\bigl(y + f_{2} + {\xi_{2}}\sqrt{x}\bigr)
\Bigr\}
\label{6.28}
\end{eqnarray}
Integrating over $z_{1}$ and $z_{2}$ we get:
\begin{eqnarray}
 \nonumber
A\bigl(\omega, \omega'\bigr) & = &
\int_{0}^{+\infty} dy \int_{-\infty}^{+\infty} \frac{dq}{2\pi}
\Ai\bigl(y + q^{2} - f_{1} + \omega + \omega' + \frac{1}{2} i q x \bigr)
\exp\bigl\{ -iq(\omega-\omega')\bigr\} \; +
\\
\nonumber
\\
\nonumber
&+&
\int_{0}^{+\infty} dy \int_{-\infty}^{+\infty} \frac{dq}{2\pi}
\Ai\bigl(y + q^{2} - f_{2} + \omega + \omega' - \frac{1}{2} i q x \bigr)
\exp\bigl\{ -iq(\omega-\omega')\bigr\} \; +
\\
\nonumber
\\
\nonumber
&-&
\int_{-\infty}^{+\infty} dy \int_{-\infty}^{+\infty} \frac{dq}{2\pi}
\int\int\int_{-\infty}^{+\infty}\frac{d{\xi_{1}} d{\xi_{2}}d{\xi_{3}}}{(2\pi)^{3/2}} \;
\Ai\bigl(y + q^{2} + \omega + \omega' - i{\xi_{3}} \sqrt{x}\bigr)
\times
\\
\nonumber
\\
\nonumber
&\times&
\exp\Bigl\{
-\frac{1}{2} \bigl({\xi_{1}} + \frac{1}{2}iq\sqrt{x}\bigr)^{2}
-\frac{1}{2} \bigl({\xi_{2}} - \frac{1}{2}iq\sqrt{x}\bigr)^{2}
-\frac{1}{2} {\xi_{3}}^{2}
- i q (\omega - \omega')
\Bigr\}
\times
\\
\nonumber
\\
&\times&
\theta\bigl(y + f_{1} + \xi_{1}\sqrt{x}\bigr) \;
\theta\bigl(y + f_{2} + \xi_{2}\sqrt{x}\bigr)
\label{6.29}
\end{eqnarray}
where $\theta(y)$ is the step function.
Redefining, $\xi_{1} = (t-\eta)/\sqrt{2}, \; $ $\xi_{2} = (t+\eta)/\sqrt{2} , \; $
$\xi_{3} = (it +\zeta)/\sqrt{2}, \; $ and integrating over $q$, $t$ and $\zeta$,
we find the following result:
\begin{eqnarray}
\nonumber
A(\omega,\omega') &=&
2^{1/3} K\Bigl[2^{1/3}\bigl(\omega - \tilde{f}_{1}\bigr) \, , \;
               2^{1/3}\bigl(\omega' - \tilde{f}_{1} \bigr)\Bigr]
\exp\Bigl\{\frac{1}{4}(\omega - \omega') x \Bigr\} \;
+
\\
\nonumber
\\
\nonumber
&+&
2^{1/3} K\Bigl[2^{1/3}\bigl(\omega - \tilde{f}_{2}\bigr) \, , \;
               2^{1/3}\bigl(\omega' - \tilde{f}_{2} \bigr) \Bigr]
\exp\Bigl\{-\frac{1}{4}(\omega - \omega') x \Bigr\} \;
-
\\
\nonumber
\\
\nonumber
&-& 2^{2/3}
\int_{-\infty}^{+\infty} dy \;
\int_{-\infty}^{+\infty} \frac{d\eta}{\sqrt{2\pi}}
\Ai\Bigl[2^{1/3}\Bigl(y + \omega - \eta \sqrt{\frac{x}{8}} \;\Bigr)\Bigr]
\Ai\Bigl[2^{1/3}\Bigl(y + \omega' + \eta \sqrt{\frac{x}{8}} \;\Bigr)\Bigr]
\times
\\
\nonumber
\\
&\times&
\exp
\Bigl\{
-\frac{1}{2} \eta^{2} \, - \, \frac{1}{2} x y \, - \, \frac{1}{4} x (\omega + \omega')
+\frac{1}{3}  \Bigl(\frac{x}{4}\Bigr)^{3}
\Bigr\}
\;
\theta\Bigl(y + \tilde{f}_{1} - \eta \sqrt{\frac{x}{8}} \; \Bigr) \;
\theta\Bigl(y + \tilde{f}_{2} + \eta \sqrt{\frac{x}{8}} \; \Bigr)
\label{6.30}
\end{eqnarray}
where $\tilde{f}_{1,2} = \frac{1}{2}\bigl(f_{1,2} - x^{2}/16 \bigr)$ and
 $K\bigl(\omega, \omega'\bigr) = \int_{0}^{\infty} dy \Ai(y + \omega)\Ai(y + \omega')$
is the Airy kernel. Note that the free energy shift $x^{2}/16$ in the above definition
of $\tilde{f}_{1,2}$ is just the rescaled (see eq.(\ref{6.13})) trivial elastic energy 
of a straight lines which start at $x=0$ at $\tau=0$ and ends up at $\pm x/2$ at $\tau = t$.

Thus the distribution function $W(f_{1}, f_{2}; x)$, eq.(\ref{6.3}),
is given the Fredholm determinant
\begin{equation}
 \label{6.31}
W(f_{1}, f_{2}; x) \; = \; \det\bigl[1 \, - \, \hat{A} \bigr]
\end{equation}
where $\hat{A}$ is the integral operator with the kernel
$A(\omega,\omega') \; \; \; (\omega,\omega' \geq 0)$
given in eq.(\ref{6.30}).

Note that using explicit expression (\ref{6.30}) one can easily test the
obtained result for three limit cases:
\begin{eqnarray}
 \label{6.32}
\lim_{f_{1}\to -\infty}
A(\omega,\omega') &=&
2^{1/3} K\Bigl[2^{1/3}\bigl(\omega - \tilde{f}_{2}\bigr) \, , \;
               2^{1/3}\bigl(\omega' - \tilde{f}_{2} \bigr) \Bigr]
\exp\Bigl\{-\frac{1}{4}(\omega - \omega') x \Bigr\}
\\
\nonumber
\\
\label{6.33}
\lim_{f_{2}\to -\infty}
A(\omega,\omega') &=&
2^{1/3} K\Bigl[2^{1/3}\bigl(\omega - \tilde{f}_{1}\bigr) \, , \;
               2^{1/3}\bigl(\omega' - \tilde{f}_{1} \bigr)\Bigr]
\exp\Bigl\{\frac{1}{4}(\omega - \omega') x \Bigr\}
\\
\nonumber
\\
\nonumber
\lim_{x\to 0}
A(\omega,\omega') &=&
2^{1/3} K\Bigl[2^{1/3}\bigl(\omega - f_{1}/2\bigr) \, , \;
               2^{1/3}\bigl(\omega' -f_{1}/2\bigr)\Bigr] \; \theta (f_{1} - f_{2})  \; +
\\
\nonumber
\\
&+&
2^{1/3} K\Bigl[2^{1/3}\bigl(\omega - f_{2}/2\bigr) \, , \;
               2^{1/3}\bigl(\omega' -f_{2}/2\bigr)\Bigr] \; \theta (f_{2} - f_{1})
\label{6.34}
\end{eqnarray}
which demonstrate that in the case $f_{1} \to -\infty$ we recover the usual GUE Tracy-Widom
distribution for $f_{2}$; in the case $f_{2} \to -\infty$ we recover the usual GUE Tracy-Widom
distribution for $f_{1}$;  while in the limit $x \to 0$ we find
the usual GUE Tracy-Widom distribution for $f_{1}$ (in the case $f_{1} > f_{2}$) and
for $f_{2}$ (in the case $f_{2} > f_{1}$), as it should be.

\subsection{$N$-point distribution}

For the fixed boundary conditions, $\phi(0) = 0; \; \phi(t) = x$, the partition function
of the model (\ref{1.1}) is given in eq.(\ref{6.1}) where in the limit $t\to \infty$
the total free energy  $F(x; t)$ can be separated into
three contributions: self-averaging linear in time part $f_{0} t$,
the elastic contribution $x^{2}/2t$, and the fluctuating part $\beta^{-1} \lambda(t) f(x)$:
\begin{equation}
\label{6.35}
 \beta F(x; t) = \beta f_{0} t + \beta x^{2}/2t + \lambda(t) f(x)
\end{equation}
where 
\begin{equation}
 \label{6.36}
\lambda(t) \; = \; \frac{1}{2} (\beta^{5}u^{2} t)^{1/3} \propto t^{1/3}
\end{equation}
Let us redefine the partition function
\begin{equation}
 \label{6.37}
 Z(x; t) \; = \;  \exp\bigl\{-\beta f_{0} t - \beta x^{2}/2t \bigr\} \, \tilde{Z}(x; t)
\end{equation}
so that
\begin{equation}
 \label{6.38}
 \tilde{Z}(x; t) \; = \; \exp\bigl\{-\lambda(t) f(x)\bigr\}
\end{equation}
The aim of this section is to  study the $N$-point free energy distribution function
\begin{equation}
\label{6.39}
W(f_{1}, ... , f_{N}; x_{1}, ... , x_{N})\; \equiv \; W({\bf f}; {\bf x}) \; = \; 
\lim_{t\to\infty} \; \mbox{Prob}\bigl[f(x_{1})>f_{1}, \, ... \, , f(x_{N})>f_{N}\bigr]
\end{equation}
which describes joint statistics of the free energies of $N$ directed polymers
coming to $N$ different endpoints. Following the usual procedure of the 
generating function approach, eqs.(\ref{2.15})-(\ref{2.19}) we get (s.f.(\ref{6.4}))
\begin{equation}
\label{6.40}
W({\bf f}; {\bf x}) = \lim_{t\to\infty}
\sum_{L_{1},...,L_{N}=0}^{\infty} 
\prod_{k=1}^{N}\Biggl[
\frac{(-1)^{L_{k}}}{L_{k}!} 
\exp\bigl(\lambda(t) L_{k} f_{k} \bigr) \Biggr]\;
\overline{\Biggl(\prod_{k=1}^{N} \tilde{Z}(x_{k}; t)\Biggr)}
\end{equation}
Performing the standard  averaging over the random potentials 
one obtains 
\begin{equation}
\label{6.41}
W({\bf f}; {\bf x}) = \lim_{\lambda\to\infty}
\sum_{L_{1},...,L_{N}=0}^{\infty} 
\prod_{k=1}^{N}\Biggl[
\frac{(-1)^{L_{k}}}{L_{k}!} 
\exp\bigl(\lambda L_{k} f_{k} \bigr) \Biggr]\;
\Psi\bigl(\underbrace{x_{1}, ...,x_{1}}_{L_{1}}, \, 
          \underbrace{x_{2}, ...,x_{2}}_{L_{1}}, \, ...\, , 
          \underbrace{x_{N}, ...,x_{N}}_{L_{N}} ; \; t\bigr)
\end{equation}
where the time dependent $n$-point wave function $\Psi(x_{1}, ..., x_{n}; t)$ 
($n = \sum_{k=1}^{N} L_{k}$)
is given in eqs.(\ref{3.15})-(\ref{3.14}).
Then, repeating the calculations of Chapter III, instead of eq.(\ref{4.2}) we obtain (s.f. eq.(\ref{6.5}))
\begin{eqnarray}
 \label{6.42}
 W({\bf f}; {\bf x}) &=& \lim_{t\to\infty} \Biggl\{
\sum_{L_{1},...,L_{N}=0}^{\infty} 
\prod_{k=1}^{N}\Biggl[
\frac{(-1)^{L_{k}}}{L_{k}!} 
\exp\bigl(\lambda(t) L_{k} f_{k} \bigr) \Biggr]\;
\times
\\
\nonumber
\\
\nonumber
&\times&
\sum_{M=1}^{\infty} \frac{1}{M!}
\prod_{\alpha=1}^{M}
\Biggl[
\sum_{n_{\alpha}=1}^{\infty}
\int_{-\infty}^{+\infty} dq_{\alpha} \frac{\kappa^{n_{\alpha}}}{2\pi} 
\mbox{\LARGE e}^{-\frac{t}{2\beta} n_{\alpha} q_{\alpha}^{2}
+ \frac{\kappa^{2} t}{24 \beta} n_{\alpha}^{3} }
\Biggr]
\; {\boldsymbol \delta}\Bigl(\sum_{\alpha=1}^{M} n_{\alpha} \; , \; \sum_{k=1}^{N}L_{k}\Bigr)
\; D_{M}({\bf q}, {\bf n})
\times
\\
\nonumber
\\
\nonumber
&\times&
\sum_{{\cal P}^{(L_{1},...,L_{N})}}  
\prod_{k=1}^{N}
\Biggl[ 
\sum_{{\cal P}^{(L_{k})}} 
\Biggr] \;
\prod_{k<l}^{N} \prod_{a_{k}=1}^{L_{k}} \prod_{a_{l}=1}^{L_{l}}
\Biggl(
\frac{Q_{{\cal P}_{a_{k}}^{(L_{k})}} - Q_{{\cal P}_{a_{l}}^{(L_{l})}} - i \kappa}{
      Q_{{\cal P}_{a_{k}}^{(L_{k})}} - Q_{{\cal P}_{a_{l}}^{(L_{l})}}}
\Biggr) \;
\exp\Bigl(i \sum_{k=1}^{N} x_{k} \sum_{a_{k}=1}^{L_{k}} Q_{{\cal P}_{a_{k}}^{(L_{k})}}\Bigr)
\Biggr\}
\end{eqnarray}
where
\begin{equation}
 \label{6.43}
D_{M}({\bf q}, {\bf n}) \; = \; 
   \det\Biggl[
   \frac{1}{\frac{1}{2}\kappa n_{\alpha} - i q_{\alpha}
          + \frac{1}{2}\kappa n_{\beta} + iq_{\beta}}\Biggr]_{\alpha,\beta=1,...M} 
\end{equation}
In the above expression the summation over permutations of $n = L_{1} + ... + L_{N}$ momenta $Q_{a}$
split into the internal permutations ${\cal P}^{(L_{k})}$ of $L_{k}$ momenta
(taken at random out of the total list $\{ Q_{a} \} \; (a=1, ..., n)$) and the permutations
${\cal P}^{(L_{1},...,L_{N})}$ of the momenta among the groups $L_{k}$.
It is evident that due to the symmetry of the expression in eq.(\ref{6.42}) the summations over
${\cal P}^{(L_{k})}$ give just the factor $L_{1}! ... L_{N}!$. 
On the other hand the structure of the 
Bethe ansatz wave functions, eq.(\ref{3.14}) is such that 
for ordered particles positions 
in the summation over permutations the momenta $Q_{a}$ belonging
to the same cluster also remain ordered (see the discussion after eq.(\ref{6.5})).
Thus, in order to perform the summation over the permutations ${\cal P}^{(L_{1},...,L_{N})}$
in eq.(\ref{6.42}) it is sufficient to split the momenta of each cluster into $N$ parts:
\begin{equation}
 \label{6.44}
 \{
\underbrace{q_{1}^{\alpha}, ...,  q_{m^{1}_{\alpha}}^{\alpha}}_{m^{1}_{\alpha}} ; \;
\underbrace{q_{m^{1}_{\alpha}+1}^{\alpha}, ...,  q_{m^{1}_{\alpha}+m^{2}_{\alpha}}^{\alpha}}_{m^{2}_{\alpha}}  ; \; ... \; ; \;
\underbrace{q_{\sum_{k=1}^{N-1}m^{k}_{\alpha}+1}^{\alpha}, ...,  q_{\sum_{k=1}^{N}m^{k}_{\alpha}}^{\alpha}}_{m^{N}_{\alpha}}  \}
\end{equation}
where the integers $m^{k}_{\alpha} = 0, 1, ... , n_{\alpha}$ are constrained by the conditions
\begin{eqnarray}
 \label{6.45}
 \sum_{k=1}^{N} m^{k}_{\alpha} &=& n_{\alpha}
 \\
 \label{6.46}
 \sum_{\alpha=1}^{M} m^{k}_{\alpha} &=& L_{k}
\end{eqnarray}
and the momenta of every group
$\{ q_{\sum_{l=1}^{k-1}m^{l}_{\alpha}+1}^{\alpha}, \; ..., \;  q_{\sum_{l=1}^{k}m^{l}_{\alpha}}^{\alpha} \}$
belong to the particles whose coordinates are all equal $x_{k}$.
Let us redefine:
\begin{equation}
 \label{6.47}
 q_{\sum_{l=1}^{k-1}m^{l}_{\alpha}+r}^{\alpha} \; \equiv \; 
 q_{k,r}^{\alpha} \; = \; q_{\alpha} + \frac{i\kappa}{2}\bigl(n_{\alpha} + 1 - 2 \sum_{l=1}^{k-1}m^{l}_{\alpha} -2r\bigr)
\end{equation}
In this way the summation over ${\cal P}^{(L_{1},...,L_{N})}$ is changed by the summation over 
the integers $\{m^{k}_{\alpha}\}$. Substituting eqs.(\ref{6.44})-(\ref{6.47}) into eq.(\ref{6.42}) after simple
algebra we find
\begin{eqnarray}
 \label{6.48}
W({\bf f}; {\bf x}) &=& \lim_{t\to\infty} 
\sum_{M=0}^{\infty} \frac{(-1)^{M}}{M!}
\prod_{\alpha=1}^{M}
\Biggl[
\sum_{\sum_{k}^{N}m^{k}_{\alpha} \geq 1} (-1)^{\sum_{k}^{N}m^{k}_{\alpha} -1}
\int_{-\infty}^{+\infty} \frac{dq_{\alpha}}{2\pi}
\times
\\
\nonumber
\\
\nonumber
&\times&
\exp\Bigl\{\lambda(t)\sum_{k=1}^{N}m^{k}_{\alpha} f_{k} +i \sum_{k=1}^{N}m^{k}_{\alpha} x_{k} q_{\alpha} 
-\frac{1}{4} \kappa \sum_{k,l=1}^{N}m^{k}_{\alpha}m^{l}_{\alpha} \big|x_{k} - x_{l}\big|
-\frac{t}{2\beta} q_{\alpha}^{2} \sum_{k=1}^{N}m^{k}_{\alpha}
+\frac{\kappa^{2} t}{24\beta} \Bigl(\sum_{k=1}^{N}m^{k}_{\alpha}\Bigr)^{3} \Bigr\}
\Biggr]
\times
\\
\nonumber
\\
\nonumber
&\times&
D_{M}\bigl({\bf q}; \{ m^{k}_{\alpha}\}\bigr) \; G_{M}\bigl({\bf q}; \{ m^{k}_{\alpha}\}\bigr)
\end{eqnarray}
where 
\begin{equation}
 \label{6.49}
 G_{M}\bigl({\bf q}; \{ m^{k}_{\alpha}\}\bigr) =
 \prod_{\alpha=1}^{M} \prod_{k<l}^{N}\prod_{r=1}^{m^{k}_{\alpha}} \prod_{r'=1}^{m^{l}_{\alpha}}
 \Biggl(
 \frac{q_{k,r}^{\alpha} - q_{l,r'}^{\alpha} - i\kappa}{
 q_{k,r}^{\alpha} - q_{l,r'}^{\alpha} } 
 \Biggr)
 \times
\prod_{\alpha<\beta}^{M} \prod_{k=1}^{N}\prod_{l=1}^{N}\prod_{r=1}^{m^{k}_{\alpha}} \prod_{r'=1}^{m^{l}_{\alpha}}
\Biggl(
 \frac{q_{k,r}^{\alpha} - q_{l,r'}^{\alpha} - i\kappa}{
 q_{k,r}^{\alpha} - q_{l,r'}^{\alpha} } 
 \Biggr)
\end{equation}

Substituting here the expressions for $q_{k,r}^{\alpha}$, eq.(\ref{6.47}), one can find an explicit 
formula for the above factor $G_{M}$ which is rather cumbersome: it contains the products of all kinds
of the Gamma functions of the type 
$\Gamma\bigl[1 +\frac{1}{2}\bigl(\sum_{k}^{N}(\pm)m^{k}_{\alpha} + \sum_{l}^{N}(\pm)m^{l}_{\beta}\bigr) 
\pm \frac{1}{\kappa}(q_{\alpha}-q_{\beta})\bigr]$. (for the example of such kind of the product see 
eq.(\ref{E.18})). We do not reproduce it here as it turns out to be irrelevant in the limit 
$t\to\infty$ (see below).

Rescaling
\begin{eqnarray}
\label{6.50}
q_{\alpha} &\to& \frac{\kappa}{2\lambda} \, q_{\alpha}
\\
\nonumber
\\
\label{6.51}
x_{k} &\to&  \frac{2 \lambda^{2}}{\kappa} \, x_{k}
\end{eqnarray}
and substituting the Airy function relation
\begin{equation}
   \label{6.52}
\exp\Bigl\{ \frac{1}{3} \lambda^{3} \Bigl(\sum_{k}^{N}m^{k}_{\alpha}\Bigr)^{3} \Bigr\} \; = \;
\int_{-\infty}^{+\infty} dy \; \Ai(y) \;
\exp\Bigl\{\lambda \Bigl(\sum_{k}^{N}m^{k}_{\alpha}\Bigr) \, y \Bigr\}
\end{equation}
into eq.(\ref{6.48}) we get
\begin{eqnarray}
 \nonumber
W({\bf f}; {\bf x}) &=& \lim_{t\to\infty} 
\sum_{M=0}^{\infty} \frac{(-1)^{M}}{M!}
\prod_{\alpha=1}^{M}
\Biggl[
\int\int_{-\infty}^{+\infty}\frac{dq_{\alpha}dy_{\alpha}}{2\pi} \Ai(y_{\alpha}+q_{\alpha}^{2})
\times
\\
\nonumber
\\
\nonumber
&\times&
\sum_{\sum_{k}^{N}m^{k}_{\alpha} \geq 1} (-1)^{\sum_{k}^{N}m^{k}_{\alpha} -1}
\exp
\Bigl\{
\lambda(t)\sum_{k=1}^{N}m^{k}_
{\alpha} \bigl(y_{\alpha} + f_{k} +i x_{k} q_{\alpha} \bigr)
-\frac{1}{2} \lambda^{2}(t) \sum_{k,l=1}^{N}m^{k}_{\alpha}m^{l}_{\alpha} \Delta_{kl}
\Bigr\}
\Biggr]
\times
\\
\nonumber
\\
&\times&
\det\hat{K} \Bigl[ 
\bigl(\sum_{k}^{N}\lambda(t) m^{k}_{\alpha}, \; q_{\alpha}\bigr) ; \; \bigl(\sum_{k}^{N}\lambda(t) m^{k}_{\beta}, \; q_{\beta}\bigr)
\Bigr]_{\alpha,\beta=1,...,M}
\times
G_{M}\bigl(\frac{\kappa {\bf q}}{2\lambda(t)}; \; \{ m^{k}_{\alpha}\}\bigr)
\label{6.53}
\end{eqnarray}
where 
\begin{equation}
 \label{6.54}
 \Delta_{kl} \; = \; \big|x_{k} - x_{l}\big|
\end{equation}
and
\begin{equation}
 \label{6.55}
 \hat{K} \Bigl[ 
\bigl(\sum_{k}^{N}\lambda m^{k}_{\alpha}, \; q_{\alpha}\bigr) ; \; \bigl(\sum_{k}^{N}\lambda m^{k}_{\beta}, \; q_{\beta}\bigr)
\Bigr] = 
\frac{1}{\bigl(\sum_{k}^{N}\lambda m^{k}_{\alpha} - i q_{\alpha}\bigr) +
         \bigl(\sum_{k}^{N}\lambda m^{k}_{\beta} + i q_{\beta}\bigr)}
\end{equation}
The quadratic in $m^{k}_{\alpha}$ term in the exponential of eq.(\ref{6.53}) can be linearized as follows:
\begin{eqnarray}
 \nonumber
 \exp
 \Bigl\{
 - \frac{1}{2} \lambda^{2}  \sum_{k,l=1}^{N}m^{k}_{\alpha}m^{l}_{\alpha} \Delta_{kl} 
 \Bigr\}
 &=&
 \exp
 \Bigl\{
 - \frac{1}{4} \lambda^{2}  \sum_{k,l=1}^{N} \Delta_{kl} \bigl( m^{k}_{\alpha} + m^{l}_{\alpha}\bigr)^{2}
 + \frac{1}{2} \lambda^{2}  \sum_{k=1}^{N} \bigl(m^{k}_{\alpha}\bigr)^{2} \sum_{l=1}^{N} \Delta_{kl}
\Bigr\}
\\
\nonumber
\\
\nonumber
&=&
\prod_{k,l=1}^{N} 
\Biggl(
\int_{-\infty}^{+\infty} \frac{d \xi_{kl}^{\alpha}}{\sqrt{2\pi}}
\exp
 \Bigl\{
 -\frac{1}{2} \bigl(\xi_{kl}^{\alpha}\bigr)^{2}
 \Bigr\} 
 \Biggr) \; 
 \prod_{k=1}^{N} 
\Biggl(
\int_{-\infty}^{+\infty} \frac{d \eta_{k}^{\alpha}}{\sqrt{2\pi}}
\exp
 \Bigl\{
 -\frac{1}{2} \bigl(\eta_{k}^{\alpha}\bigr)^{2}
 \Bigr\} 
 \Biggr)
 \times
\\
\nonumber
\\
&\times&
 \exp
 \Bigl\{
\lambda  \sum_{k}^{N}\Bigl[ 
\frac{i}{\sqrt{2}}\sum_{l=1}^{N} \sqrt{\Delta_{kl}} \, \bigl(\xi_{kl}^{\alpha} + \xi_{lk}^{\alpha}\bigr)
-\sqrt{\gamma_{k}} \, \eta_{k}^{\alpha}
\Bigr] m^{k}_{\alpha}
\Bigr\}
\label{6.56}
\end{eqnarray}
where
\begin{equation}
 \label{6.57}
 \gamma_{k} \; = \; \sum_{l=1}^{N} \Delta_{kl} \; = \; 
 \sum_{l=1}^{N}  \big|x_{k} - x_{l}\big|
\end{equation}
Substituting the representation (\ref{6.56}) into eq.(\ref{6.53}) and redefining the integration parameters
\begin{equation}
 \label{6.58}
 \eta_{k}^{\alpha} \; \to \; 
 \eta_{k}^{\alpha} + \frac{i}{\sqrt{\gamma_{k}}} \, q_{\alpha} x_{k} + 
 i \sum_{l=1}^{N} \sqrt{\frac{\Delta_{kl}}{2\gamma_{k}}} \; \bigl(\xi_{kl}^{\alpha} + \xi_{lk}^{\alpha}\bigr)
\end{equation}
we get
\begin{eqnarray}
 \nonumber
W({\bf f}; {\bf x}) &=&
\sum_{M=0}^{\infty} \frac{(-1)^{M}}{M!}
\prod_{\alpha=1}^{M}
\Biggl[
\int\int_{-\infty}^{+\infty}\frac{dq_{\alpha}dy_{\alpha}}{2\pi} \Ai(y_{\alpha}+q_{\alpha}^{2})
\prod_{k,l=1}^{N} 
\Biggl(
\int_{-\infty}^{+\infty} \frac{d \xi_{kl}^{\alpha}}{\sqrt{2\pi}}
\Biggr)
\prod_{k=1}^{N} 
\Biggl(
\int_{-\infty}^{+\infty} \frac{d \eta_{k}^{\alpha}}{\sqrt{2\pi}}
\Biggr)
\times
\\
\nonumber
\\
&\times&
\exp
 \Biggl\{
 -\frac{1}{2} \sum_{k,l=1}^{N} \bigl(\xi_{kl}^{\alpha}\bigr)^{2}
 -\frac{1}{2} \sum_{k=1}^{N}
 \Bigl[
 \eta_{k}^{\alpha} + \frac{i}{\sqrt{\gamma_{k}}} \, q_{\alpha} x_{k} + 
 i \sum_{l=1}^{N} \sqrt{\frac{\Delta_{kl}}{2\gamma_{k}}} \; \bigl(\xi_{kl}^{\alpha} + \xi_{lk}^{\alpha}\bigr)
 \Bigr]^{2}
 \Biggr\}
 \Biggr]
 \times
 {\cal S}\bigl({\bf f}, {\bf y}, {\bf q}, \{\eta_{k}\} \bigr)
 \label{6.59}
\end{eqnarray}
where
\begin{eqnarray}
 \nonumber
 {\cal S}\bigl({\bf f}, {\bf y}, {\bf q}, \{\eta_{k}\} \bigr) &=&
 \lim_{t\to\infty}
 \prod_{\alpha=1}^{M}
 \Biggl[
 \sum_{\sum_{k}^{N}m^{k}_{\alpha} \geq 1} (-1)^{\sum_{k}^{N}m^{k}_{\alpha} -1}
\exp
\Bigl\{
\lambda(t) \sum_{k=1}^{N} m^{k}_{\alpha} \bigl(y_{\alpha} + f_{k} - \sqrt{\gamma_{k}} \eta_{k} \bigr)
\Bigr\} 
\times
\\
\nonumber
\\
&\times&
\det\hat{K} \Bigl[ 
\bigl(\sum_{k}^{N}\lambda(t) m^{k}_{\alpha}, \; q_{\alpha}\bigr) ; \; \bigl(\sum_{k}^{N}\lambda(t) m^{k}_{\beta}, \; q_{\beta}\bigr)
\Bigr]_{\alpha,\beta=1,...,M}
\times
G_{M}\bigl(\frac{\kappa {\bf q}}{2\lambda(t)}; \; \{ m^{k}_{\alpha}\}\bigr)
\label{6.60}
\end{eqnarray}
The summations over $m^{k}_{\alpha}$ in the above expression can be performed as follows
\begin{eqnarray}
 \nonumber
 {\cal S}\bigl({\bf f}, {\bf y}, {\bf q}, \{\eta_{k}\} \bigr) &=&
 \lim_{t\to\infty}
 \prod_{\alpha=1}^{M}
 \Biggl[
 \prod_{k=1}^{N} 
 \Biggl(
 \sum_{m^{k}_{\alpha}=0}^{\infty} \, \delta_{m^{k}_{\alpha}, \, 0}
 \Biggr)
 - (-1)^{N}
\prod_{k=1}^{N} 
 \Biggl(
 \sum_{m^{k}_{\alpha}=0}^{\infty} (-1)^{m^{k}_{\alpha}-1}
 \exp
\Bigl\{
\lambda(t)  m^{k}_{\alpha} \bigl(y_{\alpha} + f_{k} - \sqrt{\gamma_{k}} \eta_{k} \bigr)
\Bigr\} 
\Biggr)
\Biggr]
\times
\\
\nonumber
\\
\nonumber
&\times&
\det\hat{K} \Bigl[ 
\bigl(\sum_{k}^{N}\lambda(t) m^{k}_{\alpha}, \; q_{\alpha}\bigr) ; \; \bigl(\sum_{k}^{N}\lambda(t) m^{k}_{\beta}, \; q_{\beta}\bigr)
\Bigr]_{\alpha,\beta=1,...,M}
\times
G_{M}\bigl(\frac{\kappa {\bf q}}{2\lambda(t)}; \; \{ m^{k}_{\alpha}\}\bigr)
\\
\nonumber
\\
\nonumber
\\
\nonumber
&=&
\lim_{t\to\infty}
 \prod_{\alpha=1}^{M}
 \Biggl[
 \prod_{k=1}^{N} 
 \Biggl(
 \int_{{\cal C}} dz^{k}_{\alpha}  \, \delta(z^{k}_{\alpha})
 \Biggr)
  - (-1)^{N}
\prod_{k=1}^{N} 
 \Biggl(
 \int_{{\cal C}} \frac{dz^{k}_{\alpha}}{2i\sin(\pi z^{k}_{\alpha})} 
 \exp
\Bigl\{
\lambda(t)  z^{k}_{\alpha} \bigl(y_{\alpha} + f_{k} - \sqrt{\gamma_{k}} \eta_{k} \bigr)
\Bigr\} 
\Biggr)
\Biggr]
\\
\nonumber
\\
&\times&
\det\hat{K} \Bigl[ 
\bigl(\sum_{k}^{N}\lambda(t) z^{k}_{\alpha}, \; q_{\alpha}\bigr) ; \; \bigl(\sum_{k}^{N}\lambda(t) z^{k}_{\beta}, \; q_{\beta}\bigr)
\Bigr]_{\alpha,\beta=1,...,M}
\times
G_{M}\bigl(\frac{\kappa {\bf q}}{2\lambda(t)}; \; \{ z^{k}_{\alpha}\}\bigr)
\label{6.61}
\end{eqnarray}
where the integration goes over the contour ${\cal C}$ shown in Fig.3.
Redefining $z^{k}_{\alpha} \to z^{k}_{\alpha}/\lambda(t)$, in the limit $t\to\infty$ we get

\begin{figure}[h]
\begin{center}
   \includegraphics[width=6.0cm]{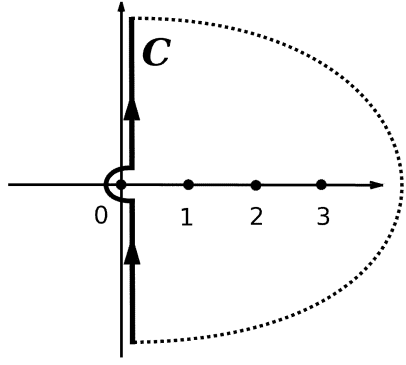}
\caption[]{The contours of integration in the complex plane used for
           summing the series eq.(\ref{6.61})}
\end{center}
\label{figure3}
\end{figure}
\begin{eqnarray}
 \nonumber
 {\cal S}\bigl({\bf f}, {\bf y}, {\bf q}, \{\eta_{k}\} \bigr) &=&
 \prod_{\alpha=1}^{M}
 \Biggl[
 \prod_{k=1}^{N} 
 \Biggl(
 \int_{{\cal C}} dz^{k}_{\alpha}  \, \delta(z^{k}_{\alpha})
 \Biggr)
  - (-1)^{N}
\prod_{k=1}^{N} 
 \Biggl(
 \int_{{\cal C}} \frac{dz^{k}_{\alpha}}{2\pi i \, z^{k}_{\alpha}} 
 \exp
\Bigl\{
 z^{k}_{\alpha} \bigl(y_{\alpha} + f_{k} - \sqrt{\gamma_{k}} \eta_{k} \bigr)
\Bigr\} 
\Biggr)
\Biggr]
\\
\nonumber
\\
&\times&
\det\hat{K} \Bigl[ 
\bigl(\sum_{k}^{N} z^{k}_{\alpha}, \; q_{\alpha}\bigr) ; \; \bigl(\sum_{k}^{N} z^{k}_{\beta}, \; q_{\beta}\bigr)
\Bigr]_{\alpha,\beta=1,...,M}
\times
\lim_{t\to\infty}
G_{M}\Bigl(\frac{\kappa {\bf q}}{2\lambda(t)}; \; \{ \frac{z^{k}_{\alpha}}{\lambda(t)}\}\Bigr)
\label{6.62}
\end{eqnarray}
Taking into account the Gamma function property $\lim_{|z|\to 0}\Gamma(1+z) = 1$ 
one can easily demonstrate (see e.g. eqs.(\ref{5.33}) or (\ref{6.21}))
that 
\begin{equation}
 \label{6.63}
 \lim_{t\to\infty}
G_{M}\Bigl(\frac{\kappa {\bf q}}{2\lambda(t)}; \; \{ \frac{z^{k}_{\alpha}}{\lambda(t)}\}\Bigr) \; = \; 1
\end{equation}
Thus in the limit $\lambda\to\infty$ the expression (\ref{6.59}) takes the form of the Fredholm determinant
\begin{eqnarray}
 \nonumber
W({\bf f}; {\bf x}) &=& 
\sum_{M=0}^{\infty} \frac{(-1)^{M}}{M!}
\prod_{\alpha=1}^{M}
\Biggl[
\int\int_{-\infty}^{+\infty}\frac{dq_{\alpha}dy_{\alpha}}{2\pi} \Ai(y_{\alpha}+q_{\alpha}^{2})
\prod_{k,l=1}^{N} 
\Biggl(
\int_{-\infty}^{+\infty} \frac{d \xi_{kl}^{\alpha}}{\sqrt{2\pi}}
\Biggr)
\prod_{k=1}^{N} 
\Biggl(
\int_{-\infty}^{+\infty} \frac{d \eta_{k}^{\alpha}}{\sqrt{2\pi}}
\Biggr)
\times
\\
\nonumber
\\
\nonumber
&\times&
\exp
 \Biggl\{
 -\frac{1}{2} \sum_{k,l=1}^{N} \xi_{kl}^{2}
 -\frac{1}{2} \sum_{k=1}^{N}
 \Bigl[
 \eta_{k}^{\alpha} + \frac{i}{\sqrt{\gamma_{k}}} \, q_{\alpha} x_{k} + 
 i \sum_{l=1}^{N} \sqrt{\frac{\Delta_{kl}}{2\gamma_{k}}} \; \bigl(\xi_{kl}^{\alpha} + \xi_{lk}^{\alpha}\bigr)
 \Bigr]^{2}
 \Biggr\}
 \times
 \\
\nonumber
\\
\nonumber
&\times&
 \prod_{k=1}^{N} 
 \Biggl(
 \int_{{\cal C}} dz^{k}_{\alpha} 
 \Biggr)
 \Biggl(
 \prod_{k=1}^{N} \, \delta(z^{k}_{\alpha}) \; - \; (-1)^{N}
 \prod_{k=1}^{N} 
 \frac{1}{2\pi i z^{k}_{\alpha}}
 \exp
\Bigl\{
 z^{k}_{\alpha} \bigl(y_{\alpha} + f_{k} - \sqrt{\gamma_{k}} \eta_{k} \bigr)
\Bigr\} 
 \Biggr)
 \Biggr]
 \times
  \label{6.64}
 \\
\nonumber
\\
&\times&
\det\hat{K} \Bigl[ 
\bigl(\sum_{k}^{N} z^{k}_{\alpha}, \; q_{\alpha}\bigr) ; \; \bigl(\sum_{k}^{N} z^{k}_{\beta}, \; q_{\beta}\bigr)
\Bigr]_{\alpha,\beta=1,...,M}
\\
\nonumber
\\
\nonumber
\\
&\equiv&
\det\bigl[\hat{1} - \hat{A}\bigr] \; = \; \exp\Bigl\{- \sum_{M=1}^{\infty} \frac{1}{M} \mbox{Tr} \hat{A}^{M}\Bigr\}
 \label{6.65}
\end{eqnarray} 
where $\hat{A}$ is the integral operator with the kernel
\begin{eqnarray}
 \nonumber
 A\Bigl[
 \Bigl(\sum_{k}^{N} z^{k}, \, q\Bigr); \; \Bigl(\sum_{k}^{N} \tilde{z}^{k}, \, \tilde{q}\Bigr)
 \Bigr]
 &=&
 \int_{-\infty}^{+\infty}\frac{dy}{2\pi} \Ai(y+q^{2})
\prod_{k,l=1}^{N} 
\Biggl(
\int_{-\infty}^{+\infty} \frac{d \xi_{kl}}{\sqrt{2\pi}}
\Biggr)
\prod_{k=1}^{N} 
\Biggl(
\int_{-\infty}^{+\infty} \frac{d \eta_{k}}{\sqrt{2\pi}}
\Biggr)
\times
 \\
\nonumber
\\
\nonumber
&\times&
\exp
 \Biggl\{
 -\frac{1}{2} \sum_{k,l=1}^{N} \xi_{kl}^{2}
 -\frac{1}{2} \sum_{k=1}^{N}
 \Bigl[
 \eta_{k} + \frac{i}{\sqrt{\gamma_{k}}} \, q_{\alpha} x_{k} + 
 i \sum_{l=1}^{N} \sqrt{\frac{\Delta_{kl}}{2\gamma_{k}}} \; \bigl(\xi_{kl} + \xi_{lk}\bigr)
 \Bigr]^{2}
 \Biggr\}
 \times
 \\
\nonumber
\\
\nonumber
&\times&
\Biggl(
 \prod_{k=1}^{N} \, \delta(z^{k}) \; - \; (-1)^{N}
 \prod_{k=1}^{N} 
 \frac{1}{2\pi i z^{k}}
 \exp
\Bigl\{
 z^{k} \bigl(y + f_{k} - \sqrt{\gamma_{k}} \eta_{k} \bigr)
\Bigr\} 
 \Biggr)
 \times
 \\
\nonumber
\\
&\times&
\frac{1}{\sum_{k}^{N} z^{k} \, - \, i q \; + \; \sum_{k}^{N} \tilde{z}^{k} \, + \, i \tilde{q}} 
\label{6.66}
\end{eqnarray}
Correspondingly, for the trace of this operator in the $M$-th power 
(in the exponential representation of the 
Fredholm determinant, eq.(\ref{6.65})) we get
\begin{eqnarray}
 \nonumber
 \mbox{Tr} \hat{A}^{M} &=&
 \prod_{\alpha=1}^{M}
 \Biggl[
 \int\int_{-\infty}^{+\infty}\frac{dy dq_{\alpha}}{2\pi} \Ai(y+q_{\alpha}^{2})
\prod_{k,l=1}^{N} 
\Biggl(
\int_{-\infty}^{+\infty} \frac{d \xi_{kl}}{\sqrt{2\pi}}
\Biggr)
\prod_{k=1}^{N} 
\Biggl(
\int_{-\infty}^{+\infty} \frac{d \eta_{k}}{\sqrt{2\pi}}
\Biggr)
\times
 \\
\nonumber
\\
\nonumber
&\times&
\exp
 \Biggl\{
 -\frac{1}{2} \sum_{k,l=1}^{N} \xi_{kl}^{2}
 -\frac{1}{2} \sum_{k=1}^{N}
 \Bigl[
 \eta_{k} + \frac{i}{\sqrt{\gamma_{k}}} \, q_{\alpha} x_{k} + 
 i \sum_{l=1}^{N} \sqrt{\frac{\Delta_{kl}}{2\gamma_{k}}} \; \bigl(\xi_{kl} + \xi_{lk}\bigr)
 \Bigr]^{2}
 \Biggr\}
 \times
 \\
\nonumber
\\
\nonumber
&\times&
\prod_{k=1}^{N}\Biggl(\int_{{\cal C}} dz^{k}_{\alpha} \Biggr)
\Biggl(
 \prod_{k=1}^{N} \, \delta(z^{k}_{\alpha}) \; - \; (-1)^{N}
 \prod_{k=1}^{N} 
 \frac{1}{2\pi i z^{k}_{\alpha}}
 \exp
\Bigl\{
 z^{k}_{\alpha} \bigl(y + f_{k} - \sqrt{\gamma_{k}} \eta_{k} \bigr)
\Bigr\} 
 \Biggr)
 \Biggr]
 \times
 \\
\nonumber
\\
&\times&
\prod_{\alpha=1}^{M}
\Biggl(
\frac{1}{\sum_{k}^{N} z^{k}_{\alpha} \, - \, i q_{\alpha} \; + \; 
         \sum_{k}^{N} z^{k}_{\alpha +1} \, + \, i q_{\alpha + 1}}
\Biggr)
\label{6.67}
\end{eqnarray}
where, by definition, $z^{k}_{M+1} \equiv z^{k}_{1}$ and $q_{M+1} \equiv q_{1}$.

Substituting
\begin{equation}
 \label{6.68}
 \frac{1}{\sum_{k}^{N} z^{k}_{\alpha} \, - \, i q_{\alpha} \; + \; 
         \sum_{k}^{N} z^{k}_{\alpha +1} \, + \, i q_{\alpha + 1}} \; = \; 
\int_{0}^{\infty} d\omega_{\alpha}
\exp\Bigl\{
-\omega_{\alpha} \Bigl(
\sum_{k}^{N} z^{k}_{\alpha} \, - \, i q_{\alpha} \; + \; 
\sum_{k}^{N} z^{k}_{\alpha +1} \, + \, i q_{\alpha + 1}
\Bigr)
\Bigr\}
\end{equation}
into eq.(\ref{6.67}) we obtain
\begin{equation}
 \label{6.69}
 \mbox{Tr} \hat{A}^{M} \; = \; 
 \int_{0}^{\infty} ... \int_{0}^{\infty} d\omega_{1} ... d\omega_{M} \; 
 \prod_{\alpha=1}^{M}
 A\bigl(\omega_{\alpha} ; \; \omega_{\alpha + 1}\bigr)
\end{equation}
where
\begin{eqnarray}
 \nonumber
 A\bigl(\omega ; \; \omega'\bigr)
 &=&
 \int\int_{-\infty}^{+\infty}\frac{dy dq}{2\pi} \Ai(y+q^{2} + \omega + \omega')
\prod_{k,l=1}^{N} 
\Biggl(
\int_{-\infty}^{+\infty} \frac{d \xi_{kl}}{\sqrt{2\pi}}
\Biggr)
\prod_{k=1}^{N} 
\Biggl(
\int_{-\infty}^{+\infty} \frac{d \eta_{k}}{\sqrt{2\pi}}
\Biggr)
\times
 \\
\nonumber
\\
\nonumber
&\times&
\exp
 \Biggl\{
 -\frac{1}{2} \sum_{k,l=1}^{N} \xi_{kl}^{2}
 -\frac{1}{2} \sum_{k=1}^{N}
 \Bigl[
 \eta_{k} + \frac{i}{\sqrt{\gamma_{k}}} \, q x_{k} + 
 i \sum_{l=1}^{N} \sqrt{\frac{\Delta_{kl}}{2\gamma_{k}}} \; \bigl(\xi_{kl} + \xi_{lk}\bigr)
 \Biggr]^{2}
 -i q (\omega - \omega')
 \Bigr\}
 \times
 \\
\nonumber
\\
&\times&
\Biggl( 1 \;  - \; (-1)^{N}
 \prod_{k=1}^{N} \int_{{\cal C}} 
 \frac{dz^{k}}{2\pi i z^{k}}
 \exp
\Bigl\{
 z^{k} \bigl(y + f_{k} - \sqrt{\gamma_{k}} \eta_{k} \bigr)
\Bigr\} 
 \Biggr)
 \label{6.70}
\end{eqnarray}
Integrating over $z^{1}, ..., z^{N}$ we finally get
\begin{eqnarray}
 \nonumber
 A\bigl(\omega ; \; \omega'\bigr)
 &=&
 \int\int_{-\infty}^{+\infty}\frac{dy dq}{2\pi} \Ai(y+q^{2} + \omega + \omega')
\prod_{k,l=1}^{N} 
\Biggl(
\int_{-\infty}^{+\infty} \frac{d \xi_{kl}}{\sqrt{2\pi}}
\Biggr)
\prod_{k=1}^{N} 
\Biggl(
\int_{-\infty}^{+\infty} \frac{d \eta_{k}}{\sqrt{2\pi}}
\Biggr)
\times
 \\
\nonumber
\\
\nonumber
&\times&
\exp
 \Biggl\{
 -\frac{1}{2} \sum_{k,l=1}^{N} \xi_{kl}^{2}
 -\frac{1}{2} \sum_{k=1}^{N}
 \Bigl[
 \eta_{k} + \frac{i}{\sqrt{\gamma_{k}}} \, q x_{k} + 
 i \sum_{l=1}^{N} \sqrt{\frac{\Delta_{kl}}{2\gamma_{k}}} \; \bigl(\xi_{kl} + \xi_{lk}\bigr)
 \Bigr]^{2}
 -i q (\omega - \omega')
 \Biggr\}
 \times
 \\
\nonumber
\\
&\times&
\Biggl[ 1 \;  - \; (-1)^{N}
 \prod_{k=1}^{N} 
 \theta \Bigl( -y - f_{k} + \eta_{k} \sqrt{\gamma_{k}} \; \Bigr)
 \Biggr]
 \label{6.71}
\end{eqnarray}
where $\Delta_{kl} = \big|x_{k} - x_{l}\big|$ and $\gamma_{k} = \sum_{l=1}^{N}\Delta_{kl}$. 
 
Thus the $N$-point free energy distribution function 
$W\bigl(f_{1}, ..., f_{N}; \; x_{1}, ..., x_{N}\bigr)$, eq.(\ref{6.39}), is given by the Fredholm determinant
\begin{equation}
 \label{6.72}
 W\bigl({\bf f}; \; {\bf x}\bigr) \; = \; \det\bigl[ \hat{1} \; - \; \hat{A} \bigr]
\end{equation}
where $\hat{A}$ is the integral operator with the kernel $A\bigl(\omega ; \; \omega'\bigr)$  
(with $\omega, \omega' \geq 0$) represented in eq.(\ref{6.71}).


\vspace{10mm}

\section{Probability distribution function of the endpoint fluctuations}

\newcounter{7}
\setcounter{equation}{0}
\renewcommand{\theequation}{7.\arabic{equation}}

In this Chapter we consider the problem in which the polymer is fixed
at the origin, $\phi(0)=0$ and it is free at $\tau = t$.
In other words, for a given realization of the random potential
$V$ the partition function of the considered system is:
\begin{equation}
\label{7.1}
   Z(t) = \int_{-\infty}^{+\infty} dx \; Z(x; t) \; = \; \exp\{-\beta F(t)\}
\end{equation}
where
\begin{equation}
\label{7.2}
   Z(x; t) = \int_{\phi(0)=0}^{\phi(t)=x}
              {\cal D} \phi(\tau)  \;  \exp\bigl\{-\beta H[\phi; V] \bigr\}
\end{equation}
is the partition function of the system with the fixed boundary condition,
$\phi(t)=x$ and $F(t)$ is the total free energy.
As it was discussed in the Introduction, besides the usual extensive   part $f_{0} t$
(where $f_{0}$ is the linear free energy density),
the total free energy $F$ of such system contains the disorder dependent
fluctuating contribution $\propto  t^{1/3}$.
In other words, in the limit of large $t$ the total (random) free energy of the system
can be represented as $\beta F(t) =  \beta f_{0} t  +  \lambda(t) f$,
where $\lambda(t) = \frac{1}{2} \bigl(\beta^{5} u^{2} t\bigr)^{1/3}$
and $f \sim 1$ is the random quantity which in the thermodynamic
limit $t\to\infty$ is described by the universal Tracy-Widom distribution 
function $P_{TW}(f)$ (see Chapters IV and V). Note that the trivial self-averaging  
contribution $f_{0}t$ to the free energy can be eliminated 
by the simple redefinition of the partition function,
$Z = \exp\{-\beta f_{0} t\} \, \tilde{Z}$, so that $\tilde{Z}  =  \exp\{-\lambda f\}$, 
(see Chapter II).

In this Chapter instead of the free energy we are going to study on the statistical
properties of the transverse fluctuations of the directed polymer itself.
The scaling properties of the typical value of the endpoint  deviations,
$\phi(t)$, at large times is well known:
$ \overline{\langle\phi(t)\rangle^{2}} \; \propto \; t^{4/3} \; \; $
 \cite{hhf_85,numer1,numer2,kardar_87}.
Much more interesting object is the probability distribution function
of the rescaled quantity $x = \phi(t)/t^{2/3}$
which  in the limit $t \to \infty$  becomes a universal function 
\cite{math1, math2, math3, end-point,math4}. This distribution function 
can be defined as follows:
\begin{equation}
\label{7.3}
W(x) \; = \; \lim_{t\to\infty} \;
\mbox{Prob}\bigl[\phi(t)t^{-2/3} \; > \; x\bigr] 
\end{equation}
In this Chapter it will be shown that 
\begin{equation}
 \label{7.4}
W(x) \; = \; \int_{-\infty}^{+\infty} df \;
F_{1}(-f)
\int_{0}^{+\infty} d\omega \int_{0}^{+\infty} d\omega'
\bigl(\hat{1} - \hat{K}\bigr)^{-1}(\omega,\omega') \;
\Phi(\omega',\omega; \; f,x)
\end{equation}
Here $\hat{K}$ is the integral operator with the kernel
$B(\omega,\omega') = \Ai(\omega+\omega'-f) \; \;  (\omega, \omega' \, > \, 0)$,
the function $F_{1}(-f) = \det\bigl[\hat{1} -  \hat{K} \bigr]$
is the GOE Tracy-Widom distribution (see Chapter V),
$\bigl(\hat{1} - \hat{K}\bigr)^{-1}(\omega,\omega')$
denotes the kernel of the inverse operator $\bigl(\hat{1} - \hat{K}\bigr)^{-1}$
in $\omega$ and $\omega'$  and finally the function $\Phi(\omega',\omega; \; f,x)$
is:
\begin{eqnarray}
 \nonumber
\Phi(\omega,\omega'; \; f,x) = -\frac{1}{2} \int_{0}^{+\infty} dy \;
&\Biggl[&
\Bigl(
\frac{\partial}{\partial \omega} - \frac{\partial}{\partial \omega'}
\Bigr)
\Psi\bigl(\omega - \frac{1}{2}f + y ; \; x \bigr)
\Psi\bigl(\omega' - \frac{1}{2}f + y ; \; -x\bigr) +
\\
\nonumber
\\
&+&
\Bigl(
\frac{\partial}{\partial \omega} + \frac{\partial}{\partial \omega'}
\Bigr)
\Psi\bigl(\omega - \frac{1}{2}f - y ; \; x\bigr)
\Psi\bigl(\omega' - \frac{1}{2}f + y ; \; -x\bigr)
\Biggr]
\label{7.5}
\end{eqnarray}
where
\begin{equation}
\label{7.6}
\Psi(\omega; x) \; = \;
2^{1/3} \mbox{Ai}\Bigl[2^{1/3}\bigl(\omega + \frac{1}{8} x^{2}\bigr)\Bigr] \, 
\exp\bigl\{-\frac{1}{2} \omega x\bigr\}
\end{equation}

Unfortunately the above expression for the distribution function is rather
complicated and its analytic properties is not so easy
to analyze although the asymptotic tail of this function  at $|x| \to \infty$
is already known to decay as $ \sim  \exp\{- |x|^{3}/12\}$ \cite{math2}.


\vspace{5mm}

In terms of the partition function $\tilde{Z}(x; t)$
the probability distribution function of the polymer's endpoint $W(x)$,
eq.(\ref{7.3}), can be defined as follows:
\begin{equation}
W(x) \, = \,
\lim_{t\to\infty} \,
\overline{\Biggl(
\int_{x}^{+\infty} dx'
\frac{\tilde{Z}(x';t)}{\int_{-\infty}^{+\infty} dx' \; \tilde{Z}(x'; t)}
\Biggr)}
\; = \;
\lim_{t\to\infty} \,
\overline{\Biggl(
\frac{\tilde{Z}^{(+)}(x; t)}{\tilde{Z}^{(+)}(x; t) + \tilde{Z}^{(-)}(x; t)}
\Biggr)}
\label{7.7}
\end{equation}
Here, as usual, $\overline{(...)}$ denotes the averaging over the disorder potential and 
\begin{eqnarray}
\label{7.8}
\tilde{Z}^{(-)}(x; t) & \equiv &
\int_{-\infty}^{x} dx' \; \tilde{Z}(x'; t)
\; = \;
\exp\{-\lambda(t) f_{(-)}\}
\\
\nonumber
\\
\tilde{Z}^{(+)}(x; t) & \equiv &
\int_{x}^{+\infty} dx' \; \tilde{Z}(x'; t)
\; = \;
\exp\{-\lambda(t) f_{(+)}\}
\label{7.9}
\end{eqnarray}
where $f_{(\pm)}$ are the free energies of the polymers with the
endpoint $\phi(t)$ located correspondingly above and below  a given
position $x$ and 
\begin{equation}
 \label{7.10}
 \lambda(t) \; = \; \frac{1}{2} \bigl(\beta^{5} u^{2} t\bigr)^{1/3} \; \propto \; t^{1/3}
\end{equation}
which implies that the limit $t \to \infty$ is equivalent to 
the limit $\lambda \to \infty$.
According to the above definitions, eqs.(\ref{7.8})-(\ref{7.9}), 
\begin{equation}
\label{7.11}
\lim_{\lambda\to\infty} \;
\frac{\exp\{-\lambda f_{(+)}\} }{
\exp\{-\lambda f_{(-)}\} \,+ \, \exp\{-\lambda f_{(+)}\} }
\; =\;  \left\{
\begin{array}{ll}
0 , \; \; \; \;  \mbox{for} \; \;  f_{(-)} < f_{(+)} \\
\\
1 , \; \; \; \;  \mbox{for} \; \;  f_{(-)} > f_{(+)}
\end{array}
\right.
\end{equation}
Let us introduce the joint probability density function
${\cal P}_{x}\bigl[f_{(+)} ; f_{(-)}\bigr]$ such that the quantity
${\cal P}_{x}\bigl[f_{(+)} ; f_{(-)}\bigr] \, df_{(+)} \, df_{(-)}$
gives the probability that the free energy of the polymer
with the endpoint located below $x$ is equal to $f_{(-)}$
(within the interval $df_{(-)}$), while
the free energy of the polymer
with the endpoint located above $x$ is equal to $f_{(+)}$
(within the interval $df_{(+)}$).
Thus, according to eq.(\ref{7.11}),
\begin{equation}
\label{7.12}
W(x) \; = \; \lim_{\lambda \to \infty} \,
\int_{-\infty}^{+\infty} df_{(+)}
\int_{f_{(+)}}^{+\infty} df_{(-)}
{\cal P}_{x}\bigl[f_{(+)} ; f_{(-)}\bigr]
\end{equation}
Let us introduce one more joint probability distribution function:
\begin{equation}
\label{7.13}
{\cal W}_{x}(f_{1}, f_{2}) \; = \; \lim_{\lambda\to\infty}
\mbox{Prob}\bigl[f_{(+)} > f_{1} ; \;  f_{(-)} > f_{2} \bigr] \; = \;
 \lim_{\lambda\to\infty}
\int_{f_{1}}^{+\infty} df_{(+)}
\int_{f_{2}}^{+\infty} df_{(-)} \,
{\cal P}_{x}\bigl[f_{(+)} ; f_{(-)}\bigr]
\end{equation}
This two-point
free energy distribution function  gives the probability
that the free energy of the polymer with the endpoint located above the position $x$
is bigger than a given value $f_{1}$, while the free energy of the polymer with
the endpoint located below the position $x$
is bigger than a given value $f_{2}$.
According to this definition,
\begin{equation}
\label{7.14}
{\cal P}_{x}\bigl[f_{1} ; f_{2}\bigr] \; = \;
\frac{\partial}{\partial f_{1}}
\,
\frac{\partial}{\partial f_{2}}
\,
{\cal W}_{x}(f_{1}, f_{2})
\end{equation}
Substituting this relation into eq.(\ref{7.11}) we find
\begin{equation}
\label{7.15}
W(x) \; = \; \lim_{\lambda\to\infty}
\int_{-\infty}^{+\infty} df_{1}
\int_{f_{1}}^{+\infty} df_{2}
\frac{\partial}{\partial f_{1}}
\,
\frac{\partial}{\partial f_{2}}
\,
{\cal W}_{x}(f_{1}, f_{2})
\end{equation}
Integrating by parts over $f_{2}$ and taking into account that
${\cal W}_{x}(f_{1}, f_{2})\big|_{f_{2}=+\infty} = 0$ we get
\begin{equation}
\label{7.16}
W(x) \; = \; \lim_{\lambda\to\infty}
- \int_{-\infty}^{+\infty} df_{1}
\Bigl(
\frac{\partial}{\partial f_{1}}
\,
{\cal W}_{x}(f_{1}, f_{2})
\Bigr)\Big|_{f_{2}=f_{1}+0}
\end{equation}
Thus, to get the distribution function $W(x)$ for the polymer's endpoint
fluctuations we have to derive the
two-point free energy distribution function ${\cal W}_{x}(f_{1}, f_{2})$ first.
Note that this function is different from the
two-point free energy distribution function derived in Chapter VI
which describes joint statistics of the free energies of the directed polymers
coming to two given endpoints.

\vspace{5mm}

According to the definition, eq.(\ref{7.12}), the probability distribution function
$V_{x}(f_{1}, f_{2})$ can be defined as follows:
\begin{equation}
\label{7.17}
{\cal W}_{x}(f_{1}, f_{2}) = \lim_{\lambda\to\infty}
\sum_{L=0}^{\infty} \sum_{R=0}^{\infty}
\frac{(-1)^{L}}{L!} \frac{(-1)^{R}}{R!}
\exp\bigl\{\lambda L f_{1} + \lambda R f_{2}\bigr\} \;
\overline{\bigl[Z^{(+)}(x)\bigr]^{L} \, \bigl[Z^{(-)}(x)\bigr]^{R}}
\end{equation}
Substituting  eqs.(\ref{7.7})-(\ref{7.8}) into 
 eq.(\ref{7.17})  we get:
\begin{equation}
\label{7.18}
{\cal W}_{x}(f_{1}, f_{2}) = \lim_{\lambda\to\infty}
\sum_{L,R=0}^{\infty}
\frac{(-1)^{L+R}}{L!R!}
\exp\bigl\{\lambda L f_{1} + \lambda R f_{2}\bigr\} \;
\int_{-\infty}^{x} dx_{1}...dx_{L}
\int_{x}^{+\infty} dy_{1}...dy_{R}
\Psi(x_{1},...,x_{L},y_{R},...,y_{1} ; t)
\end{equation}
where the wave function
\begin{equation}
\label{7.19}
\Psi(x_{1}, ..., x_{N} ; t) \; \equiv \;
\overline{Z(x_{1}) \, Z(x_{2}) \, ... \, Z(x_{N})} 
\end{equation}
is given in eqs.(\ref{3.14})-(\ref{3.15}).

Further calculations of ${\cal W}_{x}(f_{1}, f_{2})$, eq.(\ref{7.18}) to a large extent 
repeats the procedure described in detail in the Chapter V 
for the GOE Tracy-Widom free energy distribution function. The only two differences of the present 
procedure from the one in Chapter V are: (1) instead of the exponential factor $\lambda N f \; = \; \lambda(K+L)f$ 
in the expression for the probability function $W(f)$ eq.(\ref{2.19}), 
we now have the factor $\lambda L f_{1} + \lambda R f_{2}$ in eq.(\ref{7.18}); (2) in the integrations 
over $x_{a}$'s and $y_{a}$'s  the regions $(-\infty; 0]$ and $[0; +\infty)$ in eqs.(\ref{5.7}) and (\ref{5.11})
are changed correspondingly for $(-\infty; x]$ and $[x; +\infty)$ in eq.(\ref{7.18})
Following the steps given in eqs.(\ref{5.7})-(\ref{5.15}) instead of eq.(\ref{5.16}) get
\begin{eqnarray}
\nonumber
{\cal W}_{x}(f_{1}, f_{2}) &=&\lim_{\lambda\to\infty}
\sum_{L,R=0}^{\infty} \;(-1)^{L+R} \;
\exp\bigl\{\lambda L f_{1} + \lambda R f_{2} \bigr\} \times
\\
 \label{7.20}
\\
\nonumber
&\times&
\sum_{M=1}^{\infty} \frac{1}{M!}
\prod_{\alpha=1}^{M} \Biggl[
\sum_{n_{\alpha}=1}^{\infty} \,  
\int_{-\infty}^{+\infty} \frac{dq_{\alpha} \, \kappa^{n_{\alpha}}}{2\pi} 
\exp\Bigl\{
-\frac{t}{2\beta} \, n_{\alpha} q_{\alpha}^{2} \, + \, 
\frac{\kappa^{2}\, t}{24\beta} n_{\alpha}^{3}
\Bigr\}
\Biggr]
D_{M}({\bf q}, {\bf n})
 J_{L,R} ({\bf q}, {\bf n}; \, x)  \; \, 
{\boldsymbol \delta}\Bigl(\sum_{\alpha=1}^{M} n_{\alpha},  K+L\Bigr)
\end{eqnarray}
where
\begin{equation}
 \label{7.21}
D_{M}({\bf q}, {\bf n}) \; = \;  
\det\Biggl[
  \frac{1}{\frac{1}{2}\kappa n_{\alpha} - i q_{\alpha}
   + \frac{1}{2}\kappa n_{\beta} + iq_{\beta}}\Biggr]_{\alpha,\beta=1,...M} 
\end{equation}
and (s.f. eq.(\ref{5.15}))
\begin{eqnarray}
\nonumber
J_{L,R} ({\bf q}, {\bf n}; \, x) &=& i^{-(L+R)} \;
\exp\bigl\{i x \sum_{\alpha=1}^{M} n_{\alpha} q_{\alpha} \bigr\}
\sum_{{\cal P}^{(L,R)}}  \; \;
\prod_{a=1}^{L} \prod_{c=1}^{R}
\Biggl[
\frac{q_{{\cal P}_a^{(L)}} - q_{{\cal P}_c^{(R)}} - i \kappa}{q_{{\cal P}_a^{(L)}} - q_{{\cal P}_c^{(R)}}}
\Biggr]
\times
\\
\nonumber
\\
&\times&
\frac{1}{\prod_{a=1}^{L} q^{(-)}_{{\cal P}_{a}^{(L)}} }
\prod_{a<b}^{L}\Biggl[\frac{q^{(-)}_{{\cal P}_a^{(L)}} + q^{(-)}_{{\cal P}_b^{(L)}}  + i \kappa }{
q^{(-)}_{{\cal P}_a^{(L)}} + q^{(-)}_{{\cal P}_b^{(L)}}}\Biggr]
\times
\frac{(-1)^{R}}{\prod_{c=1}^{R} q^{(+)}_{{\cal P}_{c}^{(R)}} }
\prod_{c<d}^{R}\Biggl[\frac{q^{(+)}_{{\cal P}_c^{(R)}} + q^{(+)}_{{\cal P}_d^{(R)}}  - i \kappa }{
q^{(+)}_{{\cal P}_c^{(R)}} + q^{(+)}_{{\cal P}_d^{(R)}}}\Biggr]
\label{7.22}
\end{eqnarray}
The only difference of the above expression for $J_{L,R} ({\bf q}, {\bf n}; \, x)$ 
from $I_{L,R} ({\bf q}, {\bf n})$ , eq.(\ref{5.15}), is the presence of the additional factor 
$\exp\bigl\{i x \sum_{\alpha=1}^{M} n_{\alpha} q_{\alpha} \bigr\}$.
Following the derivation given in eqs.(\ref{5.17})-(\ref{5.23}) we obtain
\begin{eqnarray}
 \nonumber
{\cal W}_{x}(f_{1},f_{2}) &=& \lim_{\lambda\to\infty}
 \sum_{M=0}^{\infty} \; \frac{(-1)^{M}}{M!} \;
\prod_{\alpha=1}^{M}
\Biggl[
\int\int_{-\infty}^{+\infty} \frac{dy_{\alpha} dp_{\alpha}}{2\pi}
\Ai\bigl(y_{\alpha} + p_{\alpha}^{2} - i x p_{\alpha}\bigr)
\sum_{m_{\alpha}+s_{\alpha}\geq 1}^{\infty} (-1)^{m_{\alpha}+s_{\alpha}-1}
\times
\\
\nonumber
\\
\nonumber
&\times&
\exp\Bigl\{
\lambda m_{\alpha} (y_{\alpha} + f_{1}) +
\lambda s_{\alpha} (y_{\alpha} + f_{2})
\Bigr\} \;
{\cal G} \Bigl(\frac{p_{\alpha}}{\lambda}, \; m_{\alpha}, \; s_{\alpha}\Bigr) \;
\Biggr]
\times
\\
\nonumber
\\
&\times&
\det \hat{K}\bigl[(\lambda m_{\alpha},\, \lambda s_{\alpha}, \, p_{\alpha});
(\lambda m_{\beta}, \, \lambda s_{\beta}, \, p_{\beta})\bigr]_{\alpha,\beta=1,...,M}
\;
{\bf G}_{M} \Bigl(\frac{{\bf p}}{\lambda}, \; {\bf m}, \; {\bf s}\Bigr)
\Biggr\}
\label{7.23}
\end{eqnarray}
where
\begin{equation}
\label{7.24}
\hat{K}\bigl[(\lambda m, \, \lambda s, \, p); (\lambda m', \, \lambda s', \, p')\bigr]
\; = \;
\frac{1}{
\lambda m + \lambda s - ip +
\lambda m' + \lambda s' + ip'}
\end{equation}
and the explicit expression for the factors ${\cal G}\bigl(\frac{p_{\alpha}}{\lambda}, m_{\alpha}, s_{\alpha}\bigr)$
and ${\cal G}_{\alpha\beta}\bigl({\bf q}, \; {\bf m}, \; {\bf s}\bigr)$
are given in eqs.(\ref{5.24}) and (\ref{E.18}) correspondingly.
Performing summations over $\{m_{\alpha}\}$ and $\{s_{\alpha} \}$ (see eq.(\ref{5.25})-(\ref{5.33})) 
in the limit $\lambda \to \infty$ we find that the expression for the probability function, eq.(\ref{7.23})
takes the form of the Fredholm determinant (s.f. eq.(\ref{5.34}))
\begin{eqnarray}
 \nonumber
{\cal W}_{x}(f_{1},f_{2}) &=&
1 + \sum_{M=1}^{\infty} \; \frac{(-1)^{M}}{M!} \;
\prod_{\alpha=1}^{M}
\Biggl[
\int\int_{-\infty}^{+\infty} \frac{dy_{\alpha} dp_{\alpha}}{2\pi}
\Ai\bigl(y_{\alpha} + p_{\alpha}^{2} - i x p_{\alpha} \bigr)
\times
\\
\nonumber
\\
\nonumber
&\times&
\int\int_{{\cal C}'}
\frac{d{z_{1}}_{\alpha}d{z_{2}}_{\alpha}}{(2\pi i)^{2}}
\Bigl(
\frac{2\pi i}{{z_{1}}_{\alpha}} \delta({z_{2}}_{\alpha}) +
\frac{2\pi i}{{z_{2}}_{\alpha}} \delta({z_{1}}_{\alpha}) -
\frac{1}{{z_{1}}_{\alpha}{z_{2}}_{\alpha}}
\Bigr)
\Bigl(
1 + \frac{{z_{1}}_{\alpha}}{{z_{2}}_{\alpha} + i p_{\alpha}^{(-)}}
\Bigr)
\Bigl(
1 + \frac{{z_{2}}_{\alpha}}{{z_{1}}_{\alpha} - i p_{\alpha}^{(+)}}
\Bigr)
\times
\\
\nonumber
\\
\nonumber
&\times&
\exp\Bigl\{
{z_{1}}_{\alpha}\bigl(y_{\alpha} + f_{1}\bigr) +
{z_{2}}_{\alpha}\bigl(y_{\alpha} + f_{2}\bigr)
\Bigr\}
\Biggr]
\det \Bigl[
\frac{1}{{z_{1}}_{\alpha} + {z_{2}}_{\alpha} - i p_{\alpha} +
{z_{1}}_{\beta} + {z_{2}}_{\beta} + i p_{\beta}}
\Bigr]_{(\alpha,\beta)=1,2,...,M}
\\
\nonumber
\\
&=&
\det\bigl[\hat{1} \, - \, \hat{B} \bigr]
\label{7.25}
\end{eqnarray}
with the kernel (s.f. eq.(\ref{5.35}))
\begin{eqnarray}
 \nonumber
\hat{B}\bigl[({z_{1}}, \, {z_{2}}, \, p); ({z_{1}}', \, {z_{2}}', \, p')\bigr]
&=&
\int_{-\infty}^{+\infty} \frac{dy}{2\pi}
\Ai\bigl(y + p^{2} - i x p \bigr)
\Bigl(
\frac{2\pi i}{z_{1}} \delta(z_{2}) +
\frac{2\pi i}{z_{2}} \delta(z_{1}) -
\frac{1}{z_{1}{z_{2}}}
\Bigr)
\times
\\
\nonumber
\\
\nonumber
&\times&
\Bigl(
1 + \frac{z_{1}}{z_{2} + i p^{(-)}}
\Bigr)
\Bigl(
1 + \frac{z_{2}}{z_{1} - i p^{(+)}}
\Bigr)
\exp\Bigl\{
z_{1}\bigl(y + f_{1}\bigr) +
z_{2}\bigl(y + f_{2}\bigr)
\Bigr\}
\times
\\
\nonumber
\\
&\times&
\frac{1}{
{z_{1}} + {z_{2}} - ip +
{z_{1}}' + {z_{2}}' + ip'}
\label{7.26}
\end{eqnarray}
In the exponential representation of this determinant we get
\begin{equation}
 \label{7.27}
{\cal W}_{x}(f_{1},f_{2}) \; = \;
\exp\Bigl[-\sum_{M=1}^{\infty} \frac{1}{M} \; \mbox{Tr} \, \hat{B}^{M} \Bigr]
\end{equation}
where
\begin{eqnarray}
 \nonumber
\mbox{Tr} \, \hat{B}^{M} &=&
\prod_{\alpha=1}^{M}
\Biggl[
\int\int_{-\infty}^{+\infty} \frac{dy_{\alpha} dp_{\alpha}}{2\pi}
\Ai\bigl(y_{\alpha} + p_{\alpha}^{2} - i x p_{\alpha} \bigr)
\times
\\
\nonumber
\\
\nonumber
&\times&
\int\int_{{\cal C}'}
\frac{d{z_{1}}_{\alpha}d{z_{2}}_{\alpha}}{(2\pi i)^{2}}
\Bigl(
\frac{2\pi i}{{z_{1}}_{\alpha}} \delta({z_{2}}_{\alpha}) +
\frac{2\pi i}{{z_{2}}_{\alpha}} \delta({z_{1}}_{\alpha}) -
\frac{1}{{z_{1}}_{\alpha}{z_{2}}_{\alpha}}
\Bigr)
\Bigl(
1 + \frac{{z_{1}}_{\alpha}}{{z_{2}}_{\alpha} + i p_{\alpha}^{(-)}}
\Bigr)
\Bigl(
1 + \frac{{z_{2}}_{\alpha}}{{z_{1}}_{\alpha} - i p_{\alpha}^{(+)}}
\Bigr)
\times
\\
\nonumber
\\
&\times&
\exp\Bigl\{
{z_{1}}_{\alpha}\bigl(y_{\alpha} + f_{1}\bigr) +
{z_{2}}_{\alpha}\bigl(y_{\alpha} + f_{2}\bigr)
\Bigr\}
\Biggr]
\;
\prod_{\alpha=1}^{M}
\Biggl[
\frac{1}{
{z_{1}}_{\alpha} + {z_{2}}_{\alpha} - i p_{\alpha} +
{z_{1}}_{\alpha +1}  + {z_{2}}_{\alpha +1}  + i p_{\alpha +1}}
\Biggr]
\label{7.28}
\end{eqnarray}
Here, by definition, it is assumed that ${z_{i_{M +1}}} \equiv {z_{i_{1}}}$ ($i=1,2$)
and $p_{M +1} \equiv p_{1}$.
Substituting
\begin{equation}
\label{7.29}
\frac{1}{
{z_{1}}_{\alpha} + {z_{2}}_{\alpha} - i p_{\alpha} +
{z_{1}}_{\alpha +1}  + {z_{2}}_{\alpha +1}  + i p_{\alpha +1}}
\; = \;
\int_{0}^{\infty} d\omega_{\alpha}
\exp\Bigl[-\bigl(
{z_{1}}_{\alpha} + {z_{2}}_{\alpha} - i p_{\alpha} +
{z_{1}}_{\alpha +1}  + {z_{2}}_{\alpha +1}  + i p_{\alpha +1}
\bigr) \, \omega_{\alpha}
\Bigr]
\end{equation}
into eq.(\ref{7.28}), we obtain
\begin{equation}
 \label{7.30}
\mbox{Tr} \, \hat{B}^{M} \; = \;
\int_{0}^{\infty} d\omega_{1} \, ... \, d\omega_{M} \,
\prod_{\alpha=1}^{M}
\Biggl[
\int\int_{-\infty}^{+\infty} \frac{dy dp}{2\pi}
\Ai\bigl(y + p^{2} + \omega_{\alpha} + \omega_{\alpha -1} - ixp \bigr)
\exp\{i p \bigl(\omega_{\alpha} - \omega_{\alpha -1}\bigr)\}
\;
S\bigl(p,  y; f_{1}, f_{2} \bigr)
\Biggr]
\end{equation}
where, by definition, $\omega_{0} \equiv \omega_{M}$, and
\begin{eqnarray}
 \nonumber
S(p, y; f_{1}, f_{2}) &=&
\int\int_{{\cal C}'}
\frac{dz_{1}dz_{2}}{(2\pi i)^{2}}
\Bigl(
\frac{2\pi i}{z_{1}} \delta(z_{2}) +
\frac{2\pi i}{z_{2}} \delta(z_{1}) -
\frac{1}{z_{1}z_{2}}
\Bigr)
\Bigl(
1 + \frac{z_{1}}{z_{2} + i p^{(-)}}
\Bigr)
\Bigl(
1 + \frac{z_{2}}{z_{1} - i p^{(+)}}
\Bigr)
\times
\\
\nonumber
\\
&\times&
\exp\Bigl\{
z_{1}\bigl(y + f_{1}\bigr) +
z_{2}\bigl(y + f_{2}\bigr)
\Bigr\}
\label{7.31}
\end{eqnarray}
Simple integrations yields the following result:
\begin{eqnarray}
 \nonumber
S(p, y; f_{1}, f_{2}) &=&
  \theta(y+f_{1})
+ \theta(y+f_{2})
- \theta(y+f_{1})\theta(y+f_{2})
- \theta(y+f_{1})\theta(y+f_{2}) \exp\bigl\{i p (f_{1} - f_{2}) - 2\epsilon y\bigr\}
\\
\nonumber
\\
\nonumber
&+&
  \frac{i}{p+i\epsilon} \delta(y + f_{2})
- \frac{i}{p-i\epsilon} \delta(y + f_{1})
\\
\nonumber
\\
\nonumber
&-&
  \frac{i}{p+i\epsilon} \delta(y + f_{2}) \theta(f_{1}-f_{2})
\Bigl[1 - \exp\{i (p+i\epsilon) (f_{1} - f_{2})\}\Bigr]
\\
\nonumber
\\
&+&
  \frac{i}{p-i\epsilon} \delta(y + f_{1}) \theta(f_{2}-f_{1})
\bigl[1 - \exp\{i (p-i\epsilon) (f_{1} - f_{2})\}\bigr]
\label{7.32}
\end{eqnarray}
According to eq.(\ref{7.16}) in what follows we will be dealing with the sector $f_{2} > f_{1}$ only.
In this case the above expression  simplifies to
\begin{eqnarray}
 \nonumber
S(p, y; f_{1}, f_{2})\big|_{f_{2} > f_{1}} &=&
\Bigl(\frac{i}{p+i\epsilon} - \frac{i}{p-i\epsilon}\Bigr) \delta(y+f_{2})
\\
\nonumber
\\
\nonumber
&+&
\frac{i}{p-i\epsilon}\Bigl[\delta(y+f_{2}) - \delta(y+f_{1}) \exp\{i p (f_{1} - f_{2})\}\Bigr]
\\
\nonumber
\\
&+&
\theta(y+f_{2}) - \theta(y+f_{1}) \exp\{i p (f_{1} - f_{2})- 2\epsilon y\}
\label{7.33}
\end{eqnarray}
Note that at edge of the sector $f_{2} > f_{1}$ in the limit $\epsilon \to 0$
\begin{equation}
 \label{7.34}
S(p, y; f_{1}, f_{2})\big|_{f_{2} = f_{1}+0} \; = \; 2\pi \delta(p)\delta(y+f_{2})
\end{equation}

Substituting eq.(\ref{7.33})) into eq.(\ref{7.30}), we find that the free energy  distribution function 
${\cal W}_{x}(f_{1},f_{2})$, eq.(\ref{7.13}),
(in the sector $f_{2} > f_{1}$) is given by the Fredholm determinant,  eq.(\ref{7.27}), with the kernel
\begin{eqnarray}
 \nonumber
B(\omega, \omega') &=&
\Ai(\omega + \omega' - f_{2}) -
\\
\nonumber
\\
\nonumber
&-&
\int_{-\infty}^{+\infty} \frac{dp}{2\pi}
\frac{
\Bigl[
\Ai(\omega + \omega' + p^{2} - ipx - f_{2}) -
\Ai(\omega + \omega' + p^{2} - ipx - f_{1})
\exp\{i p (f_{1} - f_{2})\}
\Bigr]}{i (p-i\epsilon)}
\exp\{i p \bigl(\omega - \omega'\bigr)\} +
\\
\nonumber
\\
\nonumber
&+&
\int_{-\infty}^{+\infty} \frac{dp}{2\pi}
\Biggl[
\int_{-f_{2}}^{+\infty} dy
\Ai(\omega + \omega' + p^{2} - ipx + y) -
\int_{-f_{1}}^{+\infty} dy
\Ai(\omega + \omega' + p^{2} - ipx + y)
\exp\{i p (f_{1} - f_{2})\}
\Biggr]
\times
\\
\nonumber
\\
&\times&
\exp\{i p \bigl(\omega - \omega'\bigr)\}
\label{7.35}
\end{eqnarray}
with $\omega, \omega' > 0$.

Finally, 
substituting the above result, eqs.(\ref{7.27}) and (\ref{7.35}), into eq.(\ref{7.16})
for the endpoint distribution function one obtains the following expression:
\begin{equation}
 \label{7.36}
W(x) \; = \; \int_{-\infty}^{+\infty} df \;
F_{1}(-f)
\int_{0}^{+\infty} d\omega\int_{0}^{+\infty}  d\omega'
\Bigl(\hat{1} - \hat{K}\Bigr)^{-1} (\omega,\omega') \;
\Phi(\omega', \omega; f,x)
\end{equation}
where
\begin{equation}
 \label{7.37}
F_{1}(-f) = \det\bigl[\hat{1} \, - \, \hat{K} \bigr] \; = \;
\exp\Bigl[-\sum_{M=1}^{\infty} \frac{1}{M} \; \mbox{Tr} \, \hat{K}^{M} \Bigr]
\end{equation}
is the GOE Tracy-Widom distribution with the kernel
\begin{equation}
 \label{7.38}
K(\omega, \omega') \; = \;
\Ai(\omega + \omega' - f), \; \; \; \; \; \; \; (\omega, \omega' \, > \, 0)
\end{equation}
(see eqs.(\ref{5.3})-(\ref{5.5})) and
\begin{eqnarray}
 \nonumber
\Phi(\omega, \omega'; f,x) &=&
i \int_{-\infty}^{+\infty} \frac{dp}{2\pi} \;
\frac{1}{p-i\epsilon} \;
\Ai'(\omega + \omega' + p^{2} - ipx - f) \;
\exp\{i p \bigl(\omega - \omega'\bigr)\}
\\
\nonumber
\\
&-&
i \int_{-f}^{+\infty} dy \; \int_{-\infty}^{+\infty} \frac{dp}{2\pi} \, p \,
\Ai(\omega + \omega' + p^{2} - ipx + y) \;
\exp\{i p \bigl(\omega - \omega'\bigr)\}
\label{7.39}
\end{eqnarray}
Using the standard integral representation of the Airy function (see Appendix B) 
one can easily reduce the above function
$\Phi(\omega, \omega'; f,x)$ to the following more simple form:
\begin{eqnarray}
 \nonumber
\Phi(\omega, \omega'; f,x) = -\frac{1}{2} \int_{0}^{+\infty} dy \;
&\Biggl[&
\Bigl(
\frac{\partial}{\partial \omega} - \frac{\partial}{\partial \omega'}
\Bigr)
\Psi\bigl(\omega - \frac{1}{2}f + y ; \; x \bigr)
\Psi\bigl(\omega' - \frac{1}{2}f + y ; \; -x\bigr) +
\\
\nonumber
\\
&+&
\Bigl(
\frac{\partial}{\partial \omega} + \frac{\partial}{\partial \omega'}
\Bigr)
\Psi\bigl(\omega - \frac{1}{2}f - y ; \; x\bigr)
\Psi\bigl(\omega' - \frac{1}{2}f + y ; \; -x\bigr)
\Biggr]
\label{7.40}
\end{eqnarray}
where 
\begin{equation}
 \label{7.41}
\Psi(\omega; x) \; = \;
2^{1/3} \mbox{Ai}\Bigl[2^{1/3}\bigl(\omega + \frac{1}{8} x^{2}\bigr)\Bigr] \, 
\exp\bigl\{-\frac{1}{2} \omega x\bigr\}
\end{equation}

Thus, eqs.(\ref{7.36})-(\ref{7.41}) complete the derivation of the
probability distribution function for the directed polymer's endpoint.
Unfortunately, we see that at present stage this final expression is rather involved,
so that the study of its analytical properties requires special efforts.




\newpage

\begin{center}
 {\bf \Large PART 2: Unsolved problems}
\end{center}

\vspace{5mm}

\section{Zero temperature limit}

\newcounter{8}
\setcounter{equation}{0}
\renewcommand{\theequation}{8.\arabic{equation}}

In this Chapter we will study the statistical properties of one-dimensional directed polymers
in {\it correlated} random potential in the zero-temperature limit. This system is defined
in terms of the  Hamiltonian
\begin{equation}
   \label{9.1}
   H[\phi(\tau), V] = \int_{0}^{t} d\tau
   \Bigl\{\frac{1}{2} \bigl[\partial_\tau \phi(\tau)\bigr]^2
   + V[\phi(\tau),\tau]\Bigr\};
\end{equation}
where 
$V(\phi,\tau)$ is the Gaussian distributed random potential with a zero mean, $\overline{V(\phi,\tau)}=0$,
and the correlation function
\begin{equation}
\label{9.2}
\overline{V(\phi,\tau)V(\phi',\tau')} = u \delta(\tau-\tau') U(\phi-\phi')
\end{equation}
Here  $U(\phi)$ is the "spatial" correlation
function characterized by some finite correlation length $R$. For simplicity we take
\begin{equation}
\label{9.3}
U(\phi) \; = \; \frac{1}{\sqrt{2\pi} \, R} \; \exp\Bigl\{-\frac{\phi^{2}}{2 R^{2}}\Bigr\}
\end{equation}

It can be argued (see below) that the model, eqs.(\ref{9.1})-(\ref{9.3}),
with finite correlation length $R$ in the high temperature limit is equivalent to the one
with the $\delta$-correlated random potentials considered in the previous chapters. 
Note that this provides the basis for the
short-length regularization of the model with the $\delta$-potentials which, in fact,
is ill defined at short scales.
The zero-temperature limit for this model is much less clear. The problem is that in the exact solution of the 
model with $\delta$-correlated potentials (Chapter IV) the fluctuating part 
of the free energy $F(T, t)$ (in the limit $t\to\infty$) is proportional to $(u/T)^{2/3} \, t^{1/3} \, f$ 
and this free energy does not reveal any finite zero-temperature limit.
The physical origin of this pathology is clear: the exact solution mentioned above is valid {\it only}
for the model with $\delta$-correlated potentials which is ill defined at short scales while it is the 
short scales which are getting the most relevant in the limit $T \to 0$. One can propose two types of the 
regularization of the model on short scales: (1) introducing a lattice, and (2) keeping continuous structure of
the "space-time" but introducing smooth finite size correlations for the random potential like in eq.(\ref{9.3}). 
In both cases, however, the solution derived for the model with $\delta$-correlated potentials becomes
not valid. Nevertheless, it is generally believed that in the zero-temperature  
limit the considered system must reveal the same TW distribution. On one hand there are exact results
for strictly zero temperature lattice models revealing main features of (1+1) directed polymers (see e.g
rigorous solution of the directed polymer lattice model with geometric disorder \cite{Johansson}).
On the other hand there are analytic indications that in the zero-temperature limit 
the continuous model with finite correlation length of the type introduced above, eqs.(\ref{9.1})-(\ref{9.3}), 
after crossover to a new regime keeps the main features of the finite-temperature solution \cite{Korshunov,Lecomte}.

In this Chapter we will try  to formulate a general 
scheme which would allow to obtain a finite  zero-temperature limit for the continuous model  (\ref{9.1})-(\ref{9.3}). 
In terms of the standard replica technique (see Chapters II and III) it will be demonstrated 
that in the limit $T \to 0$ the considered system 
reveals  the one-step replica symmetry breaking structure  similar to the one which 
takes place in the low-temperature phase of the Random Energy Model (REM) \cite{REM}. 
Of course, the considered system (unlike REM) reveal no phase transition: here at the temperature
$T_{*} \sim (u R)^{1/3}$ we observe only a {\it crossover}
from the high- to the low-temperature regime. Namely, at $T \gg T_{*}$ the fluctuating part of the free energy is
the same as in the model with the $\delta$-interactions, $F(T, t) \simeq (u/T)^{2/3} \, t^{1/3} \, f$,
while at $T \ll T_{*}$ the non-universal prefactor of the free energy saturates at finite
value, $F(T=0, t) \simeq (u^{2}/R)^{1/9} \, t^{1/3} \, f$. The probability distribution function of the
random quantity $f$ is, of course, expected to be the TW one, although at present stage this is not proved yet.

\vspace{5mm}

For simplicity let us consider the case of the zero boundary conditions: $\phi(0) = \phi(t) = 0$.
Following the procedure described in Chapters II and III, 
the free energy probability distribution function
$P(F)$ of this system can be studied in terms of the replica  partition function:
\begin{equation}
\label{9.4}
Z(N,t) \; = \; \int_{-\infty}^{+\infty} dF \, P(F) \, \exp\bigl\{-\beta N F\bigr\}
\end{equation}
where 
\begin{equation}
\label{9.5}
Z(N,t) \; = \; \prod_{a=1}^{N}\int_{\phi_{a}(0)=0}^{\phi_{a}(t)=0}
        {\cal D}\phi_{a}(\tau) \;
        \exp\Bigl\{-\beta  H_{N}[\boldsymbol{\phi}(\tau)]\Bigr\}
\end{equation}
with the replica Hamiltonian
\begin{equation}
   \label{9.6}
   \beta H_{N}[\boldsymbol{\phi}] =  \int_{0}^{t} d\tau
   \Biggl[
   \frac{1}{2} \beta \sum_{a=1}^{N} \Bigl(\partial_\tau \phi_{a}(\tau)\Bigr)^2
   \; - \;
   \frac{1}{2} \beta^{2} \, u \,
    \sum_{a,b=1}^{N} U\bigl(\phi_{a}(\tau) - \phi_{b}(\tau)\bigr)
   \Biggr];
\end{equation}
which describes $N$ elastic strings
$\bigl\{\phi_{1}(\tau), \, \phi_{2}(\tau), \, ... \, , \phi_{N}(\tau)\bigr\}$
with the finite width attractive interactions $U\bigl(\phi_{a} - \phi_{b}\bigr)$,
eq.(\ref{9.3}).

Mapping this system to the $N$-particle quantum boson model (see Chapter III)
one introduces the wave function
\begin{equation}
\label{9.7}
\Psi(x_{1}, \, x_{2}, \, ... \, x_{N}; \; t) \; = \;
        \prod_{a=1}^{N}\int_{\phi_{a}(0)=0}^{\phi_{a}(t)=x_{a}}
        {\cal D}\phi_{a}(\tau) \;
        \exp\Bigl\{-\beta  H_{N}[\boldsymbol{\phi}(\tau)]\Bigr\}
\end{equation}
such that $Z(N,t) \, = \, \Psi(x_{1},  x_{2},  ...  x_{N}; \, t)\big|_{x_{a}=0}$.
The above wave function can be obtained as the solution of the 
the imaginary time Schr\"odinger equation
\begin{equation}
\label{9.8}
\beta \frac{\partial}{\partial t} \Psi({\bf x}; \, t) \; = \;
\frac{1}{2}\sum_{a=1}^{N} \, \frac{\partial^{2}}{\partial x_{a}^{2}} \Psi({\bf x}; \, t)
\; + \; \frac{1}{2} \, \beta^{3} u \, \sum_{a,b=1}^{N} U(x_{a} - x_{b}) \, \Psi({\bf x}; \, t)
\end{equation}
with the initial condition $\Psi({\bf x}; \, 0) \; = \; \prod_{a=1}^{N}\, \delta(x_{a})$. 
In the high temperature limit ($\beta \to 0$) the typical distance between the particles (defined
by the wave function $\Psi({\bf x}; \, t)$) is much larger that the size $R$ of the potential
$U(x)$, eq.(\ref{9.3}) (see below). In this case the potential $U(x)$ can be approximated by
the $\delta$-function, so that a generic solution of the Schr\"odinger equation (\ref{9.9})
(with $U(x) = \delta(x)$) is obtained in terms of the Bethe ansatz eigenfunctions,
eqs.(\ref{3.12})-(\ref{3.18}).
In particular, the ground state wave function (which in eqs.(\ref{3.12})-(\ref{3.14}) 
corresponds to $M = 1; \; Q_{a} = q - \frac{i}{2} \kappa (N + 1 - 2a)$) of this system is:
\begin{equation}
\label{9.9}
\Psi_{q}({\bf x}) \; \propto \;
   \exp\Bigl\{-\frac{1}{4} \, \kappa \sum_{a,b=1}^{N} |x_{a} - x_{b}| \, + \, i q \sum_{a=1}^{N} x_{a}
   \Bigr\}
\end{equation}
For the excited states (with $M > 1$) the generic wave function can be represented as a linear combination 
of various products of the "cluster" wave functions which have the structure similar to the one
in eq.(\ref{9.9}). We see that in any case the typical distance between the particles
is of the order of $\kappa^{-1} \, = \, (\beta^{3} u)^{-1}$. Thus, the approximation of the potential
$U(x)$, eq.(\ref{9.3}), by the $\delta$-function is justified provided $(\beta^{3} u)^{-1} \gg R$
which for a given $R$ and $u$ is valid only in the high temperature region,
\begin{equation}
\label{9.10}
\beta \, \ll \, (R u)^{-1/3}
\end{equation}
At temperatures of the order of the $(R u)^{1/3}$ and below the typical distance between particles becomes
comparable with the size $R$ of the potential $U(x)$, and therefore its approximation by the
$\delta$-function is no longer valid, which makes the considered model unsolvable (at least for
 the time being). It turns out, however, that in the zero temperature limit,
at $T \ll (R u)^{1/3}$ the situation somewhat simplifies again (see below).

\vspace{5mm}

One can easily show that the parameters of the considered system can be redefined in such a way
that  in the limit $t \to \infty$  the properties of the system would depend on the
only parameter, which is the reduced temperature $\tilde{T} \, = \, T/T_{*}$ where
\begin{equation}
\label{9.11}
T_{*} \, = \, \Bigl(\frac{uR}{\sqrt{2\pi}}\Bigr)^{1/3}
\end{equation}
Indeed, redefining
\begin{eqnarray}
\nonumber
\phi &\to & R \, \phi
\\
\beta & = & T_{*}^{-1} \, \tilde{\beta}
\label{9.12}
\\
\nonumber
\tau & \to & \tau_{*} \, \tau
\end{eqnarray}
with
\begin{equation}
\label{9.13}
\tau_{*} \, = \, \Bigl(\sqrt{2\pi} \, R^{5} u^{-1} \Bigr)^{1/3}
\end{equation}
instead of the replica Hamiltonian, eq.(\ref{9.6}), one gets
\begin{equation}
   \label{9.14}
   \beta \tilde{H}_{N}[\boldsymbol{\phi}] =  \int_{0}^{t/\tau_{*}} d \tau
   \Biggl[
   \frac{1}{2} \, \tilde{\beta} \sum_{a=1}^{N} \Bigl(\partial_\tau \phi_{a}(\tau)\Bigr)^2
   \; - \;
   \frac{1}{2} \, \tilde{\beta}^{2}
    \sum_{a,b=1}^{N} U_{0}\bigl(\phi_{a}(\tau) - \phi_{b}(\tau)\bigr)
   \Biggr];
\end{equation}
where
\begin{equation}
\label{9.15}
U_{0}(\phi) \; = \;  \exp\Bigl\{-\frac{1}{2} \, \phi^{2} \Bigr\}
\end{equation}
Accordingly, the  wave function $\tilde{\Psi}({\bf x}; t)$ defined by eq.(\ref{9.7}) with
$\beta H_{N}[\boldsymbol{\phi}]$ replaced by $\beta \tilde{H}_{N}[\boldsymbol{\phi}]$
is given by the solution of the Schr\"odinger equation
\begin{equation}
\label{9.16}
2 \tilde{\beta} \frac{\partial}{\partial \tilde{t}} \tilde{\Psi}({\bf x}; \, \tilde{t}) \; = \;
\sum_{a=1}^{N} \, \frac{\partial^{2}}{\partial x_{a}^{2}} \tilde{\Psi}({\bf x}; \, \tilde{t})
\; + \;  {\tilde{\beta}}^{3} \sum_{a,b=1}^{N} U_{0}(x_{a} - x_{b}) \,
\tilde{\Psi}({\bf x}; \, \tilde{t})
\end{equation}
where $\tilde{t} = t/\tau_{*}$. The corresponding eigenvalue equation for the eigenfunctions
$\psi({\bf x})$, defined by the relation
$\tilde{\Psi}({\bf x}; \, \tilde{t}) \, = \, \psi({\bf x}) \exp\{- \tilde{t} E\}$,
reads:
\begin{equation}
\label{9.17}
- 2 E \, \tilde{\beta} \, \psi({\bf x}) \; = \;
\sum_{a=1}^{N} \, \frac{\partial^{2}}{\partial x_{a}^{2}} \psi({\bf x})
\; + \;  {\tilde{\beta}}^{3} \sum_{a,b=1}^{N} U_{0}(x_{a} - x_{b}) \, \psi({\bf x})
\end{equation}
As in what follows we will be interested only in the limit
$t \to \infty$ the rescaling  of the time $t$  by the factor $\tau_{*}$ does not change anything.
In other words, in the limit $t \to \infty$ our problem
(defined by the replica Hamiltonian (\ref{9.14}) or the Schr\"odinger equation (\ref{9.16}))
is controlled by the only parameter $\tilde{\beta}$.

\vspace{5mm}

In the  zero temperature limit ($\tilde{\beta} \to \infty$),
according to the discussion at the end of the previous section,
one may naively suggest that the typical distance between particles defined by the
eigenfunctions $\psi({\bf x})$, eq.(\ref{9.17}), becomes small compared to size ($\sim 1$)
of the potential $U_{0}(x)$. In other words, all the particles are expected to be localized
near the bottom of this potential so that one could approximate
\begin{equation}
\label{9.18}
U_{0}(\phi) \; \simeq \;  1 \; - \; \frac{1}{2} \, x^{2}
\end{equation}
For the  corresponding eigenvalue equation
\begin{equation}
\label{9.19}
- 2 E \, \tilde{\beta} \, \psi({\bf x}) \; = \;
\sum_{a=1}^{N} \, \frac{\partial^{2}}{\partial x_{a}^{2}} \psi({\bf x})
\; + \;  {\tilde{\beta}}^{3} N^{2} \psi({\bf x}) \; - \;
\frac{1}{2} \, {\tilde{\beta}}^{3} \sum_{a,b=1}^{N} (x_{a} - x_{b})^{2} \, \psi({\bf x})
\end{equation}
one finds simple (exact) ground state solution
\begin{equation}
\label{9.20}
\psi_{0}({\bf x}) \; = \;
C \, \exp\Biggl\{
-\frac{1}{4} \, \sqrt{\frac{{\tilde{\beta}}^{3}}{N}} \sum_{a,b=1}^{N} (x_{a} - x_{b})^{2}
\Biggr\}
\end{equation}
where $C$ is the normalization constant, and
\begin{equation}
\label{9.21}
E_{0}(\tilde{\beta}, N) \; = \; -\frac{1}{2} \, \bigl(\tilde{\beta} \, N\bigr)^{2}
              +\frac{1}{2} \, (N-1) \, \sqrt{\tilde{\beta} \, N}
\end{equation}
Using the explicit form of the wave function (\ref{9.20}) one can easily estimate the average distance
$\Delta x$ between arbitrary two particles in this $N$-particle system:
\begin{equation}
\label{9.22}
\Delta x \; \sim \; \sqrt{\frac{1}{N(N-1)} \sum_{a,b=1}^{N}\overline{(x_{a} - x_{b})^{2}}}
\; \sim \;
\bigl(\tilde{\beta}^{3} N\bigr)^{-1/4}
\end{equation}
In the limit $\tilde{\beta} \to \infty$ (at fixed $N$)
$\Delta x \sim \tilde{\beta}^{-3/4} \to 0$. Therefore, in the zero temperature limit
for any fixed value of $N$ the wave function
$\psi_{0}({\bf x})$ is indeed nicely localized neat the "bottom" of the potential well
$U_{0}(x)$ which justifies the approximation (\ref{9.18}).
On the other hand, one can easily see that both the wave function (\ref{9.20}) and the
ground state energy (\ref{9.21}) demonstrate completely
pathological behavior in the limit $N\to 0$ which is crucial for the reconstruction of the
free energy distribution function in the limit $t \to \infty$.
The typical size $\Delta x$ (\ref{9.22}) of the wave function (\ref{9.20}) grows with decreasing
$N$ and become of the order of one at $N \sim  \tilde{\beta}^{-3}$. Therefore at smaller values
of $N$ the ground state wave function must have essentially different form.
The simplest way to let the particles enjoy their mutual attraction
while keeping their number in the well finite consists in splitting them into several well separated
groups, such that each group would consist of finite number of particles
This idea (similar to the one step replica symmetry breaking solution of the Random Energy Model
\cite{REM}) has been proposed by Sergei Korshunov many years ago \cite{Korshunov}.
More specifically,
let us split $N$ particles into $K$ groups each consisting of $m$ particles, so that $K = N/m$.
In this way, instead of the coordinates of the particles $\{x_{a}\} \; \; (a=1,...,N)$
we introduce the coordinates of the center of masses of the groups
$\{X_{\alpha}\} \; \; (\alpha = 1, ..., K)$ and the deviations
$\xi_{i}^{\alpha} \; \; (i = 1, ... , m)$ of the particles of a given  group $\alpha$
from the position of its center of mass:
\begin{equation}
\label{9.23}
x_{a} \; \to \; X_{\alpha} \; + \; \xi_{\alpha}^{i} \, , \; \; \; \; \alpha = 1, ..., K
                                                    \, , \; \; \; \; i = 1, ..., m
                                                    \, , \; \; \; \; K = N/m
\end{equation}
where $\sum_{i=1}^{m} \xi_{\alpha}^{i} \, = \, 0$. It is supposed that in the zero temperature limit
 ($\tilde{\beta} \to \infty$)  the typical value
of the particle's deviations inside the groups are small,
$\langle \bigl(\xi_{i}^{\alpha}\bigr)^{2}\rangle\big|_{\tilde{\beta} \to \infty} \, \to \, 0$,
while the value of the typical distance between the groups remains finite.

In terms of the above ansatz the original replica partition function
\begin{equation}
\label{9.24}
Z(N,\tilde{t}) \; = \; \prod_{a=1}^{N}\int_{\phi_{a}(0)=0}^{\phi_{a}(\tilde{t})=0}
        {\cal D}\phi_{a}(\tau) \;
        \exp\Biggl\{
        -\frac{1}{2}\int_{0}^{\tilde{t}} d \tau
   \Biggl[
   \tilde{\beta} \sum_{a=1}^{N} \Bigl(\partial_\tau \phi_{a}\Bigr)^2
   \; - \;
   \tilde{\beta}^{2}
    \sum_{a,b=1}^{N} U_{0}\bigl(\phi_{a} - \phi_{b}\bigr)
   \Biggr]
        \Biggr\}
\end{equation}
factorizes into two independent contributions: (a) the "internal" partition functions
$Z_{0}\bigl(\tilde{\beta}, m, \tilde{t}\bigr)$  of tightly bound groups
of $m$ polymers for which one can use the approximation (\ref{9.18}), and (b)
the "external" partition function ${\cal Z}\bigl(\tilde{\beta}, N/m, m, \tilde{t}\bigr)$
of $N/m$ "complex" polymers (each consisting of $m$ tightly bound original polymers):
\begin{equation}
\label{9.25}
Z(N,\tilde{t}) \; \simeq \;
\Bigl[Z_{0}\bigl(\tilde{\beta}, m, \tilde{t}\bigr)\Bigr]^{\frac{N}{m}} \times
{\cal Z}\bigl(\tilde{\beta}, N/m, m, \tilde{t}\bigr)
\end{equation}
In the limit $\tilde{t} \to \infty$ the "internal" partition functions
$Z_{0}\bigl(\tilde{\beta}, m, \tilde{t}\bigr)$ is dominated by the ground state,
eqs.(\ref{9.20})-(\ref{9.21}):
\begin{equation}
\label{9.26}
Z_{0}\bigl(\tilde{\beta}, m, \tilde{t}\bigr) \; \simeq \;
\exp\bigl\{- E_{0}(\tilde{\beta}, m) \, \tilde{t} \bigr\} \times
\psi_{0}(\boldsymbol{\xi})\big|_{\boldsymbol{\xi}=0}
\; \propto \;
\exp\Biggl\{
\frac{1}{2} \,\Bigl[ \bigl(\tilde{\beta} m\bigr)^{2}
              +(1-m) \, \sqrt{\tilde{\beta} m} \; \Bigr] \, \tilde{t}
\Biggr\}
\end{equation}
According to the definition (\ref{9.24}), the "external" partition function
${\cal Z}\bigl(\tilde{\beta}, N/m, m, \tilde{t}\bigr)$
can be represented as follows:
\begin{equation}
\label{9.27}
{\cal Z}\bigl(\tilde{\beta}, N/m, m, \tilde{t}\bigr) \; = \; \prod_{\alpha=1}^{N/m}\int_{\phi_{\alpha}(0)=0}^{\phi_{\alpha}(\tilde{t})=0}
        {\cal D}\phi_{\alpha}(\tau) \;
        \exp\Biggl\{
        -\frac{1}{2}\int_{0}^{\tilde{t}} d \tau
   \Biggl[
   (\tilde{\beta}m) \, \sum_{\alpha=1}^{N/m} \Bigl(\partial_\tau \phi_{\alpha}\Bigr)^2
   \; - \;
   (\tilde{\beta} m)^{2}
    \sum_{\alpha\not=\beta}^{N/m} U_{0}\bigl(\phi_{\alpha} - \phi_{\beta}\bigr)
   \Biggr]
        \Biggr\}
\end{equation}

Now, to make the present approach self-consistent one has to formulate the procedure
which would define the value of the parameter $m$ which for the moment remains arbitrary.
In fact, the algorithm which would fix the value of this parameter can be borrowed 
from the standard procedure of replica symmetry breaking (RSB) scheme of spin glasses.
In particular, it is successfully used in the one-step RSB solution of the Random Energy Model (REM)
\cite{REM}. In simple words, this procedure is in the following. Originally the parameter $m$ is introduced 
as an  integer bounded by the condition $1 \leq m \leq N$ and  such that the replica parameter $N$
must be a multiple of $m$ as the number of the groups of particles $K = N/m$ must be an integer. 
After analytic continuation of the replica parameters $N$ and $m$ to arbitrary real values in the 
limit $N \to 0$ the above restriction $1 \leq m \leq N$ (interpreted now as $m$ is bounded 
{\it between} $1$ and $N$) turns into $N \leq m \leq 1$. Then the physical value of the parameter $m_{*}$
is fixed by the condition that the extensive part of the free energy of the considered
system $F_{0}(m)$ (considered as a function of $m$ at the interval $N \leq m \leq 1$) 
has the {\it maximum} at $m = m_{*}$. Of course, from the mathematical point of view this 
procedure is ill grounded (or better to say not grounded at all), but somehow in all know
cases (where the solution can also be obtained in some other methods) it works perfectly well. 
In any case, for the moment we don't have any other method anyway.

In the present case the situation is even worse as 
the solution of the  problem defined by the replica partition function (\ref{9.27})
(which also gives a contribution to the extensive part of the free energy)
is not known. Nevertheless,
it would be natural to suggest that its generic structure is similar to the one
with the  $\delta$-interactions (instead of  $U_{0}(\phi)$).
Namely, let us suppose (like in the problem with the  $\delta$-interactions) that
the free energy of the system which is defined by the replica partition function
(\ref{9.27}) contains both the linear in time (self-averaging) part $f_{0}(\tilde{\beta}m)$ and
the fluctuating part which scales with time as $t^{1/3}$. In this case, due to factorization,
eq.(\ref{9.25}), in the limit $\tilde{t} \to \infty$, the total free energy of the system $F$
can be represented as the sum of two contributions:
\begin{equation}
\label{9.28}
F \; = \; F_{0} \, t \; + \; {\cal F} \, t^{1/3}\
\end{equation}
where the fluctuating part ${\cal F}$ is defined by the solution of the problem (\ref{9.27}),
while the linear part $F_{0}$ is given by the sum of the "internal" contribution
$E_{0}(\tilde{\beta}, m)$, eq.(\ref{9.21}), and the contribution $f_{0}(\tilde{\beta}m)$ coming
from the "external" partition function (\ref{9.27}). In the limit $t \to \infty$ 
the fluctuating contribution of the free energy $\sim t^{1/3}$ can be neglected compared with 
its linear in $t$ part so that 
we gets
\begin{eqnarray}
 \nonumber
F_{0}(\tilde{\beta}, m) &=& - \lim_{t\to\infty} \, \frac{1}{\beta N t} \, \ln\Bigl[
Z_{0}^{N/m} \, {\cal Z} \Bigr]
\\
\nonumber
\\
\nonumber
&=&  
- \lim_{t\to\infty} \, \frac{T_{*}}{\tilde{\beta} N t} \, \Bigl[
\frac{N}{m} \, E_{0}(\tilde{\beta}, m) \, \frac{t}{\tau_{*}} \; + \; 
\frac{N}{m} \, f_{0}(\tilde{\beta}m) \frac{t}{\tau_{*}} 
\Bigr]
\\
\nonumber
\\
&=&
\frac{T_{*}}{2 \tau_{*}} \Bigl[
-(\tilde{\beta}m) \; - \; (1-m) \, (\tilde{\beta}m)^{-1/2} \; + \; 
\frac{2}{(\tilde{\beta}m)} f_{0}(\tilde{\beta}m) 
\Bigr]
 \label{9.29}
\end{eqnarray}
where the quantities $T_{*}$, $\tau_{*}$ and $E_{0}(\tilde{\beta}m)$ 
are defined in eqs.(\ref{9.11}), (\ref{9.13}) and (\ref{9.21}).
The parameter $m$ is defined by the solution of the equation
\begin{equation}
 \label{9.30}
\frac{\partial}{\partial m} \, F_{0}(\tilde{\beta}, m) \; = \; 0
\end{equation}
which corresponds to the {\it maximum} of the function $F_{0}(m)$ at the interval $0 \leq m \leq 1$
(as in the limit $t\to\infty$ the relevant values of the replica parameter $N$ which define
the statistics of the free energy fluctuations are of the order of $t^{-1/3} \to 0$).
Let us introduce  a new parameter 
\begin{equation}
 \label{9.31}
\zeta \; = \; \tilde{\beta} m
\end{equation}
which is supposed to remain {\it finite} in the zero-temperature limit,  $\tilde{\beta} \to \infty$.
In terms of this parameter the expression for the self-averaging free energy $F_{0}$, eq.(\ref{9.29}),
reduces to
\begin{equation}
 \label{9.32}
F_{0}(\zeta) \; \simeq \; \frac{T_{*}}{2 \tau_{*}} \Bigl[
-\zeta  \; - \;  \zeta^{-1/2} \; + \; 
\frac{2}{\zeta} \, f_{0}(\zeta) 
\Bigr]
\end{equation}
where  for a finite value of $\zeta$ the factor $(1-m)$ in second term in eq.(\ref{9.29}) 
in the limit $\tilde{\beta} \to \infty$ can be approximated
as $(1-m) \; = \; \bigl(1 \, - \, \zeta/\tilde{\beta}\bigr) \; \simeq \; 1$. 
Then, according to eq.(\ref{9.32}) in the zero-temperature  limit the  
saddle-point equation (\ref{9.30}) reads
\begin{equation}
 \label{9.33}
-1 \; + \; \frac{1}{2} \, \zeta^{-3/2} \; + \; \frac{d}{d \zeta} \, \Bigl[ \frac{2}{\zeta} f_{0}(\zeta) \Bigr] \; = \; 0
\end{equation}
which contains no parameters and correspondingly its solution defines a {\it finite} value of $\zeta_{*}$ (which is 
just a number). As a matter of illustration, if we approximate the solution of the "external problem", eq.(\ref{9.27}),
by the well known result for the model with the $\delta$-interactions, namely
$f_{0}(\tilde{\beta} m) \; = \; \frac{1}{24} \bigl(\tilde{\beta} m\bigr)^{5}$ (see e.g. \cite{BA-TW3}), 
the above equation  reduces to 
\begin{equation}
 \label{9.34}
-1 \; + \; \frac{1}{2} \, \zeta^{-3/2} \; + \; \frac{1}{3} \zeta^{3} \; = \; 0
\end{equation}
The solution of this equation is $\zeta_{*} \; \simeq \; 0.68$.

Thus, in the zero temperature limit the partition function of the "external problem", eq.(\ref{9.27}), as the function
of a new replica parameter $N/m \equiv K$ becomes temperature independent:
\begin{equation}
\label{9.35}
{\cal Z}\bigl(\zeta_{*}; K, \tilde{t}\bigr) \; = \; \prod_{\alpha=1}^{K}\int_{\phi_{\alpha}(0)=0}^{\phi_{\alpha}(\tilde{t})=0}
        {\cal D}\phi_{\alpha}(\tau) \;
        \exp\Biggl\{
        -\frac{1}{2}\int_{0}^{\tilde{t}} d \tau
   \Biggl[
   \zeta_{*} \, \sum_{\alpha=1}^{K} \Bigl(\partial_\tau \phi_{\alpha}\Bigr)^2
   \; - \;
   \zeta_{*}^{2}
    \sum_{\alpha\not=\beta}^{K} U_{0}\bigl(\phi_{\alpha} - \phi_{\beta}\bigr)
   \Biggr]
        \Biggr\}
\end{equation}
Let us extract from this partition function the explicit contribution containing the linear in time  free energy 
$f_{0}(\zeta_{*})\tilde{t}$ and redefine
\begin{equation}
 \label{9.36}
{\cal Z}\bigl(\zeta_{*}; K, \tilde{t}\bigr) \; = \; \exp\bigl\{- K \, f_{0}(\zeta_{*})\, \tilde{t} \bigr\} \, \times \,
\tilde{{\cal Z}}\bigl(\zeta_{*}; K \tilde{t}^{1/3}\bigr)
\end{equation}
Here by analogy with the solution of the corresponding
problem with the $\delta$-interactions in the limit $\tilde{t} \to \infty$ the  function 
$\tilde{{\cal Z}}\bigl(\zeta_{*}; K \tilde{t}^{1/3}\bigr)$ 
is expected to depend on the replica parameter $K$ and the time $\tilde{t}$ 
in the combination $K \tilde{t}^{1/3}$. It is this function which defines the probability distribution function 
$P\bigl({\cal F}\bigr)$   of the fluctuating part of the free energy, eq.(\ref{9.28}). According to the 
relations, eq.(\ref{9.4}), (\ref{9.25}) and (\ref{9.36}), and the definition (\ref{9.28}) the  
probability distribution function $P\bigl({\cal F}\bigr)$ 
and the partition function $\tilde{{\cal Z}}\bigl(K \tilde{t}^{1/3}\bigr)$
are related via the  Laplace transform:
\begin{equation}
 \label{9.37}
\int_{-\infty}^{+\infty} d{\cal F} \,P\bigl({\cal F}\bigr)\, \exp\bigl\{- K \tilde{t}^{1/3} \, {\cal F}  \bigr\} \; = \; 
\tilde{{\cal Z}}\bigl(\zeta_{*}; K \tilde{t}^{1/3}\bigr)
\end{equation}
which, at least formally, allows to reconstruct the probability distribution function $P\bigl({\cal F}\bigr)$
via the inverse Laplace transform
\begin{equation}
 \label{9.38}
P\bigl({\cal F}\bigr) \; = \; \int_{-i\infty}^{+i\infty} \frac{ds}{2\pi i} \, \tilde{{\cal Z}}(\zeta_{*}; s) \, 
\exp\bigl\{ s \, {\cal F}  \bigr\}
\end{equation}

\vspace{5mm}

Summarizing all the above speculations, the systematic solution of the considered problem in the zero-temperature limit
consists of three steps.

{\bf First}. For a given integer $K$  and for a given real positive $\zeta$ one has to compute the partition
function ${\cal Z}(\zeta; K, t)$, eq.(\ref{9.35}), 
where $U_{0}\bigl(\phi\bigr) \; = \; \exp\{-\frac{1}{2}\phi^{2}\}$. This partition function in the limit $t \to \infty$
(similar to the case of the $\delta$-interactions)  is expected to factorize into two essentially different contributions:
(a) the one which explicitly reveal the linear in time non-random (self-averaging) free energy part $f_{0}(\zeta)$; and 
(b) a function $\tilde{{\cal Z}}\bigl(\zeta; K t^{1/3} \bigr)$ which depends on the replica parameter $K$ and time $t$
in the combination $K t^{1/3}$, eq.(\ref{9.36}).

{\bf Second}. The physical value of the parameter $\zeta_{*}$ is defined by the solution of the equation (\ref{9.33}).
This solution has to be substituted into the function  $\tilde{{\cal Z}}\bigl(\zeta_{*}; K t^{1/3} \bigr)$.

{\bf Third}. The probability distribution function $P\bigl({\cal F}\bigr)$ of the free energy fluctuating part ${\cal F}$
is defined by the Laplace transform relation, eq.(\ref{9.37}). Here the function 
$\tilde{{\cal Z}}\bigl(\zeta_{*}; K t^{1/3} \bigr)$
has to be analytically continued for arbitrary complex values of the replica parameter $K$. 
Then, introducing the Laplace transform 
parameter $s = K t^{1/3}$ the probability distribution function $P\bigl({\cal F}\bigr)$ is obtained by the inverse 
Laplace transform, eq.(\ref{9.38}). 

As in the considered zero-temperature limit the partition function (\ref{9.35}) does not depend on the temperature,
the free energy probability distribution function $P\bigl({\cal F}\bigr)$ must be  temperature independent too.
The only "little" problem in the above derivation scheme is that for the moment the solution for  the 
partition function ${\cal Z}(\zeta; K, t)$, eq.(\ref{9.35}) (the First step) is not known.
Nevertheless even at present (somewhat speculative) stage we can claim that  in the limit $T \to 0$ 
the considered system reveals  the one-step replica symmetry breaking structure, eqs.(\ref{9.23}) and (\ref{9.25}),
similar to the one which takes place in the Random Energy Model. Besides,  at the temperature
$T_{*} \sim (u R)^{1/3}$ we observe a crossover from the high- to the low-temperature regime. 
Namely,  at high temperatures, $T \gg T_{*}$, the model is equivalent to the one with the 
$\delta$-correlated potential in which the non-universal prefactor of the fluctuating part of the free 
energy is proportional to $(u/T)^{2/3}$,  
while at $T \ll T_{*}$ this non-universal prefactor  saturates 
at a finite (temperature independent) value $\sim (u^{2}/R)^{1/9}$. 
The formal proof that the zero temperature limit free energy distribution function
of the considered system is indeed the Tracy-Widom one requires further investigation.


\vspace{10mm}

\section{Random force Burgers turbulence}

\newcounter{9}
\setcounter{equation}{0}
\renewcommand{\theequation}{9.\arabic{equation}}

In this Chapter we will consider the possibility to apply the ideas and technique developed in the previous
Chapters for the  problem of the randomly forced Burgers turbulence ("burgulence") which is formally 
equivalent to the KPZ problem.
Let us consider a velocity field $v(x,t)$ governed by the Burgers equation
\begin{equation}
   \label{10.1}
\partial_{t} v(x,t) + v(x,t)\partial_{x} v(x,t) = 
\nu  \partial^{2}_{x} v(x,t) +  \zeta(x,t)
\end{equation}
where the parameter $\nu$ is the viscosity and $\zeta(x,t)$ is the Gaussian distributed random force 
which is $\delta$-correlated in time and which is characterized by finite correlation length $\xi$ in space: 
$\overline{\zeta(x,t) \zeta(x',t')} = u \delta(t-t') U[(x-x')/\xi]$. Here U(x) is a smooth function decaying to zero fast
enough at large arguments and the parameter $u$ is the injected energy density. In this problem one 
would like to derive e.g. the the probability distribution functions of the velocity gradients 
$P[\partial_{x} v(x,t)]$ or two-points distribution function $P[v(x,t), v(x',t)]$ at scales smaller
than the length scale $\xi$ of the stirring force $\zeta$
(see e.g. \cite{burgers_74,Sinai,Bouch-Mez-Par,turbulence} and references there in).

Formally the above problem is equivalent to the KPZ equation as well as to the (1+1) directed
polymers. Indeed, redefining $v(x,t) = -\partial_{x} F(x,t)$
and $\zeta(x,t) = -\partial_{x} V(x,t)$ one gets the KPZ equation for the interface profile
$F(x,t)$ (which is the free energy of (1+1) directed polymers):
\begin{equation}
   \label{10.2}
\partial_{t} F(x,t) = \frac{1}{2} \bigl(\partial_{x} F(x,t)\bigr)^{2} 
- \frac{1}{2} T \partial^{2}_{x} F(x,t) -  V(x,t)
\end{equation}
where $T = 2\nu$ is the temperature parameter of the directed polymer problem and $V(x,t)$ is the Gaussian
distributed random potential.
The idea of a new approach to the Burgulence problem which I would like to discuss  in this Chapter
is in the following. According to the above definitions the velocity field can be
represented as 
\begin{equation}
\label{10.3}
v(x,t)=\lim_{\epsilon\to 0} [F(x+\epsilon,t) - F(x,t)]/\epsilon
 \end{equation}
Thus, deriving the four-point KPZ probability distribution function
${\cal P}[F(x+\epsilon,t), F(x,t), F(x'+\epsilon,t'),F(x',t')]$ and taking the limit $\epsilon \to 0$
one could hopefully obtain the result for $P[v(x,t),v(x',t')]$.
The only "little problem" is that unlike the usual KPZ studies operating with the $\delta$-correlated in space
random potential, in the Burgulence problem one is mainly interested in the spatial scales comparable
or much smaller than the random potential correlation length $\xi$. In other words, in this  
approach, first one has to study KPZ problem with random potentials having {\it finite} correlation length  (see Chapter VIII).
In this Chapter (as a matter of "warming up" exercise) I'm going to consider another "extreme case" 
in which the random potential $V(x,t)$ of the KPZ problem is changed by it's linear approximation: 
$V(x,t) \to \zeta(t) x$ where $\zeta(t)$ is Gaussian distributed random force. In this case we obtain 
the model introduced by Larkin \cite{Larkin, Larkin-Ovchinnikov} long time ago
to study small scale displacements of directed polymers. In this approximation the model becomes Gaussian
and therefore exactly solvable. Nevertheless, the statistical properties of its free energy (as well as some others
quantities)  turn out to be rather non-trivial (see e.g \cite{Gorochov-Blatter,gaussian,replicas}).
For that reason this model hopefully could serve as a good ground for testing various approaches developed 
in the recent KPZ studies.

\subsection{The model}

Let us consider the model of one-dimensional directed polymers 
defined in terms of an elastic string $\phi(\tau)$ which passes through a random medium
described by a random potential $V(\phi,\tau) \; = \; \zeta(\tau) \phi$. The energy of a given polymer's trajectory
$\phi(\tau)$ is given by the Hamiltonian
\begin{equation}
   \label{10.4}
   H[\phi(\tau), V] = \int_{0}^{t} d\tau
   \Bigl\{\frac{1}{2} \bigl[\partial_\tau \phi(\tau)\bigr]^2
   + \zeta(\tau) \phi(\tau) \Bigr\};
\end{equation}
where the random force $\zeta(\tau)$ is described by the Gaussian distribution 
with a zero mean $\overline{\zeta(\tau)}=0$ 
and the $\delta$-correlations: 
\begin{equation}
   \label{10.5}
{\overline{\zeta(\tau)\zeta(\tau')}} = u \delta(\tau-\tau')
\end{equation}
For the fixed boundary conditions, $\phi(0) = y; \; \phi(t) = x$, the partition function
of the model (\ref{10.4}) is
\begin{equation}
\label{10.6}
   Z(x|y; t) = \int_{\phi(0)=y}^{\phi(t)=x}
              {\cal D} \phi(\tau)  \;  \mbox{\Large e}^{-\beta H[\phi]}
\; = \; \exp\bigl[-\beta F(x|y; t)\bigr]
\end{equation}
where $F(x|y; t)$ is the free energy. In the replica approach (see Chapter II) 
one consider the average of the $N$-th power of the 
above partition function:
\begin{equation}
 \label{10.7}
 \overline{Z^{N} (x|y; t)} \; \equiv \; Z(N; x|y ; t) \; = \; \overline{\exp\bigl[-\beta N F(x|y; t)\bigr]}
\end{equation}
Simple Gaussian averaging yields:
\begin{equation}
   \label{10.8}
Z(N; x|y ; t)  \; = \; \prod_{a=1}^{N} \Biggl[
\int_{\phi_{a}(0)=y}^{\phi_{a}(t)=x} {\cal D} \phi_{a}(\tau) \Biggr] \;
   \exp\Bigl[-\beta H_{N}[{\boldsymbol \phi}]  \Bigr]
\end{equation}
where
\begin{equation}
 \label{10.9}
    H_{N}[{\boldsymbol \phi}] \; = \;
   \frac{1}{2} \int_{0}^{t} d\tau \Biggl(
   \sum_{a=1}^{N} \bigl[\partial_\tau \phi_{a}(\tau)\bigr]^2 
   - \beta u \sum_{a, b}^{N} \phi_{a}(\tau) \phi_{b}(\tau) \Biggr) 
\end{equation}
is the Gaussian replica Hamiltonian.
Introducing the free energy distribution function, $P_{x|y; t}(F)$, the relation (\ref{10.7})
can be represented as follows:
\begin{equation}
 \label{10.10}
 Z(N; x|y ; t) \; = \; \int_{-\infty}^{+\infty} dF \; P_{x|y; t}(F) \exp\bigl[-\beta N F\bigr]
\end{equation}
which is the Laplace transform of the distribution function, $P_{x|y; t}(F)$ with respect to 
the parameter $\beta N$. 
In the lucky case when the moments of the partition function $Z(N; x|y ; t)$ allows an analytic continuation
from integer to arbitrary complex values of the replica parameter $N$ the above relation makes possible
to reconstruct the probability distribution function $P_{x|y; t}(F)$ via the inverse Laplace transform:
\begin{equation}
 \label{10.11}
 P_{x|y; t}(F) \; = \; \int_{-i\infty}^{+i\infty} \frac{ds}{2\pi i} \; Z\bigl(\frac{s}{\beta}; x|y ; t\bigr) \; 
                       \exp\bigl(s F\bigr)
\end{equation}
In the present model this distribution can be computed explicitly and the resulting
function $ P_{x|y; t}(F)$ turns out to be rather non-trivial \cite{Gorochov-Blatter,gaussian,replicas}.

Mapping the considered system to quantum bosons (see Chapter III) one introduces
the $N$-particle wave function:
\begin{equation}
 \label{10.12}
 \Psi_{N}\bigl[{\bf x}|{\bf y}; t\bigr] \; = \; \prod_{a=1}^{N} 
 \Biggl[
\int_{\phi_{a}(0)=y_{a}}^{\phi_{a}(t)=x_{a}} 
   {\cal D} \phi_{a}(x)
   \Biggr] \;
   \exp\Bigl[-\beta H_{N}[{\boldsymbol \phi}]  \Bigr]
\end{equation}
with the Hamiltonian $H_{N}[{\boldsymbol \phi}]$ is given in eq.(\ref{10.9}).
This wave function can be obtained as the solution of 
the imaginary time Schr\"odinger equation
\begin{equation}
\label{10.13}
\beta \frac{\partial}{\partial t} \Psi({\bf x}|{\bf y}; \, t) \; = \;
\frac{1}{2}\sum_{a=1}^{N} \, \frac{\partial^{2}}{\partial x_{a}^{2}} \Psi({\bf x}|{\bf y}; \, t)
\; + \; \frac{1}{2} \, \beta^{3} u \, \sum_{a,b=1}^{N} x_{a} \, x_{b} \; \Psi({\bf x}|{\bf y}; \, t)
\end{equation}
with the initial condition $\Psi_{N}\bigl[{\bf x}|{\bf y}; t=0\bigr] \; = \; \prod_{a=1}^{N}\, \delta(y_{a}- x_{a})$. 
One can easily check that this solution is \cite{replicas}:
\begin{equation}
\label{10.14}
 \Psi_{N}\bigl[{\bf x}|{\bf y}; t\bigr] = C\bigl(N, t\bigr)  
 \exp\Biggl\{
 -\frac{\beta}{2t} \sum_{a=1}^{N} \bigl(x_{a} - y_{a}\bigr)^{2}  
 -\frac{\beta}{2t} A(N,t) \Biggl[\sum_{a=1}^{N} \bigl(x_{a} - y_{a}\bigr) \Biggr]^{2} 
 +\frac{\beta}{t} B(N,t) \sum_{a,b=1}^{N} x_{a} y_{b} 
\Biggr\}
\end{equation}
where
\begin{eqnarray}
 \label{10.15}
 C(N,t) &=& \Bigl(\frac{\beta}{2\pi t}\Bigr)^{N/2} \; 
            \sqrt{\frac{\sqrt{\beta N u t^{2}}}{\sin\bigl( \sqrt{\beta N u t^{2}}\bigr)}}
            \\
            \nonumber
            \\
 \label{10.16}
 A(N,t) &=& \frac{1}{N} \Biggl( 
            \frac{\sqrt{\beta N u t^{2}}}{\tan\bigl( \sqrt{\beta N u t^{2}}\bigr)} \; - \; 1
            \Biggr)
            \\
            \nonumber
            \\
 \label{10.17}
 B(N,t) &=& \frac{1}{N} \; \frac{\sqrt{\beta N u t^{2}}}{\sin\bigl( \sqrt{\beta N u t^{2}}\bigr)} \; 
                           \bigl[ 1 \; - \; \cos\bigl(\sqrt{\beta N u t^{2}}\bigr) \bigr]
\end{eqnarray}
Note that the above expression for the wave function eq.(\ref{10.14}) is valid
only at finite time interval: $t \; < \; \frac{\pi}{2} (\beta N u)^{-1/2} \equiv t_{c}(N) $. 
The reason is that due to specific form of the
interaction potentials in the replica Hamiltonian (\ref{10.9}) the directed polymers trajectories
go to infinity at {\it finite time} $t_{c}(N)$.  For the original physical system this phenomenon
is explained by the presence of the slowly decaying left tail 
of the free energy distribution function 
$ P_{x|y; t}(F\to -\infty) \sim \exp\bigl\{-\frac{\pi^{2}}{4 u t^{2}} \, |F|\bigr\}$ 
\cite{Gorochov-Blatter,gaussian,replicas} 
which (according to the relation (\ref{10.10})) results in the divergence of all partition function moments   
$Z(N; x|y ; t)$ with $N > N_{c}(t) = \frac{\pi^{2}}{4} (\beta u t^{2})^{-1}$.

\subsection{Two-time free energy distribution function}

For simplicity, let us consider the directed polymer problem with the zero boundary conditions:
$\phi(0) = \phi(t) = 0$. For a given realization of the random function $\zeta(\tau)$
let us denote by $F_{1}$ the free energy of the directed polymers with the (zero) ending point
at time $t$ and by $F_{2}$ the one of the directed polymers with the (zero) ending point
at time $t \, + \, \Delta t$.
According to the definition, eq.(\ref{10.6}), for the difference of these free energies, $F = F_{2} - F_{1}$
one has:
\begin{equation}
 \label{10.18}
 \exp\{-\beta F \} \; = \; Z^{-1}(0|0; \, t) \; Z(0|0; \, t+\Delta t) 
\end{equation}
Taking $n$-th power of the above relation and performing the averaging over quenched randomness we gets
\begin{equation}
 \label{10.19}
 \overline{\exp\{-\beta n F \}} \; = \; \overline{Z^{-n}(0|0; \, t) \, Z^{n}(0|0; \, t+\Delta t) }
\end{equation}
Introducing the two-time free energy difference probability distribution function
$P_{t,\Delta t}(F)$ in terms of the replica approach the above relation can be represented as 
\begin{equation}
 \label{10.20}
 \int_{-\infty}^{+\infty} dF \, P_{t,\Delta t}(F) \, \exp\{-\beta n F \} \; = \; 
 \lim_{N\to 0} \, \overline{Z^{N-n}(0|0; \, t) \, Z^{n}(0|0; \, t+\Delta t) }
\end{equation}
where according to the usual replica formalism, in the first step, the r.h.s of the above relation 
is computed for an arbitrary {\it integers} $N > n$ and then, at the second step, the obtained result is analytically
continued for arbitrary real $N$ and $n$, and finally the limit $N\to 0$ has to be  taken.

According to the definition of the partition function, eq.{\ref{10.6}),
\begin{equation}
 \label{10.21}
Z(0|0; \, t+\Delta t) \; = \; \int_{-\infty}^{+\infty} dx \, Z(0|x; \, t) \, Z^{*}(x|0; \, \Delta t)  
\end{equation}
where
\begin{equation}
 \label{10.22}
Z^{*}(x|0; \, \Delta t) = \int_{\phi(t+\Delta t)=0}^{\phi(t)=x}
              {\cal D} [\phi(\tau)]  \;  \exp\Bigl\{-\beta H[\phi(\tau); V]\Bigr\}
\end{equation}
is the partition function of the directed polymer system
in which time goes backwards, from $(t+\Delta t)$ to $t$. 
Thus, for an integer $N > n$ in terms of the wave function, eqs.(\ref{10.12}), the product of the
partition functions in the r.h.s. of eq.(\ref{10.20}) can be represented as follows:
\begin{equation}
 \label{10.23}
 \overline{Z^{N-n}(0|0;  t)  Z^{n}(0|0;  t+\Delta t) } \equiv
 Z(N, n; t, \Delta t)  =  
 \int_{-\infty}^{+\infty} dx_{1} ... dx_{n}  
 \Psi_{N}\bigl[\underbrace{0, ..., 0}_{N-n}, x_{1}, ..., x_{n}|{\bf 0}; t\bigr] \,
 \Psi^{*}_{n}\bigl[x_{1}, ..., x_{n}|{\bf 0}; \Delta t\bigr]
\end{equation}
where the second (conjugate) wave function represent the "backward" propagation
from the time moment $(t + \Delta t)$ to the previous time moment $t$.
Schematically the above expression is represented in Figure 1.
\begin{figure}[h]
\begin{center}
   \includegraphics[width=12.0cm]{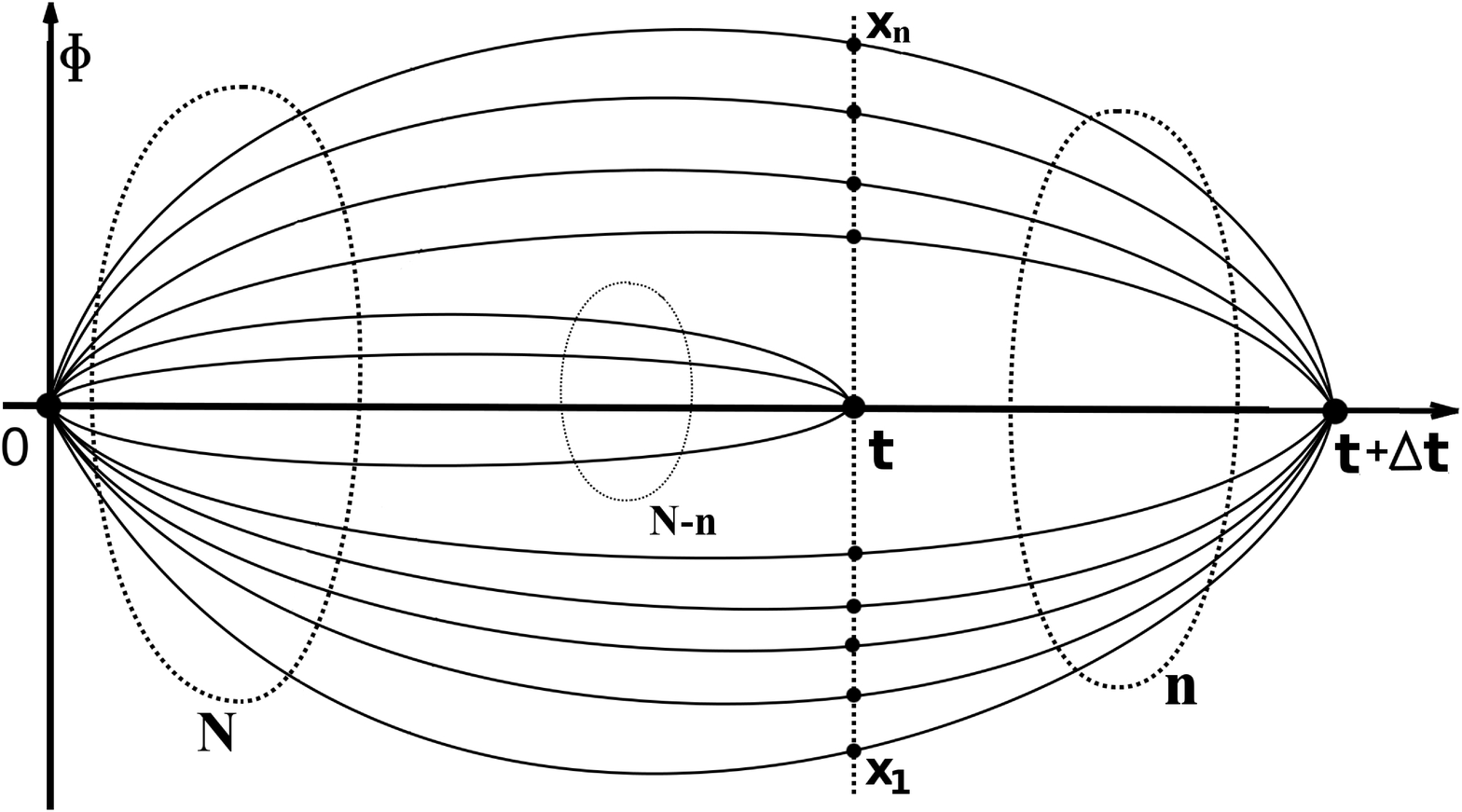}
\caption[]{Schematic representation of the directed polymer paths
corresponding to eq.(\ref{10.23})}
\end{center}
\label{figure5}
\end{figure}
Substituting eq.(\ref{10.14}) into eq.(\ref{10.23}) we get:
\begin{equation}
 \label{10.24}
 Z(N, n; t, \Delta t)  = 
 C(N,t) C(n,\Delta t) \int_{-\infty}^{+\infty} dx_{1} ... dx_{n} 
 \exp\Biggl\{
 -\frac{1}{2}\beta \sum_{a,b=1}^{n} 
 \Bigl[
 \frac{t + \Delta t}{t \Delta t} \delta_{ab} + 
 \frac{1}{t} A(N,t) +\frac{1}{\Delta t} A(n,\Delta t)
 \Bigr] x_{a} x_{b} 
 \Biggr\}
\end{equation}
Simple integration yields:
\begin{equation}
 \label{10.25}
 Z(N, n; t, \Delta t)  =  C(N,t) C(n,\Delta t)  
 \Bigl[\frac{2\pi t \Delta t}{\beta (t + \Delta t)}\Bigr]^{n/2} \; 
 \Biggl[
 1 \; + \; n\frac{t \Delta t}{t + \Delta t} \Bigl(\frac{1}{t} A(N,t) +\frac{1}{\Delta t} A(n,\Delta t)\Bigr)
 \Biggr]^{-1/2}
\end{equation}
Substituting here the explicit expressions (\ref{10.15}) and (\ref{10.16}), and taking the limit $N\to 0$ we obtain
\begin{equation}
 \label{10.26}
\lim_{N\to 0}  Z(N, n; t, \Delta t)  \; = \; 
\Bigl(\frac{t}{t + \Delta t}\Bigr)^{n/2}
\sqrt{\frac{
\sqrt{\beta n u (\Delta t)^{2}} \; (t + \Delta t)}{
\sin\bigl( \sqrt{\beta n u (\Delta t)^{2}}\bigr) 
\Bigl[
\Delta t - \frac{1}{3}\beta n u (\Delta t) t^{2} + 
t \frac{\sqrt{\beta n u (\Delta t)^{2}}}{\tan\bigl( \sqrt{\beta n u (\Delta t)^{2}}\bigr)}
\Bigr]}}
\end{equation}
Substituting eq.(\ref{10.26}) into eq.(\ref{10.20}) and redefining:
\begin{eqnarray}
 \label{10.27}
 \Delta t \; = \; \xi \, t 
 \\
 \nonumber
 \\
 \label{10.28}
 \beta n u (\Delta t)^{2} \; = \; \omega
 \\
 \nonumber
 \\
 \label{10.29}
 F \; = \; u \xi^{2} t^{2} \, f
\end{eqnarray}
we get the following relation 
for the probability distribution function $P_{t,\xi} (f)$  
of the rescaled free energy $f$:
\begin{equation}
 \label{10.30}
 \int_{-\infty}^{+\infty} df \, P_{t,\xi}(f) \, \exp\{-\omega f \} \; = \; 
 \bigl(1 + \xi\bigr)^{-\frac{\omega}{2\beta u \xi^{2} t^{2}}}
 \frac{\omega^{1/4} \sqrt{1+ \xi}}{
 \sqrt{
\Bigl(\xi - \frac{\omega}{3\xi}\Bigr) \sin(\sqrt{\omega}) \; + \; \sqrt{\omega} \cos(\sqrt{\omega})}}
 \end{equation}
By inverse Laplace transform in the limit when both $t \to \infty$ and $\Delta t \to \infty$ (such that
the parameter ($\xi = \Delta t/t$ remains finite) we get the following universal 
result for the limiting two-time free energy distribution function:
\begin{equation}
\label{10.31}
\lim_{t\to\infty} P_{t,\xi}(f) \equiv {\cal P}_{\xi} (f) \; = \; 
\sqrt{1+\xi} \, 
\int_{-i\infty}^{+i\infty} \frac{d\omega}{2\pi i} \; 
\frac{\omega^{1/4} \exp\{\omega f\}}{
 \sqrt{
\Bigl(\xi - \frac{\omega}{3\xi}\Bigr) \sin(\sqrt{\omega}) \; + \; \sqrt{\omega} \cos(\sqrt{\omega})}}
 \end{equation}
It is interesting to note that this function (like its one-time  counterpart 
\cite{Gorochov-Blatter,gaussian,replicas})
is identically equal to zero at $f > 0$. 
Indeed, since at $f>0$ the function under the integral in the r.h.s of eq.(\ref{10.31}) quickly
goes to zero at $w \to -\infty$, the contour of integration in the complex plane 
can be safely shifted to $-\infty$, which means that $ {\cal P}_{\xi}(f > 0) \equiv 0$.

\subsection{Two-time velocity distribution function}

Velocity in the Burgers problem is given by the derivative of the free energy
of the directed polymer problem:
\begin{equation}
 \label{10.32}
 v(x,t) \; = \; -\frac{\partial F(x,t)}{\partial x} \; = \; 
             - \lim_{\epsilon \to 0} \frac{F(x+\epsilon) - F(x,t)}{\epsilon}
\end{equation}
Thus, to compute the two-point velocity distribution function in terms of the directed polymers, first, keeping 
$\epsilon$ finite we consider specially constructed four-point object (see below), and only in the final 
stage of calculations we take the limit $\epsilon \to 0$.

According to the relation (\ref{10.10})
\begin{equation}
 \label{10.33}
 \exp\{-\beta\bigl[F(x+\epsilon,t) - F(x,t)\bigr]\} \; = \; 
 \exp\{\beta\epsilon v(x,t)\} \; = \; 
 \frac{Z(x+\epsilon|0; \, t)}{Z(x|0; \, t)}
\end{equation}
Following the procedure described in the previous section we have:
\begin{eqnarray}
 \label{10.34}
 &&
\overline{
\exp\bigl\{ \beta n_{1} \epsilon v(x_{1},t) + \beta n_{2} \epsilon v(x_{2},t+\Delta t)\bigr\} } \; = \; 
\\
\nonumber
 \\
 \nonumber
&& \hspace{20mm} = \;
\lim_{N_{1}, N_{2} \to 0}
\overline{ 
Z^{n_{1}}(x_{1}+\epsilon|0; \, t) \, Z^{N_{1} - n_{1}}(x_{1}|0; \, t) 
Z^{n_{2}}(x_{2}+\epsilon|0; \, t + \Delta t) \, Z^{N_{2} - n_{2}}(x_{2}|0; \, t + \Delta t) }
\end{eqnarray}
Introducing two-time velocity distribution function $P_{x_{1},x_{2} t, \Delta t} (v_{1}, v_{2})$ 
the above relation can be represented as follows:
\begin{equation}
 \label{10.35}
 \int\int_{-\infty}^{+\infty} d v_{1} d v_{2} \; P_{x_{1},x_{2} t, \Delta t} (v_{1}, v_{2})
 \exp\bigl\{ \beta n_{1} \epsilon v_{1} + \beta n_{2} \epsilon v_{2}\bigr\} 
 \; = \;
 \lim_{N_{1}, N_{2} \to 0} \; {\cal Z}_{\epsilon}\bigl( N_{1}, n_{1}, N_{2}, n_{1}, x_{1}, x_{2}, t, \Delta t \bigr)
\end{equation}
where 
\begin{eqnarray}
 \label{10.36}
&& {\cal Z}_{\epsilon}\bigl( N_{1}, n_{1}, N_{2}, n_{2}, x_{1}, x_{2}, t, \Delta t \bigr) \; = 
 \\
\nonumber
 \\
 \nonumber
&& \hspace{20mm} = \; 
 \overline{ 
Z^{n_{1}}(x_{1}+\epsilon|0; \, t) \, Z^{N_{1} - n_{1}}(x_{1}|0; \, t) 
Z^{n_{2}}(x_{2}+\epsilon|0; \, t + \Delta t) \, Z^{N_{2} - n_{2}}(x_{2}|0; \, t + \Delta t) }
\end{eqnarray}
In terms of the wave function, eqs.(\ref{10.12})-(\ref{10.14}),
\begin{eqnarray}
 \nonumber
&&
{\cal Z}_{\epsilon}\bigl( N_{1}, n_{1}, N_{2}, n_{2}, x_{1}, x_{2}, t, \Delta t \bigr) \; = 
 \\
\nonumber
 \\
 \nonumber
&&  = \; 
\int_{-\infty}^{+\infty} \, Dy \; 
\int_{-\infty}^{+\infty} \, Dz \; 
\Psi_{N_{1}+N_{2}}\bigl[\underbrace{x_{1}, ..., x_{1}}_{N_{1}-n_{1}}, 
                        \underbrace{x_{1}+\epsilon, ..., x_{1}+\epsilon}_{n_{1}},
                        y_{1}, ..., y_{N_{2}-n_{2}},
                        z_{1}, ..., z_{n_{2}} 
                        \, | \,{\bf 0}; 
                        \; t \bigr] \times
 \\
\label{10.37}
 \\
 \nonumber
&&  \hspace{35mm}  \times                       
 \Psi^{*}_{N_{2}}\bigl[y_{1}, ..., y_{N_{2}-n_{2}},
                       z_{1}, ..., z_{n_{2}} 
                       \,| \,
                       \underbrace{x_{2}, ..., x_{2}}_{N_{2}-n_{2}}, 
                       \underbrace{x_{2}+\epsilon, ..., x_{2}+\epsilon}_{n_{2}} ;
                       \, \Delta t\bigr]
\end{eqnarray}
where
\begin{eqnarray}
 \label{10.38}
 \int_{-\infty}^{+\infty} \, Dy \; &\equiv& \; \prod_{a=1}^{N_{2}-n_{2}} \int_{-\infty}^{+\infty} dy_{a}
 \\
\nonumber
 \\
 \nonumber
 \int_{-\infty}^{+\infty} \, Dz \; &\equiv& \; \prod_{a=1}^{n_{2}} \int_{-\infty}^{+\infty} dz_{a}
\end{eqnarray}
Schematically the expression in eq.(\ref{10.37}) is represented in Figure 2.

 \begin{figure}[h]
\begin{center}
   \includegraphics[width=13.0cm]{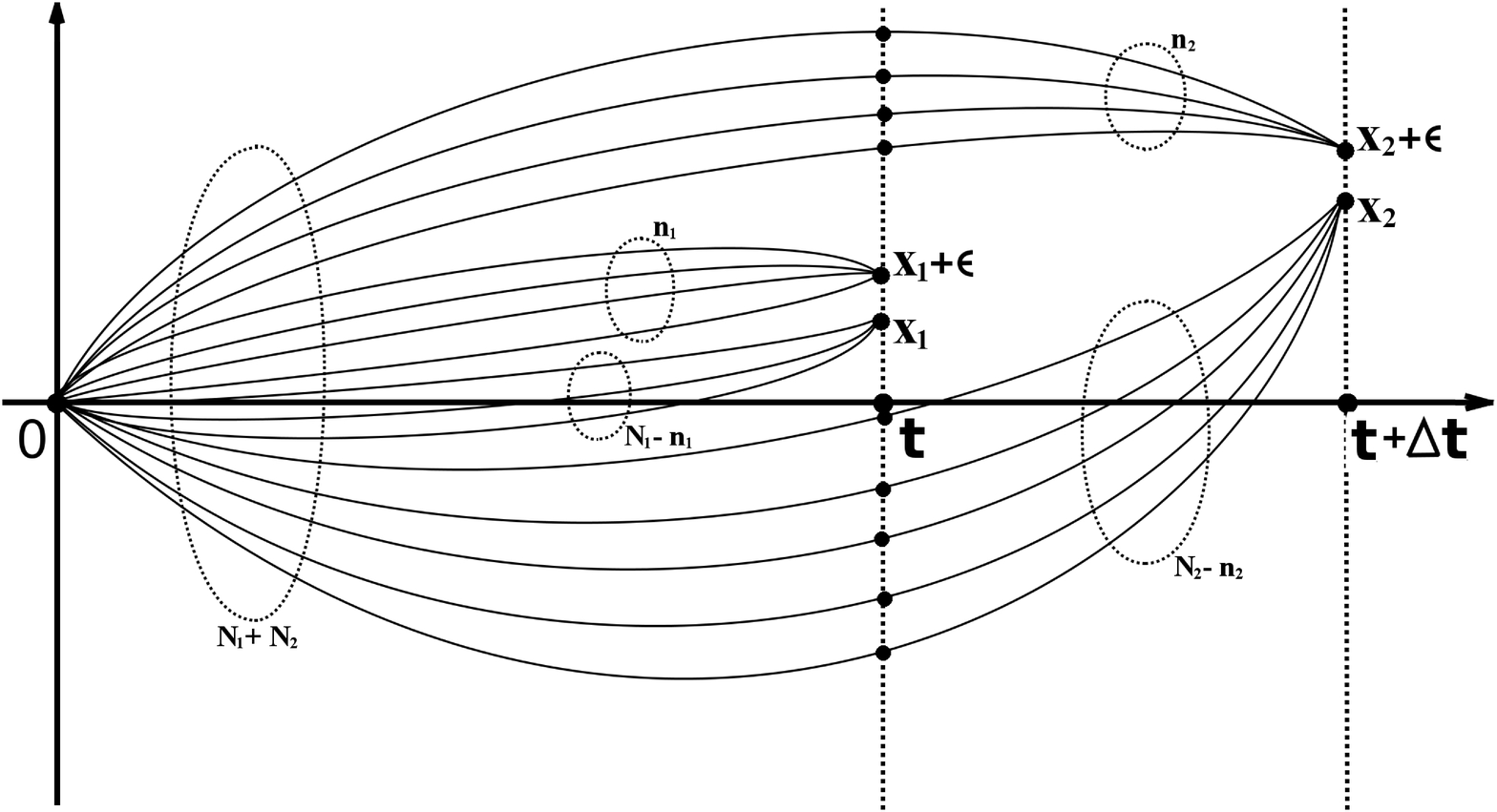}
\caption[]{Schematic representation of the directed polymer paths
corresponding to eq.(\ref{10.37})}
\end{center}
\label{figure6}
\end{figure}

Substituting the explicit expressions for the wave function (\ref{10.14}) into eq.(\ref{10.37}) 
 we get:
\begin{eqnarray}
\nonumber
&&
{\cal Z}_{\epsilon}\bigl( N_{1}, n_{1}, N_{2}, n_{2}, x_{1}, x_{2}, t, \Delta t \bigr) \; = \;
C(N_{1}+N_{2},t) \, C(N_{2}, \Delta t) \times
 \\
\nonumber
 \\
 \nonumber
&&  \times
\int_{-\infty}^{+\infty} \, Dy \; 
\int_{-\infty}^{+\infty} \, Dz \; 
\exp\Biggl\{
-\frac{\beta}{2t}
\Biggl[
(N_{1}-n_{1})x_{1}^{2} + n_{1} (x_{1}+\epsilon)^{2} + \sum_{a=1}^{N_{2}-n_{2}} y_{a}^{2} + \sum_{a=1}^{n_{2}} z_{a}^{2} 
\Biggr] \, -
 \\
\nonumber
 \\
 \nonumber
&& \hspace{23mm}
-\frac{\beta}{2t} A(N_{1}+N_{2}, t)
\Biggl[
(N_{1}-n_{1})x_{1} + n_{1} (x_{1}+\epsilon) + \sum_{a=1}^{N_{2}-n_{2}} y_{a} + \sum_{a=1}^{n_{2}} z_{a} 
\Biggr]^{2} \, -
 \\
\nonumber
 \\
 \nonumber
&& \hspace{23mm}
-\frac{\beta}{2\Delta t}
\Biggl[
\sum_{a=1}^{N_{2}-n_{2}} (y_{a}-x_{2})^{2} + \sum_{a=1}^{n_{2}} (z_{a}-x_{2}-\epsilon)^{2} 
\Biggr] \, -
 \\
\nonumber
 \\
 \nonumber
&& \hspace{23mm}
-\frac{\beta}{2\Delta t} A(N_{2}, \Delta t)
\Biggl[
\sum_{a=1}^{N_{2}-n_{2}} (y_{a}-x_{2}) + \sum_{a=1}^{n_{2}} (z_{a}-x_{2}-\epsilon) 
\Biggr]^{2} \, +
 \\
\nonumber
 \\
&& \hspace{23mm}
+ \frac{\beta}{\Delta t} B(N_{2}, \Delta t)
\Bigl(
\sum_{a=1}^{N_{2}-n_{2}} y_{a} + \sum_{a=1}^{n_{2}} z_{a}
\Bigr)
\Bigl[
(N_{2}-n_{2})x_{2} + n_{2} (x_{2}+\epsilon)
\Bigr]
\Biggr\}
\label{10.39}
\end{eqnarray}
where $A(N,t)$, $B(N,t)$ and $C(N,t)$ are given in eqs.(\ref{10.15})-(\ref{10.17}).
Introducing $N_{2}$-component vector ${\boldsymbol \chi} = \{y_{1}, ..., y_{N_{2}-n_{2}}, z_{1}, ..., z_{n_{2}}\}$
after simple algebra we get
\begin{eqnarray}
\nonumber
{\cal Z}_{\epsilon}\bigl( N_{1}, n_{1}, N_{2}, n_{2}, x_{1}, x_{2}, t, \Delta t \bigr) &=&
C(N_{1}+N_{2},t) \, C(N_{2}, \Delta t) 
\times
 \\
\nonumber
 \\
&\times&
\Biggl[\prod_{a=1}^{N_{2}} \int_{-\infty}^{+\infty} d\chi_{a}
\Biggr] \; 
\exp\Biggl\{
-\frac{1}{2} \sum_{a,b=1}^{N_{2}} T_{ab} \chi_{a} \chi_{b} \; + \; \sum_{a=1}^{N_{2}} L_{a} \chi_{a}
-\frac{1}{2} \beta  G
\Biggr\}
 \label{10.40}
\end{eqnarray} 
where
\begin{eqnarray}
\nonumber
 G &=&      \frac{1}{t} \bigl[ n_{1}(x_{1}+\epsilon)^{2} + (N_{1} - n_{1}) x_{1}^{2}\bigr] \; + \;
     \frac{1}{\Delta t} \bigl[ n_{2}(x_{2}+\epsilon)^{2} + (N_{2} - n_{2}) x_{2}^{2}\bigr] \; +
\\
\nonumber
 \\
&+& \frac{1}{t} A(N_{1} + N_{2},t) \bigl(N_{1} x_{1} + n_{1} \epsilon\bigr)^{2} \; + \;
    \frac{1}{\Delta t} A(N_{2},\Delta t) \bigl(N_{2} x_{2} + n_{2} \epsilon\bigr)^{2}  
\label{10.41}
\end{eqnarray}
and 
\begin{eqnarray}
 \label{10.42}
 T_{ab} &=& \gamma \, \delta_{ab} \; + \; \kappa
\\
\nonumber
\\ 
\label{10.43}
\gamma &=& \beta \frac{t + \Delta t}{t \Delta t}
\\
\nonumber
\\ 
\label{10.44}
\kappa &=& \frac{\beta}{t} A(N_{1}+N_{2}, t) \; + \; \frac{\beta}{\Delta t} A(N_{2}, \Delta t)
\\
\nonumber
\\ 
\label{10.45}
L_{a} &=& L \; + \; X_{a}
\\
\nonumber
\\ 
\label{10.46}
L &=& -\frac{\beta}{t} A(N_{1}+N_{2}, t) \bigl(N_{1} x_{1} + n_{1} \epsilon\bigr) \; + \; 
       \frac{\beta}{\Delta t} \bigl[ A(N_{2},\Delta t) + B(N_{2},\Delta t) \bigr] \bigl(N_{2} x_{2} + n_{2} \epsilon)
\\
\nonumber
\\ 
\label{10.47}
X_{a} &=& \left\{ \begin{array}{ll}
                 x_{2} + \epsilon,  & \mbox{for $a = 1, ..., n_{2}$}
 \\
                 x_{2},             & \mbox{for $a = n_{2}+1, ..., N_{2}$}
                            \end{array}
                            \right.       
\end{eqnarray}
Simple integration over $\chi$'s in eq.(\ref{10.40}) yields:
\begin{eqnarray}
\nonumber
{\cal Z}_{\epsilon}\bigl( N_{1}, n_{1}, N_{2}, n_{2}, x_{1}, x_{2}, t, \Delta t \bigr) \; &=& \;
C(N_{1}+N_{2},t) \, C(N_{2}, \Delta t) 
\times
 \\
\nonumber
 \\
&\times&
\exp\Biggl\{-\frac{1}{2} \beta  G
-\frac{1}{2} \mbox{Tr} \,\ln \hat{T} \; + \; 
 \frac{1}{2} \sum_{a,b=1}^{N_{2}} L_{a} L_{b} \hat{T}^{-1}_{ab}
 \Biggr\}
\label{10.48}
\end{eqnarray}
where
\begin{eqnarray}
 \label{10.49}
\mbox{Tr} \,\ln \hat{T} &=& N_{2} \ln \gamma \; + \; \ln\Bigl(1 \; + \; N_{2} \frac{\kappa}{\gamma}\Bigr)
 \\
\nonumber
 \\
\hat{T}^{-1}_{ab} &=& \frac{1}{\gamma} \delta_{ab} \; - \; \frac{\kappa}{\gamma\bigl(\gamma + N_{2} \kappa\bigr)}
\label{10.50}
\end{eqnarray}
Simple calculations yield:
\begin{eqnarray}
\nonumber
&&{\cal Z}_{\epsilon}\bigl( N_{1}, n_{1}, N_{2}, n_{2}, x_{1}, x_{2}, t, \Delta t \bigr) \; = \;
C(N_{1}+N_{2},t) \, C(N_{2}, \Delta t) \; 
\times
 \\
\nonumber
 \\
 \nonumber
&& \hspace{10mm} \times \exp\Biggl\{
 -\frac{1}{2} \beta G
 -\frac{1}{2} N_{2} \ln \gamma  -  \frac{1}{2} \ln\Bigl(1 + N_{2} \frac{\kappa}{\gamma}\Bigr)
 -\frac{L^{2} N_{2}}{2(\gamma + N_{2}\kappa)} 
+ \frac{\beta L}{\gamma\Delta t}\bigl(N_{2} x_{2} + n_{2}\epsilon\bigr) -
\\
\nonumber
 \\
&& \hspace{10mm} - \frac{\beta\kappa N_{2}(N_{2} x_{2} + n_{2}\epsilon)}{
                         \gamma \Delta t (\gamma + N_{2} \kappa)} 
                 - \frac{\beta^{2}\kappa (N_{2} x_{2} + n_{2}\epsilon)^{2}}{
                        2 \gamma (\Delta t)^{2} (\gamma + N_{2} \kappa)}
         + \frac{\beta^{2}}{2\gamma (\Delta t)^{2}} \Bigl[(N_{2}-n_{2})x_{2}^{2} + n_{2}(x_{2}+\epsilon)^{2}\Bigr]
\Biggr\}
\label{10.51}
\end{eqnarray}

Next step of the calculations is to take the limits $N_{1,2}\to 0$. Using explicit expressions
(\ref{10.15})-(\ref{10.17}), (\ref{10.41}), (\ref{10.44}) and (\ref{10.46}), one easily finds:\
\begin{eqnarray}
 \label{10.52}
 \lim_{N\to 0} C(N,t) &=& 1
 \\
\nonumber
 \\
  \label{10.53}
 \lim_{N\to 0} A(N,t) &=& - \frac{1}{3}\beta u t^{2}
  \\
\nonumber
 \\
 \label{10.54}
 \lim_{N\to 0} B(N,t) &=&  \frac{1}{2}\beta u t^{2}
   \\
\nonumber
 \\
 \label{10.55}
  \lim_{N_{1},N_{2}\to 0} \beta G &=& 
           \frac{\beta n_{1}\epsilon}{t} (2x_{1} + \epsilon)
          +\frac{\beta n_{2}\epsilon}{\Delta t} (2x_{2} + \epsilon)
          -\frac{1}{3} (\beta n_{1}\epsilon)^{2} u t
          -\frac{1}{3} (\beta n_{2}\epsilon)^{2} u \Delta t
 \\
\nonumber
 \\
 \label{10.56}
 \lim_{N_{1},N_{2}\to 0} L &=& \frac{1}{3} \beta^{2} n_{1} \epsilon u t 
                             + \frac{1}{6} \beta^{2} n_{2} \epsilon u \Delta t
 \\
\nonumber
 \\
 \label{10.57}
\lim_{N_{1},N_{2}\to 0} \kappa &=& -\frac{1}{3} \beta^{2} u (t + \Delta t) 
\end{eqnarray}
Substituting the above limiting values into eq.(\ref{10.51}) we get
\begin{eqnarray}
\nonumber
&& \lim_{N_{1,2}\to 0} {\cal Z}_{\epsilon}\bigl( N_{1}, n_{1}, N_{2}, n_{2}, x_{1}, x_{2}, t, \Delta t \bigr)
   \; \equiv \; 
   Z_{\epsilon}\bigl(n_{1}, n_{2}, x_{1}, x_{2}, t, \Delta t \bigr) \; =
 \\
\nonumber
 \\
&&=\exp\Biggl\{ 
          -\frac{\beta n_{1}\epsilon}{2t} (2x_{1} + \epsilon)
          -\frac{\beta n_{2}\epsilon}{2(t + \Delta t)} (2x_{2} + \epsilon)
          +\frac{1}{6} (\beta n_{1}\epsilon)^{2} u t
          +\frac{1}{6} (\beta n_{2}\epsilon)^{2} u (t + \Delta t)
          + \frac{(\beta n_{1}\epsilon)(\beta n_{2} \epsilon) u t^{2}}{3(t + \Delta t)} 
\Biggr\}
\label{10.58}
\end{eqnarray}
Substituting the above result into eq.(\ref{10.35}) and introducing notations
$\beta n_{1,2} \epsilon \; = \; s_{1,2}$ in the limit $\epsilon \to 0$ we obtain:
\begin{eqnarray}
 \nonumber
&& \int\int_{-\infty}^{+\infty} d v_{1} d v_{2} \; P_{x_{1},x_{2} t, \Delta t} (v_{1}, v_{2})
 \exp\bigl\{ s_{1} v_{1} + s_{2} v_{2}\bigr\} 
 \; = \;
 \\
\nonumber
 \\
\label{10.59}
&&=\exp\Biggl\{ 
 -\frac{x_{1}}{t} s_{1} - \frac{x_{2}}{t + \Delta t} s_{2} 
 +\frac{1}{6} u t s_{1}^{2} + \frac{1}{6} u (t + \Delta t) s_{2}^{2}
 +\frac{u t^{2}}{3(t + \Delta t)} s_{1} s_{2}\
\Biggr\}
\end{eqnarray}
Redefining
\begin{eqnarray}
 \label{10.60}
 s_{1} &=& \sqrt{\frac{3}{ut}} \; \omega_{1}
\\
\nonumber
 \\
\label{10.61} 
 s_{2} &=& \sqrt{\frac{3}{u(t+\Delta t)}} \; \omega_{2}
\\
\nonumber
 \\
\label{10.62}  
v_{1} &=& -\frac{x_{1}}{t} \; + \; \sqrt{\frac{1}{3} u t} \; \tilde{v}_{1}
\\
\nonumber
 \\
\label{10.63}  
v_{2} &=& -\frac{x_{2}}{t+\Delta t} \; + \; \sqrt{\frac{1}{3} u (t+\Delta t)} \; \tilde{v}_{2}
\\
\nonumber
 \\
\label{10.64}
\Delta t &=& \xi \, t
\end{eqnarray}
we get the following relation for the probability distribution function 
${\cal P}_{\xi}\bigl(\tilde{v}_{1}, \, \tilde{v}_{2}\bigr)$ for the rescaled
velocities $\tilde{v}_{1}$ and $\tilde{v}_{2}$, eqs.(\ref{10.62})-(\ref{10.63}):
\begin{equation}
 \label{10.65}
\int\int_{-\infty}^{+\infty} d\tilde{v}_{1} \tilde{v}_{2} {\cal P}_{\xi}\bigl(\tilde{v}_{1}, \, \tilde{v}_{2}\bigr)
\exp\bigl\{ \omega_{1} \tilde{v}_{1} + \omega_{2} \tilde{v}_{2}\bigr\} 
 \; = \; 
\exp\Bigl\{
\frac{1}{2} \omega_{1}^{2} + \frac{1}{2} \omega_{2}^{2} + \frac{\omega_{1} \omega_{2}}{(1+\xi)^{3/2}}
\Bigr\}
\end{equation}
Performing simple inverse Laplace transformation
\begin{equation}
 \label{10.66}
{\cal P}_{\xi}\bigl(\tilde{v}_{1}, \, \tilde{v}_{2}\bigr) \; = \;
\int\int_{-i\infty}^{+i\infty} \frac{d\omega_{1} d\omega_{2}}{(2\pi i)^{2}}
\; 
\exp\Bigl\{
\frac{1}{2} \omega_{1}^{2} + \frac{1}{2} \omega_{2}^{2} + \frac{\omega_{1} \omega_{2}}{(1+\xi)^{3/2}}
-\omega_{1} \tilde{v}_{1} - \omega_{2} \tilde{v}_{2}
\Bigr\}
\end{equation}
one eventually obtain the following very simple result for the two-time velocities distribution function:
\begin{equation}
 \label{10.67}
{\cal P}_{\xi}\bigl(\tilde{v}_{1}, \, \tilde{v}_{2}\bigr) \; = \;
\frac{1}{2\pi} \sqrt{\frac{(1+\xi)^{3}}{(1+\xi)^{3} - 1}} \; 
\exp\Biggl\{
-\frac{(1+\xi)^{3}}{2\bigl[(1+\xi)^{3} - 1\bigr]}
\Bigl(
\tilde{v}_{1}^{2} -2 \frac{\tilde{v}_{1}\tilde{v}_{2}}{(1+\xi)^{3/2}} + \tilde{v}_{2}^{2}
\Bigr)
\Biggr\}
\end{equation}
where $\xi = \Delta t/t$ is the reduced separation time parameter.

One can easily check that in the limit of infinite separation time, $\xi \to \infty$,
the distributions of two velocities are getting independent:
\begin{equation}
 \label{10.68}
\lim_{\xi\to\infty} {\cal P}_{\xi}\bigl(\tilde{v}_{1}, \, \tilde{v}_{2}\bigr) \; = \;
\frac{1}{2\pi} \; 
\exp\Bigl\{ -\frac{1}{2} \tilde{v}_{1}^{2} - \frac{1}{2} \tilde{v}_{2}^{2}
\Bigr\}
\end{equation}
while in the opposite limit of coinciding times, $\xi \to 0$, one finds
\begin{equation}
 \label{10.69}
{\cal P}_{0}\bigl(\tilde{v}_{1}, \, \tilde{v}_{2}\bigr) \; = \;
\frac{1}{\sqrt{2\pi}} \; 
\exp\Bigl\{ -\frac{1}{2} \tilde{v}_{1}^{2} \Bigr\} \; 
\delta\bigl(\tilde{v}_{1} - \tilde{v}_{2}\bigr)
\end{equation}
as it should be.

Besides, using the exact result, eq.(\ref{10.67}) one can easily compute the time 
dependence of the two velocities correlation function:
\begin{equation}
 \label{10.70}
\langle \tilde{v}_{1} \tilde{v}_{2} \rangle \; = \; (1 + \xi)^{-3/2}
\end{equation}
as well as the probability distribution function for the velocities difference
$\tilde{v} \equiv \tilde{v}_{2} - \tilde{v}_{1}$:
\begin{equation}
 \label{10.71}
{\cal P}_{\xi}\bigl(\tilde{v}\bigr) \; = \;
\frac{1}{2\pi} \sqrt{\frac{(1+\xi)^{3/2}}{4\pi \bigl[(1+\xi)^{3/2} - 1\bigr]}} \; 
\exp\Biggl\{
-\frac{(1+\xi)^{3/2}}{4\bigl[(1+\xi)^{3/2} - 1\bigr]} \, \tilde{v}^{2}
\Biggr\}
\end{equation}

 \subsection{Conclusions}

In this Chapter we have considered the problem of velocity distribution functions 
in the Burgulence problem in terms of the toy Gaussian model of (1+1) directed polymers.
In particular the exact result for the two-time free energy, eq.(\ref{10.31}),
and two-time velocity distribution functions, eq.(\ref{10.67}) has been derived. 
Of course the considered system is too far from the realistic one. Nevertheless, it has one important
advantage: being exactly solvable,  some of its statistical properties are 
rather non-trivial. 
All that, in my view, makes this model to be rather useful tool for testing new ideas and various 
technical aspects of the calculations (like the replica technique considered in this paper).
Following the proposed route, the next step  would be to consider the model
with finite range correlations of the random potentials (see Chapter VIII). 
In terms of the replica approach here one is facing the problem 
of $N$-particle  quantum bosons with attractive {\it finite range} interactions 
whose solution is not known. Nevertheless even the qualitative understanding of the
structure of the $N$-particle wave function of this system
(which at the qualitative level might be not so much different from that of the Bethe ansatz solution 
for the $\delta$-correlated potentials)
could hopefully be sufficient to get some understanding of the velocity statistics 
in the Burgulence problem.


\vspace{10mm}

\section{Joint distribution function of free energies at two different temperatures}

\newcounter{10}
\setcounter{equation}{0}
\renewcommand{\theequation}{10.\arabic{equation}}

In this Chapter we will consider  one more "direction" in the studies of directed polymers
with $\delta$-correlated random potential, eqs.(\ref{1.1})-(\ref{1.2}), namely, 
joint statistics of the free energies (or the interfaces, in the KPZ-language) at two different temperatures 
defined for the same quenched disorder. In other words, we are going to study the joint probability distribution function
of the free energies at two (or more) different temperatures for a given realization of the disorder potential $V[\phi,\tau]$.
Some years ago similar kind  of problem (under the name "temperature chaos") has been investigated for spin glass like
systems \cite{SG-T-chaos1,SG-T-chaos2,SG-T-chaos3,SG-T-chaos4} 
as well as for directed polymers on a hierarchical lattice \cite{DP-T-chaos}.
Here in terms of the standard replica formalism 
we will derive  the general scaling dependence for the 
reduced free energy difference 
${\cal F} = F(T_{1})/T_{1} - F(T_{2})/T_{2}$ 
on the two temperatures $T_{1}$ and $T_{2}$ (eqs.(\ref{11.32}) and (\ref{11.27})
In particular, it will be shown that if the two temperatures $T_{1} \, < \, T_{2}$ are close to each other,
so that $(1-T_{1}/T_{2}) \; \ll \; 1$, the difference
of the two free energies scales as $(1-T_{1}/T_{2})^{1/3} \, t^{1/3}$ ( eq.(\ref{11.39})).  
It will also be shown that the left tail asymptotics of this free energy difference probability
distribution function coincides with corresponding tail of the TW distribution (eq.(\ref{11.34})).

\vspace{10mm}

For the fixed boundary conditions, $\phi(0) = \phi(t) = 0$, and for a given realization of disorder 
the partition function of the model defined in eqs.(\ref{1.1})-(\ref{1.2}) is
\begin{equation}
\label{11.1}
   Z(\beta, t) = \int_{\phi(0)=0}^{\phi(t)=0}
              {\cal D} \phi(\tau)  \;  \mbox{\Large e}^{-\beta H[\phi; \, V]}
\; = \; \exp\bigl(-\beta F(\beta, t)\bigr)
\end{equation}
In the limit $t\to\infty$ the free energy $F(\beta, t)$ scales as
\begin{equation}
\label{11.2}
F(\beta, t) = f_{0}(\beta) \, t +  \frac{1}{2} \, (\beta u)^{2/3} \, t^{1/3} \, f \, ,
\end{equation}
where $f_{0}(\beta)$ is the (non-random) selfaveraging free energy density, and $f$
is a random quantity described by the GUE Tracy-Widom distribution (see Chapters II-IV).

For a given realization of the disorder potential $V[\phi,\tau]$ let us consider the above system at two
different temperatures $T_{1} \not= T_{2}$. More specifically, we are going to study
how the the two free energies $F(\beta_{1}, t)$ and $F(\beta_{2}, t)$
of the same system are related to each other. In other words, we are going to study
the statistical and scaling properties of the quantity
\begin{equation}
\label{11.3}
{\cal F}(\beta_{1}, \beta_{2}; \, t) \; = \;
\beta_{1} F(\beta_{1}, t) \, - \, \beta_{2} F(\beta_{2}, t)
\end{equation}
where, in what follows it will be assumed that $\beta_{1} \, > \, \beta_{2}$ (or $T_{1} \, < \, T_{2}$).
According to the definition (\ref{11.1})
\begin{equation}
\label{11.4}
\exp\bigl\{- {\cal F}(\beta_{1}, \beta_{2}; \, t) \bigr\} \; = \;
Z(\beta_{1}, t) \, Z^{-1}(\beta_{2}, t)
\end{equation}
Taking $N$-th power of the the both sides of the above relation and averaging over the disorder
we get
\begin{equation}
\label{11.5}
\int  d{\cal F}  P_{\beta_{1},\beta_{2},t}({\cal F})  \exp\bigl\{- N {\cal F}\bigr\}  = 
\overline{Z^{N}(\beta_{1}, t)  Z^{-N}(\beta_{2}, t)}
\end{equation}
where $\overline{(...)}$ denotes the averaging over the random potential $V$ and
$P_{\beta_{1},\beta_{2},t}({\cal F})$
is the probability distribution function of the random quantity ${\cal F}$, eq.(\ref{11.3}).
Introducing the replica partition function
\begin{equation}
\label{11.6}
{\cal Z}(M,N; \, \beta_{1}, \beta_{2}; \, t) \; = \;
\overline{Z^{N}(\beta_{1}, t) \, Z^{M-N}(\beta_{2}, t)}
\end{equation}
the relation (\ref{11.5}) can be formally represented as
\begin{equation}
\label{11.7}
\int  d{\cal F}  P_{\beta_{1},\beta_{2},t}({\cal F})  \exp\bigl\{- N {\cal F}\bigr\}  = 
\lim_{M\to 0} {\cal Z}(M,N;  \beta_{1}, \beta_{2};  t) \, .
\end{equation}
Following the standard "logic" of the replica technique, first it will be assumed that
both $M$ and $N$ are integers such that $M > N$. Next, after computing the replica partition
function ${\cal Z}(M,N; \, \beta_{1}, \beta_{2}; \, t)$ an analytic continuation for
arbitrary (complex) values of the parameters $M$ and $N$ has to be performed and the
limit $M \to 0$ has to be taken. After that, the relation (\ref{11.7}) can be considered
as the Laplace transform of the the probability distribution function
$P_{\beta_{1},\beta_{2},t}({\cal F})$ over the parameter $N$. In the case the function
${\cal Z}(0,N; \, \beta_{1}, \beta_{2}; \, t)$ would have "good" analytic properties
in the complex plane of the argument $N$, this relation, at least formally, allows
to reconstruct by inverse Laplace transform the probability distribution function
$P_{\beta_{1},\beta_{2},t}({\cal F})$. At present, for the considered problem
it is possible to derive an explicit expression for the function
${\cal Z}(0,N; \, \beta_{1}, \beta_{2}; \, t)$ only in the limit $N \gg 1$. Nevertheless,
using the relation (\ref{11.7}) this allows to reconstruct the left tail (${\cal F} \to -\infty$)
of the distribution function $P_{\beta_{1},\beta_{2},t}({\cal F})$. Moreover, it also allows to
derive the scaling dependence of free energy difference ${\cal F}$ on $\beta_{1},\beta_{2}$ and $t$.
Indeed, in the case the replica partition function has an exponential asymptotics
\begin{equation}
\label{11.8}
{\cal Z}(0,N \to \infty ;  \beta_{1}, \beta_{2};  t\to\infty)  \sim 
\exp\bigl\{A(\beta_{1}, \beta_{2})  t  N^{\alpha}\bigr\}  ,
\end{equation}
the left tail of the probability distribution function assumes the stretched-exponential form
\begin{equation}
\label{11.9}
P_{\beta_{1},\beta_{2},t}({\cal F}\to -\infty) \; \sim \;
\exp\bigl\{-B(\beta_{1}, \beta_{2}; \, t) \, |{\cal F}|^{\omega}\bigr\} \, .
\end{equation}
Then the saddle-point estimate of the integral in the l.h.s of eq.(\ref{11.7}) yields:
\begin{equation}
\label{11.10}
\int  d{\cal F}  \exp\bigl\{-B\, |{\cal F}|^{\omega}+ N |{\cal F}|\bigr\} \; \sim \; 
\exp\bigl\{(\omega-1)  \omega^{-\frac{\omega}{\omega-1}} B^{-\frac{1}{\omega-1}}
 N^{\frac{\omega}{\omega-1}}\bigr\} 
\end{equation}
so that
\begin{equation}
\label{11.10a}
\exp\bigl\{(\omega-1)  \omega^{-\frac{\omega}{\omega-1}} B^{-\frac{1}{\omega-1}}
 N^{\frac{\omega}{\omega-1}}\bigr\} \; \sim \;
\exp\bigl\{A \, t \, N^{\alpha}\bigr\}
\end{equation}
From this relation we find that
\begin{equation}
\label{11.11}
\omega \, = \, \alpha/(\alpha-1)
\end{equation}
and
\begin{equation}
\label{11.12}
B \; = \; (\alpha - 1) \, \alpha^{-\frac{\alpha}{\alpha-1}} \, \bigl(A \, t\bigr)^{-\frac{1}{\alpha-1}}
\end{equation}
Substituting this into eq.(\ref{11.9}) we get
\begin{equation}
\label{11.13}
P_{\beta_{1},\beta_{2},t}({\cal F}\to -\infty) \sim
\exp\Bigl\{
-(\alpha-1) \,
\Biggl(\frac{|{\cal F|}}{\alpha \bigl(A t \bigr)^{1/\alpha} }\Biggr)^{\frac{\alpha}{\alpha-1}}
\Bigr\} .
\end{equation}
If we assume that the (unknown) entire probability distribution function has a universal shape
the above asymptotic behavior implies that the considered quantity ${\cal F}$ scales as follows
\begin{equation}
\label{11.14}
{\cal F} \; = \; \bigl(A(\beta_{1}, \beta_{2})\bigr)^{1/\alpha} t^{1/\alpha} \, f
\end{equation}
where the random quantity $f\sim 1$ is described by some (unknown) probability distribution
function ${\cal P}(f)$ with the left asymptotics
${\cal P}(f\to -\infty)\sim \exp\bigl\{- (\mbox{const}) |f|^{\alpha/(\alpha - 1)}\bigr\}$.

Thus, the above speculations demonstrates that even if we know the replica partition function only in the limit $N \gg 1$,
we can still derive not only the left tail of the distribution function, but (supposing that the entire 
distribution function is universal) the general scaling of the free energy.
Let us consider now how this replica scheme can be applied for the concrete system under consideration.

\vspace{5mm}

Performing  the averaging over the random potential in eq.(\ref{11.6})
we get
\begin{equation}
\label{11.15}
 {\cal Z}(M,N;\beta_{1},\beta_{2}; t)=
\prod_{a=1}^{M} \int_{\phi_{a}(0)=0}^{\phi_{a}(t)=0} {\cal D}\phi_{a}(\tau) 
\exp\bigl\{-H_{M}[\boldsymbol{\phi}]\bigr\}
\end{equation}
where $H_{M}[\boldsymbol{\phi}]$ is the replica Hamiltonian
\begin{equation}
\label{11.16}
H_{M}[\boldsymbol{\phi}] \; = \;
\int_{0}^{t} d\tau
   \Biggl[
\frac{1}{2} \sum_{a=1}^{M} \beta_{a} \Bigl(\partial_\tau \phi_{a}(\tau)\Bigr)^2
-
\frac{1}{2} \, u^{2} \sum_{a\not b=1}^{M} \beta_{a} \beta_{b} \, \delta(\phi_{a} - \phi_{b})
\Biggr]
\end{equation}
with
\begin{equation}
\label{11.17}
\beta_{a} \; = \;
\left\{
                          \begin{array}{ll}
\beta_{1}\,
\; \;
\mbox{for} \; a = 1, ..., N
\\
\\
\beta_{2}  \,
\; \;
\mbox{for} \; a = N+1, ..., M,
                          \end{array}
\right.
\end{equation}
Introducing:
\begin{equation}
\label{11.18}
\Psi(x_{1}, ..., x_{M} ; t) \; \equiv \;
\prod_{a=1}^{M} \int_{\phi_{a}(0)=0}^{\phi_{a}(t)=x_{a}} {\cal D}\phi_{a}(\tau) \;
\exp\bigl\{-H_{M}[\boldsymbol{\phi}]\bigr\}
\end{equation}
one can easily show that $\Psi({\bf x}; t)$ is the wave function of $M$-particle
boson system with attractive $\delta$-interaction defined by the Schr\"odinger equation:
\begin{equation}
\label{11.19}
 -\partial_t \Psi({\bf x}; t) \; = \;
\sum_{a=1}^{M} \frac{1}{2\beta_{a}} \partial_{x_a}^2 \Psi({\bf x}; t) 
+
\frac{1}{2} u^{2} \sum_{a\not=b}^{M}\beta_{a}\beta_{b}\delta(x_{a}-x_{b}) \Psi({\bf x}; t)
\end{equation}
with the initial condition $\Psi({\bf x}; 0) = \Pi_{a=1}^{M} \delta(x_a)$.
According to the definitions (\ref{11.15}) and (\ref{11.18}),
\begin{equation}
\label{11.20}
 {\cal Z}(M,N; \, \beta_{1}, \beta_{2}; \, t) \, = \,
 \Psi(x_{1}, ..., x_{M} ; t)\Big|_{x_{a}=0}
\end{equation}
The time dependent wave function  $\Psi({\bf x}; \, t)$ of the above quantum problem can be represented
in terms of the linear combination of the eigenfunctions $\Psi({\bf x})$ defined by the solutions of the
eigenvalue equation
\begin{equation}
   \label{11.21}
2E\Psi({\bf x}) =
\sum_{a=1}^{M} \frac{1}{\beta_{a}}\partial_{x_a}^2 \Psi({\bf x}) + 
 u^{2} \sum_{a\not=b}^{M} \beta_{a}\beta_{b} \delta(x_{a}-x_{b})\Psi({\bf x})
\end{equation}
Unlike the case with all $\beta$'s equal \cite{Lieb-Liniger,McGuire,Yang}, for the time being, 
the general solution of this equation
is not known. However, if we do not pretend to derive the exact result for the entire probability
distribution function $P_{\beta_{1},\beta_{2},t}({\cal F})$ but we want to know only its left tail
asymptotics  in the limit $t \to \infty$ then it would be enough
to get the behavior of the replica partition function
${\cal Z}(0,N \to \infty; \, \beta_{1}, \beta_{2}; \, t \to \infty)$
which is defined by the ground state solution only:
\begin{equation}
\label{11.22}
\Psi({\bf x}; t\to \infty) \; \sim \; \exp\bigl\{ - E_{g.s.}t\bigr\} \,  \Psi_{g.s.}({\bf x})
\end{equation}
One can easily check that the ground state solution of eq.(\ref{11.21}) is given by the eigenfunction
\begin{equation}
 \label{11.23}
\Psi_{g.s.}({\bf x}) \; \propto \;
\exp\Biggl\{
-\frac{1}{2} \, u \sum_{a,b=1}^{M} \gamma_{ab} \, \big|x_{a} - x_{b}\big|
\Biggr\}
\end{equation}
where
\begin{equation}
 \label{11.24}
\gamma_{ab} \; = \; \frac{\beta_{a}^{2} \, \beta_{b}^{2}}{\beta_{a} + \beta_{b}}
\end{equation}
The corresponding ground state energy is
\begin{equation}
 \label{11.25}
E_{g.s.}(M, N,\beta_{1},\beta_{2}) = -\frac{1}{2} u^{2}
\sum_{a=1}^{M} \frac{1}{\beta_{a}} \Biggl(\sum_{b=1}^{a-1}\gamma_{ab} - \sum_{b=a+1}^{M}\gamma_{ab}\Biggr)^{2}
\end{equation}
Note that in the trivial case $\beta_{1} = \beta_{2} = \beta$, using eqs.(\ref{11.23})-(\ref{11.25}),  
one easily recovers the well known ground state solution
$\psi_{g.s.} \propto \exp\bigl\{ -\frac{1}{4} \, u \beta^{3} \sum_{a,b=1}^{M} \big|x_{a} - x_{b}\big| \bigr\}$
and $E_{g.s.} \; = \; -\frac{1}{24} u^{2}\beta^{5} (M^{3} - M)$.
Substituting eqs.(\ref{11.17}) and (\ref{11.24}) into eq.(\ref{11.25}) after simple algebra in the limit $M\to 0$ we obtain
\begin{equation}
 \label{11.26}
E_{g.s.}(0, N,\beta_{1},\beta_{2}) = -\frac{u^{2}}{24} \lambda(\beta_{1},\beta_{2}) N^{3}
+ \frac{u^{2}}{24}\bigl(\beta_{1}^{5}-\beta_{2}^{5}\bigr) \, N
\end{equation}
where
\begin{eqnarray}
 \nonumber
\lambda(\beta_{1},\beta_{2}) &=& 
4\bigl(\beta_{1}^{5}-\beta_{2}^{5}\bigr) 
- 
6 \bigl(\beta_{1}-\beta_{2}\bigr) 
\frac{2\beta_{1}^{4}\beta_{2}+2\beta_{2}^{4}\beta_{1} +\beta_{1}^{5} +\beta_{2}^{5}}{\beta_{1}+\beta_{2}} 
\\
\nonumber
\\
&+&
3 \bigl(\beta_{1}-\beta_{2}\bigr)^{2} 
\frac{\beta_{1}^{3}(2\beta_{2}+\beta_{1})^{2} - \beta_{2}^{3}(2\beta_{1}+\beta_{2})^{2}}{(\beta_{1}+\beta_{2})^{2}}
 \label{11.27}
\end{eqnarray}
According to eqs.(\ref{11.20}) and (\ref{11.22}) we find
\begin{equation}
 \label{11.28}
{\cal Z}(0,N\to\infty;\beta_{1},\beta_{2}; t\to\infty) \; \sim \; 
\exp\Bigl\{
\frac{u^{2}}{24}  \lambda(\beta_{1},\beta_{2}) \, N^{3}  t  
- 
\frac{ u^{2}}{24} \bigl(\beta_{1}^{5} - \beta_{2}^{5}\bigr)  N  t
\Bigr\}
\end{equation}
The second (linear on $N$ term) in the exponential of the above relation provides  the contribution to
the selfaveraging (non-random) linear in time part of the free energy variance ${\cal F}$.
Substituting eq.(\ref{11.28}) into eq.(\ref{11.7}) and redefining
\begin{equation}
 \label{11.29}
{\cal F} \; = \; \frac{1}{24} u^{2} \, \bigl(\beta_{1}^{5} - \beta_{2}^{5}\bigr) \,  t \; + \; \tilde{{\cal F}}
\end{equation}
we find that in the limits $t\to\infty$ and $N\to\infty$ the left tail of the probability distribution function
for the random quantity $\tilde{{\cal F}}$ (as $\tilde{{\cal F}} \to -\infty$) is defined by the relation
\begin{equation}
\label{11.30}
\int  d\tilde{{\cal F}} \, P_{\beta_{1},\beta_{2},t}(\tilde{{\cal F}})  \exp\bigl\{- N \tilde{{\cal F}} \bigr\} \sim
\exp\Bigl\{
\frac{u^{2}}{24}  \, \lambda(\beta_{1},\beta_{2})  N^{3}  t
\Bigr\} .
\end{equation}
Redefining
\begin{equation}
\label{11.31}
N \; = \; 2 (u^{2} \lambda)^{-1/3} \, s
\end{equation}
we find that the free energy difference $\tilde{{\cal F}}$ scales as
\begin{equation}
\label{11.32}
\tilde{{\cal F}} \; = \; 
\frac{1}{2} \, u^{2/3} \, \bigl(\lambda(\beta_{1},\beta_{2})\bigr)^{1/3} \, t^{1/3} \, f
\end{equation}
where the left tail of the {\it universal} probability distribution function ${\cal P}(f)$ 
of the random quantity $f$
is defined by the relation
\begin{equation}
\label{11.33}
\int \, df \, {\cal P}(f) \, \exp\bigl\{- s \, f \bigr\} \; \sim \;
\exp\Bigl\{ \frac{1}{3} \, s^{3} \Bigr\} .
\end{equation}
Simple saddle-point estimate of the above integral (for $s \gg 1$ and $|f| \gg 1$) yields
\begin{equation}
\label{11.34}
{\cal P}(f \to -\infty) \; \sim \; 
\exp\Bigl\{ -\frac{2}{3} \, |f|^{2/3} \Bigr\} .
\end{equation}
Note that this tail coincides with the corresponding asymptotics 
of the usual free energy TW distribution \cite{TW-GUE}
(both GUE, GOE and GSE) but this tells nothing about its entire exact shape
which at present stage remains unknown.

\vspace{5mm}

Let us consider in more detail the scaling relation (\ref{11.32}) which demonstrate the dependence
of the typical value of the fluctuating part of the reduced free energy difference, eq.(\ref{11.3}), 
on the strength of disorder $u$, on  the inverse temperatures $\beta_{1}$ and $\beta_{2}$,  
and on time $t$. First of all, one notes that the disorder scaling $\sim u^{2/3}$ as well as  
and time scaling $\sim t^{1/3}$ coincide with the ones  of the usual free energy scaling in 
$(1+1)$ directed polymers, which of course is not surprising. 
On the other hand, the dependence on the inverse temperatures $\beta_{1}$ and $\beta_{2}$ 
turns out to be   less trivial.

First of all, using explicit expression (\ref{11.27}) one easily finds that in the limit
$\beta_{1} \gg \beta_{2}$  (or $T_{1} \ll T_{2}$)
\begin{equation}
 \label{11.35}
\lambda\bigl(\beta_{1}, \beta_{2}\bigr)\big|_{\beta_{1}\gg \beta_{2}} \, \simeq \, \beta_{1}^{5} \; ,
\end{equation}
so that in this limit the scaling relation (\ref{11.32}) turns into the usual one-temperature
free energy scaling 
\begin{equation}
 \label{11.36}
\tilde{{\cal F}} \; \simeq \; \beta_{1} \tilde{F}_{1} \, = \, \frac{1}{2} \bigl(u^{2}\beta_{1}^{5}\bigr)^{1/3} \, t^{1/3} \, f
\end{equation}
In other words, in this case the free energy $F_{1}$ of the polymer with the temperature $T_{1}$ is much lower 
than that of the polymer with the temperature $T_{2} \gg T_{1}$, and the free energy difference
$\tilde{{\cal F}}$ is dominated by the free energy $F_{1}$ as it should be.

Let us consider now what happens if the two temperature parameters $\beta_{1}$ and $\beta_{2}$ are
close to each other. Introducing a small (positive) parameter 
\begin{equation}
 \label{11.37}
\epsilon \; = \; \frac{\beta_{1} \, - \, \beta_{2}}{\beta_{1}} \, \ll \, 1
\end{equation}
and substituting $\beta_{2} \, = \, (1-\epsilon) \beta_{1}$ into eq.(\ref{11.27}) in the leading order in $\epsilon \ll 1$
we get
\begin{equation}
 \label{11.38}
\lambda \; \simeq \; 2\beta_{1}^{5} \, \epsilon
\end{equation}
Substituting this into eq.(\ref{11.32}) we find that in this case   the fluctuating part of the 
the corresponding free energy difference $\tilde{{\cal F}}$, eq.(\ref{11.3}), scales as
\begin{equation}
\label{11.39}
\tilde{{\cal F}}\; \simeq \; 
\frac{1}{2} \bigl(2 u^{2}\beta_{1}^{5}\bigr)^{1/3} \, 
\Biggl(\frac{\beta_{1} \, - \, \beta_{2}}{\beta_{1}}\Biggr)^{1/3} \, t^{1/3} \, f
\end{equation}
where the random quantity $f$ is described by a universal distribution function ${\cal P}(f)$ whose
left tail asymptotics is given in eq.(\ref{11.34}). 
The above eq.(\ref{11.39}) constitutes the main result of the present study and it implies "one more 1/3"
exponent in these type of systems.

\newpage

\vspace{10mm}

\section{Conclusions}

\newcounter{11}
\setcounter{equation}{0}
\renewcommand{\theequation}{11.\arabic{equation}}

In this rather technical review we considered various aspects of the methods and ideas which
allows to derive various universal statistical properties of 
one-dimensional directed polymers in a random potential.
The special emphasis was made on the detailed description of the so called replica
Bethe ansatz technique which is the combination of the usual replica method 
(used for systems with quenched disorder) 
and the Bethe ansatz wave function solution for one-dimensional $N$-particle quantum boson
system (which is equivalent to the replicated representation of the original 
directed polymer problem).
In the thermodynamic limit (the limit of infinite system size) it has been demonstrated how 
in terms of this technique the exact results for different types of the free energy
probability distribution functions can be obtained. Besides, we derive the explicit 
expression for the probability distribution function for the polymer's endpoint fluctuations.
In all these cases the final (universal) results are expressed in terms of the standard Fredholm determinants
which deffer from each other only by  the different types of kernels.

\vspace{3mm}

The second part of the review was devoted to the problems which are still waiting for their solutions: 

\vspace{3mm}

$\bullet$  The zero-temperature limit for the directed polymers  in a random potential
with {\it finite} correlation length (Chapter VIII). Here the problem is that the exact solution of the 
model with $\delta$-correlated potentials (considered in Chapters IV-VII)  
does not reveal any finite zero-temperature limit. 
The physical origin of this pathology is clear:  the model with $\delta$-correlated potentials 
is ill defined at short scales while it is the 
short scales which are getting the most relevant in the limit $T \to 0$. 
Introducing smooth finite size correlations for the random potential 
we can cure this pathology but in this case  the solution derived for the model with 
$\delta$-correlated potentials becomes not valid. 
In Chapter VIII in terms of the standard replica technique have formulated a general 
scheme which hopefully would allow to obtain a finite  zero-temperature limit solution for the 
directed polymer model with finite size correlations of the random potential.

\vspace{3mm}


$\bullet$  The problem of velocity distribution functions in the random force Burgers
turbulence (Chapter IX). Formally, this problem is equivalent to the 
considered directed polymers in a random potential, but 
 unlike the usual KPZ studies with the $\delta$-correlated 
random potential, in the Burgulence problem one is mainly interested in the spatial scales comparable
or much smaller than the random potential correlation length . 
In other words, (like Chapter VIII) here   
one has to deal with the model with finite size correlations of the random potential 
whose solution is not known.

\vspace{3mm}


$\bullet$ Joint distribution function of the directed polymers free energies 
at {\it different}  temperatures with {\it the same} realization of a random potential (Chapter X).
Here it is possible to derive the general scaling dependence for the difference of these free energies 
on the two temperatures $T_{1}$ and $T_{2}$. In particular, if the two temperatures $T_{1} \, < \, T_{2}$ 
are close to each other, so that $(1-T_{1}/T_{2}) \; \ll \; 1$, the difference
of the two free energies can be shown to scale as $(1-T_{1}/T_{2})^{1/3} \, t^{1/3}$. 
On the other hand,  a general joint distribution function of these free energies
remains to be computed.

\newpage


\vspace{10mm}

\begin{center}

\appendix{\large \bf Appendix A: GUE Tracy-Widom distribution function}

\end{center}

\newcounter{A}
\setcounter{equation}{0}
\renewcommand{\theequation}{A.\arabic{equation}}

\vspace{5mm}

Originally the Tracy-Widom distribution function has been derived in the context 
of the statistical properties of the Gaussian Unitary Ensemble (GUE) 
of random Hermitian matrices
\cite{TW-GUE}. GUE is the set of $N\times N$ 
random complex Hermitian matrices $G_{ij}$
(such that $G_{ij} = G_{ji}^{*}$) whose elements are drawn independently from the Gaussian
distribution
\begin{equation}
\label{A.1}
{\cal P}[{\bf G}] \; = \; B_{N} \, \exp\Bigl\{ -\frac{1}{2} \mbox{Tr} \, \hat{G}^{2} \Bigr\}
\end{equation}
where $B_{N}$ is the normalization constant. The joint probability density of $N$ eigenvalues
$\{\lambda_{1}, \lambda_{2}, ..., \lambda_{N}\} $
of such matrices has rather compact form \cite{Wigner}:
\begin{equation}
\label{A.2}
{\cal P}[\lambda_{1}, \lambda_{2}, ..., \lambda_{N}] \; = \; C_{N} \, 
\prod_{i\not= j}^{N} \Big|\lambda_{i} - \lambda_{j}\Big|^{2} 
\exp\Bigl\{ -\sum_{i=1}^{N} \lambda_{i}^{2} \Bigr\}
\end{equation}
where $C_{N}$ is the normalization constant. Using this joint probability 
density one can calculate various averaged characteristics of the 
eigenvalue statistics. For example, one can introduce the average
density of the eigenvalues 
$\rho(\lambda, N) \; = \; \frac{1}{N} \sum_{i=1}^{N}
 \langle \delta(\lambda - \lambda_{i})\rangle$ where the averaging 
$\langle .....\rangle$ is performed with the probability distribution,
eq.(\ref{A.2}). Using the symmetry of this distribution one gets
\begin{equation}
\label{A.3}
\rho(\lambda, N) \; = \; 
\prod_{i=2}^{N} \Biggl[
\int_{-\infty}^{+\infty} \; d\lambda_{i} 
\Biggr]
{\cal P}[\lambda, \lambda_{2}, ..., \lambda_{N}] 
\end{equation}
It can be shown \cite{Wigner} that in the limit of large $N$
\begin{equation}
\label{A.4}
\rho(\lambda, N) \; = \; \sqrt{\frac{2}{N \pi^{2}} \; 
\biggl(1 - \frac{\lambda^{2}}{2 N} \biggr) }
\end{equation}
We see that on average the eigenvalues lie within the 
finite interval $\bigl[ -\sqrt{2N} < \lambda < \sqrt{2N} \bigr]$
where, according to eq.(\ref{A.4}), their density has the semi-circular form.
This is one of the central results of the random matrix theory which is 
called the Wigner semi-circular law. In particular this result tells that
on average the maximum eigenvalue $\lambda_{max}$ is equal to $\sqrt{2N}$. 
However, at large but finite $N$ the value of $\lambda_{max}$ is the random quantity
which fluctuates from sample to sample. One may ask, what is the full probability
distribution  of the largest eigenvalue  $\lambda_{max}$?
This distribution can be computed in terms of the general probability density,
eq.(\ref{A.2}). Introducing standard notations of the random matrix theory
we define the function 
$F_{2}(s) \equiv \mbox{Prob}\bigl[\lambda_{max} < s\bigr]$ which gives the probability that
$\lambda_{max}$ is less than a given value $s$.
In these notations the functions $F_{1}(s)$, $F_{2}(s)$ and $F_{4}(s)$ 
denote the probability distributions of the largest eigenvalues in the 
Gaussian Orthogonal Ensemble (GOE), Gaussian Unitary Ensemble (GUE) 
and Gaussian Symplectic Ensemble (GSE) correspondingly \cite{TW2}.
By definition
\begin{equation}
\label{A.5}
F_{2}(s) \;  = \; 
\prod_{i=i}^{N} \Biggl[
\int_{-\infty}^{s} \; d\lambda_{i} 
\Biggr]
{\cal P}[\lambda_{1}, \lambda_{2}, ..., \lambda_{N}] \; \equiv \; 
 \int_{-\infty}^{s} d\lambda \, P_{TW}(\lambda) 
\end{equation}
It is this problem which has been solved by Tracy and Widom in 1994 \cite{TW-GUE}. 
It has been shown that at large $N$ the typical fluctuations of $\lambda_{max}$ around
its mean value $\sqrt{2N}$ scale as $N^{-1/6}$, namely 
\begin{equation}
\label{A.6}
\lambda_{max} = \sqrt{2N} \; + \; \frac{1}{\sqrt{2} \, N^{1/6} } \, s
\end{equation}
where the random quantity $s$ is described by $N$-independent distribution
$P_{TW}(s) = dF_{2}(s)/ds$. The function $F_{2}(x)$, has the following 
explicit form\
\begin{equation}
\label{A.7}
F_{2}(s) \; = \; \exp\biggl(-\int_{s}^{\infty} dt \; (t-s) \; q^{2}(t)\biggr)
\end{equation}
or
\begin{equation}
 \label{A.8}
P_{TW}(s) \; = \; \frac{d}{ds} F_{2}(s) \; = \; 
\exp\biggl\{
-\int_{s}^{\infty} dt \, (t-s) q^{2}(t) 
\biggr\} \times \int_{s}^{\infty} dt \, q^{2}(t)
\end{equation}
where the function $q(t)$ is the solution of 
the Panlev\'e II equation$^{(1)}$\footnotetext[1]{There exist six Panlev\'e differential
equations which were discovered about a hundred years ago\cite{Panleve} 
(for the recent review
see e.g. \cite{Clarkson}). It is proved that the general solutions 
of the Panleve\'e equations 
are transcendental in a sense that they can not be expressed in terms 
of any of the previously 
known function including all classical special functions. At present the Panlev\'e 
equations have many applications in various parts of modern physics including
statistical mechanics, plasma physics, nonlinear waves, quantum field theory
and general relativity}, 
\begin{equation}
 \label{A.9}
q'' = t q + 2 q^{3}
\end{equation}
with the boundary condition, $q(t\to +\infty) \sim Ai(t)$. 
\begin{figure}[h]
\begin{center}
   \includegraphics[width=8.0cm]{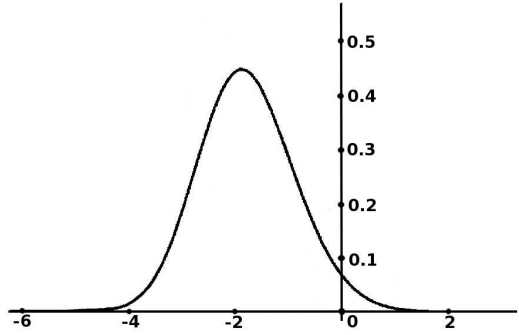}
\caption[]{Tracy-Widom distribution function $P_{TW}(x)$, eq.(\ref{A.8}).}
\end{center}
\label{figure7}
\end{figure}
The shape of the function $P_{TW}(s)$ is shown in Figure 6. Note that the asymptotic tails 
of this function are strongly asymmetric. While its right tail coincides the Airy function
asymptotic $P_{TW}(s \to +\infty) \sim \exp\bigl[-\frac{4}{3} s^{3/2}\bigr]$, the left tail
exhibits much faster decay 
$P_{TW}(s\to -\infty) \sim  \exp\bigl[-\frac{1}{12} |s|^{3}\bigr]$


\vspace{10mm}

\begin{center}

\appendix{\large \bf Appendix B: The Airy function integral relations}

\end{center}

\newcounter{B}
\setcounter{equation}{0}
\renewcommand{\theequation}{B.\arabic{equation}}

\vspace{5mm}

The Airy function $\Ai(x)$ is the solution of the differential equation
\begin{equation}
\label{B.1}
y''(x) \; = \; x\, y(x)
\end{equation}
with the boundary condition $y(x\to +\infty) = 0$. At $x\to +\infty$ this function
goes to zero exponentially fast
\begin{equation}
\label{B.2}
\Ai(x\to +\infty) \; \simeq \; \frac{1}{2\sqrt{\pi} x^{1/4}} \exp\Bigl(-\frac{2}{3} x^{3/2}\Bigr)
\end{equation}
while at $x\to -\infty$ it oscillates and decays much more slowly:
\begin{equation}
\label{B.3}
\Ai(x\to -\infty) \; \simeq \; \frac{1}{\sqrt{\pi} |x|^{1/4}} 
\sin\Bigl(\frac{2}{3} |x|^{3/2} + \frac{1}{4}\pi\Bigr)
\end{equation}
The Airy function can also be represented in the integral form:
\begin{equation}
\label{B.4}
\Ai(x) \; = \; \int_{{\cal C}} \frac{dz}{2\pi i} \; \exp\Bigl(\frac{1}{3} z^{3} - z x\Bigr)
\end{equation}
where the integration path in the complex plane starts at a point at infinity with the argument
$-\pi/2 < \theta_{(-)} < -\pi/3$ and ends up at a point at infinity with the argument
$\pi/3 < \theta_{(+)} < \pi/2$. Choosing the argument of the staring point 
$\theta_{(-)} = -\pi/2 + \epsilon$ and that of the ending point 
$\theta_{(+)} = \pi/2 - \epsilon$ where the positive parameter $\epsilon \to 0$ is introduced 
just to provide the convergence of the integration, the integration 
path in eq.(\ref{B.4}) may  coincide with the imaginary axes $z = i y$.

\vspace{5mm}

Below we give several useful integral relations for the Airy function which can be easily proved using 
the above definitions, eqs.(\ref{B.1}) and (\ref{B.4}) .

\vspace{3mm}

1. Orthogonality:
\begin{equation}
 \label{B.5}
I_{1} \; = \; \int_{-\infty}^{+\infty} \, dx \, \Ai(x + u) \, \Ai(x + v) \; = \; \delta(u-v)
\end{equation}

\vspace{3mm}

2. Just like in the well known Hubbard-Stratonovich transformation the Gaussian function
is used to linearize quadratic expressions in the exponential,
the Airy function can be used to linearize the {\it cubic} exponential terms:
\begin{equation}
\label{B.6} 
I_{2} \; = \; \int_{-\infty}^{+\infty} \, dx \; \Ai(x) \, \exp\bigl( F x \bigr) 
\; = \; \exp\Bigl\{\frac{1}{3} F^{3} \Bigr\}
\end{equation}
where the quantity $F$ is assumed to be non negative. 

\vspace{3mm}

3. Fourier-like integration:
\begin{equation}
\label{B.7}
I_{3} \; = \; \int_{-\infty}^{+\infty} \frac{dp}{2\pi} \;
\Ai\bigl(ap^{2} + b\bigr) \; 
\exp\bigl\{i p \,c\bigr\}
\; = \; 
2^{-1/3} \, a^{-1/2} \, \Ai\Bigl[2^{-2/3}(b + a^{-1/2} c)\Bigr]  
                        \Ai\Bigl[2^{-2/3}(b - a^{-1/2} c)\Bigr] 
\end{equation}
In the special case $a=1$, $\; b = u + v$ and $c = u - v$,
\begin{equation}
\label{B.8}
I_{3}' \; = \; \int_{-\infty}^{+\infty} \frac{dp}{2\pi} \;
\Ai\bigl(p^{2} + u + v\bigr) \; 
\exp\bigl\{i p \, (u-v)\bigr]
\; = \; 
2^{-1/3}  \Ai\bigl(2^{1/3} u\bigr)  \Ai\bigl(2^{1/3} v\bigr) 
\end{equation}

\vspace{3mm}

4. The Airy kernel:
\begin{equation}
\label{B.9}
I_{4} \; = \; \int_{0}^{\infty} dx \; 
\Ai\bigl(x + u\bigr) \Ai\bigl(x + v\bigr) 
\; = \; 
\frac{\Ai\bigl(u\bigr) \Ai'\bigl(v\bigr) \; - \; 
\Ai'\bigl(u\bigr) \Ai\bigl(v\bigr)}{u - v}
\end{equation}

\vspace{3mm}

5. The integral which transform the product of two Airy functions into one Airy function:
\begin{equation}
 \label{B.10}
I_{5} \; = \; \int_{-\infty}^{\infty} dx \;
\Ai(x + u) \, \Ai(-x/c + v) \; = \; \frac{c}{(1 + c^{3})^{1/3}} \, \Ai\Bigl[\frac{u + c v}{(1 + c^{3})^{1/3}} \Bigr]
\end{equation}

\vspace{3mm}

6. Finally, the differential operator relation:
\begin{equation}
 \label{B.11}
\exp\Bigl\{a \frac{\partial}{\partial x} \Bigr\} \, \Ai(x) \; = \; \Ai(x + a)
\end{equation}


\vspace{10mm}

\begin{center}

\appendix{\large \bf Appendix C: Fredholm determinant with the Airy kernel}

\end{center}

\newcounter{C}
\setcounter{equation}{0}
\renewcommand{\theequation}{C.\arabic{equation}}

\vspace{5mm}

In simplified terms the Fredholm determinant $\det \bigl(1 - \hat{K}\bigr)$
can be defined as follows 
(for a rigorous mathematical definition see e.g. \cite{Mehta}):
\begin{equation}
\label{D1}
\det \bigl(1 - \hat{K}\bigr) \; = \; 
1 \; + \; \sum_{n=1}^{\infty} \frac{(-1)^{n}}{n!}
\int\int ...\int_{a}^{b} dt_{1} dt_{2} ... dt_{n} \; 
\det\Bigl[K(t_{i}, t_{j})\Bigr]_{(i,j) = 1, ..., n}
\end{equation}
where the  kernel $\hat{K} \equiv K(t,t')$ is a function of two
variables defined in a region $a \leq (t,t') \leq b$. Equivalently 
the Fredholm determinant can also be represented in the 
exponential form
\begin{equation}
\label{D2}
\det \bigl(1 - \hat{K}\bigr) \; = \; 
\exp\biggl[-\sum_{n=1}^{\infty} \frac{1}{n} \, \mbox{Tr} \, \hat{K}^{n} \biggr]
\end{equation}
where
\begin{equation}
\label{D3}
\mbox{Tr} \, \hat{K}^{n} \; \equiv \;
\int\int ...\int_{a}^{b} dt_{1} dt_{2} ... dt_{n} \;
K(t_{1}, t_{2}) K(t_{2}, t_{3}) ... K(t_{n}, t_{1})
\end{equation}

In this Appendix following the original derivation of  Tracy and Widom \cite{TW-GUE} it will be 
demonstrated that the function $F_{2}(s)$
defined as the Fredholm determinant with the Airy kernel can be expressed in terms of the 
solution of the Panlev\'e II differential equation, namely
\begin{equation}
 \label{D4}
 F_{2}(s) \equiv \det\bigl[1 - \hat{K}_{A}\bigr] \; = \; 
\exp\biggl[
-\int_{s}^{\infty} dt \, (t-s) q^{2}(t) 
\biggr]
\end{equation}
where $\hat{K}_{A}$ is the Airy kernel defined on semi-infinite interval $[s, \infty)$:  
\begin{equation}
 \label{D5}
K_{A}(t_{1},t_{2}) \; = \; 
\frac{\Ai(t_{1}) \Ai'(t_{2}) - \Ai'(t_{1}) \Ai(t_{2})}{t_{1} - t_{2}}
\end{equation}
and the function $q(t)$ is the solution of the Panlev\'e II differential equation,
\begin{equation}
 \label{D6}
q'' = t q + 2 q^{3}
\end{equation}
with the boundary condition, $q(t\to +\infty) \sim \Ai(t)$.

\vspace{5mm}

Let us introduce a new function $R(t)$ such that
\begin{equation}
 \label{D7}
 F_{2}(s) \; = \; 
\exp\biggl[
-\int_{s}^{\infty} dt R(t) 
\biggr]
\end{equation}
or, according to the definition, Eq.(\ref{D4}),
\begin{equation}
 \label{D8}
R(s) \; = \; \frac{d}{ds} \ln\Bigl[\det\bigl(1 - \hat{K}_{A}\bigr) \Bigr]
\end{equation}
Here the logarithm of the determinant can be expressed in terms of the trace:
\begin{eqnarray}
 \label{D9}
\ln\Bigl[\det\bigl(1 - \hat{K}_{A}\bigr) \Bigr] 
&=&
- \sum_{n=1}^{\infty} \frac{1}{n} \, Tr \, \hat{K}_{A}^{n} 
\\
\nonumber
\\
\nonumber
&\equiv& 
- \sum_{n=1}^{\infty} \frac{1}{n}
\int_{s}^{\infty} dt_{1} \int_{s}^{\infty} dt_{2} \, ... \, \int_{s}^{\infty} dt_{n} \; 
 K_{A}(t_{1},t_{2}) K_{A}(t_{2},t_{3}) \, ... \, K_{A}(t_{n},t_{1}) 
\end{eqnarray}
Taking derivative of this expression we gets
\begin{eqnarray}
 \label{D10}
R(s) &=& - \int_{s}^{\infty} dt \, \bigl(1 - \hat{K}_{A}\bigr)^{-1} (s,t) \, K_{A}(t,s)
\\
\nonumber
\\
\nonumber
&\equiv& 
 - K_{A}(s,s) 
- \sum_{n=2}^{\infty} 
\int_{s}^{\infty} dt_{1} \int_{s}^{\infty} dt_{2} \, ... \, \int_{s}^{\infty} dt_{n-1} \; 
K_{A}(s,t_{1}) K_{A}(t_{1},t_{2}) \, ... \, K_{A}(t_{n-1},s) 
\end{eqnarray}
Substituting here the integral representation of the Airy kernel, Eq.(\ref{D5}),
\begin{equation}
 \label{D11}
K_{A}(t_{1},t_{2}) \; = \; 
\int_{0}^{\infty} dz \Ai(t_{1} + z) \, \Ai(t_{2} + z)
\end{equation}
after  simple algebra one gets
\begin{equation}
 \label{D12}
R(s) \; = \; 
\int_{s}^{\infty} dt_{1} \int_{s}^{\infty} dt_{2} \, 
\Ai(t_{1}) \, \bigl(1 - \hat{K}_{A}\bigr)^{-1} (t_{1},t_{2}) \, \Ai(t_{2})
\end{equation}
Taking the derivative of this expression after  somewhat cumbersome
algebra, we obtain
\begin{equation}
 \label{D13}
\frac{d}{ds} R(s) \; = \; - q^{2}(s)
\end{equation}
where
\begin{equation}
 \label{D14}
q(s) \; = \; \int_{s}^{\infty} dt \, \bigl(1 - \hat{K}_{A}\bigr)^{-1} (s,t) \, \Ai(t)
\end{equation}
According to Eq.(\ref{D13}),
\begin{equation}
 \label{D15}
 R(s) \; = \; \int_{s}^{\infty} dt \, q^{2}(t)
\end{equation}
Let us introduce two more functions
\begin{eqnarray}
 \label{D16}
v(s) &=& \int_{s}^{\infty} dt_{1} \int_{s}^{\infty} dt_{2} \, 
\Ai(t_{1}) \, \bigl(1 - \hat{K}_{A}\bigr)^{-1} (t_{1},t_{2}) \, \Ai'(t_{2})
\\
\nonumber
\\
\label{D17}
p(s) &=& \int_{s}^{\infty} dt \, \bigl(1 - \hat{K}_{A}\bigr)^{-1} (s,t) \, \Ai'(t)
\end{eqnarray}
Taking derivatives of the above three functions $q(s)$, $v(s)$ and $p(s)$, 
Eqs.(\ref{D14}), (\ref{D16}) and (\ref{D17}),
after somewhat painful algebra one finds the following three relations:
 \begin{eqnarray}
 \label{D18}
q' &=& p \; - \; R \, q
\\
\label{D19}
p' &=& s \, q \; - \; p \, R \; - \; 2 q \, v
\\
\label{D20}
v' &=& - p \, q
\end{eqnarray}
Taking derivative of the combination $\bigl(R^{2} - 2v\bigr)$ and using Eqs.(\ref{D13}) and (\ref{D20}),
we get
\begin{equation}
 \label{D21}
\frac{d}{ds} \bigl(R^{2} - 2v\bigr) \; = \; 2q \, (p \; - \; R\, q)
\end{equation}
On the other hand, multiplying Eq.(\ref{D18}) by $2q$ we find
\begin{equation}
 \label{D22}
\frac{d}{ds} q^{2} \; = \; 2q \, (p \; - \; R\, q)
\end{equation}
Comparing Eqs.(\ref{D21}) and (\ref{D22}) and taking into account that 
the value of all the above functions at $s\to\infty$ is zero, we obtain the following relation
\begin{equation}
 \label{D23}
R^{2} - 2v \; = \; q^{2}
\end{equation}
Finally, taking the derivative of Eq.(\ref{D18}) and using 
Eqs.(\ref{D13}), (\ref{D18}), (\ref{D19})
and (\ref{D23}) we easily find
\begin{equation}
 \label{D24}
q'' \; = \; 2q^{3} \; + \; s q
\end{equation}
which is the special case of the Panlev\'e II differential equation 
\cite{Panleve,Clarkson,Iwasaki}.
Thus, substituting Eq.(\ref{D15}) into Eq.(\ref{D7}) we obtain Eq.(\ref{D4}).

In the limit $s\to\infty$ the function $q(s)$, according to its definition, eq.(\ref{D14}),
must go to zero, and in this case Eq.(\ref{D24})
turns into the Airy function equation, $q'' = s q$. Thus
\begin{equation}
 \label{D27}
q(s\to\infty) \; \simeq \; \Ai(s) \; \sim \; \exp\Bigl[-\frac{2}{3} s^{3/2}\Bigr]
\end{equation}
It can be proved \cite{Hastings} that in the opposite limit, $s\to -\infty$, the asymptotic form 
of the solution of the Panleve\'e equation (\ref{D24}) (which has the right tail Airy function
limit, Eq.(\ref{D27})) is
\begin{equation}
 \label{D28}
q(s\to -\infty) \; \simeq \; \sqrt{-\frac{1}{2}s}
\end{equation}

\vspace{5mm}

The GUE Tracy-Widom probability density  distribution function $P_{GUE}(t)$ is defined as 
\begin{equation}
 \label{D25}
P_{GUE}(s) \; = \;  \frac{d}{ds} \, F_{2}(s) 
\end{equation}
Substituting  here  Eq.(\ref{D4}), we find
\begin{equation}
 \label{D26}
P_{GUE}(s) \; = \; 
\exp\biggl\{
-\int_{s}^{\infty} dt \, (t-s) q^{2}(t) 
\biggr\}
\times \int_{s}^{\infty} dt \, q^{2}(t)
\end{equation}
where the function $q(t)$ is the solution of the differential equation (\ref{D24})
Substituting the two asymptotics, eqs.(\ref{D27}) and (\ref{D28}), into Eq.(\ref{D26}), 
we can estimate the asymptotic behavior
for the right and the left tails:
\begin{eqnarray}
 \label{D29}
P_{GUE}(s\to +\infty) &\sim&
 \exp\Bigl[-\frac{4}{3} s^{3/2}\Bigr]
\\
\nonumber
\\
\label{D30}
P_{GUE}(s\to -\infty) &\sim&
 \exp\Bigl[-\frac{1}{12} |s|^{3}\Bigr]
\end{eqnarray}


\vspace{10mm}

\begin{center}

\appendix{\large \bf Appendix D: Useful combinatorial identities}

\end{center}

\newcounter{D}
\setcounter{equation}{0}
\renewcommand{\theequation}{D.\arabic{equation}}

\vspace{5mm}

In this appendix we give (without proof) several remarkable combinatorial identities.

\vspace{5mm}

1. Cauchy double alternant identity. 
In can be proved \cite{Cauchy} (see also \cite{Muir}) that for a given two sets of arbitrary parameters
$\{x_{1}, x_{2}, ..., x_{N}\}$ and $\{y_{1}, y_{2}, ..., y_{N}\}$
\begin{equation}
 \label{D.1}
 \frac{\prod_{1<a<b}^{N} \,(x_{a}-x_{b})(y_{a}-y_{b})}{\prod_{a,b=1}^{N} \,(x_{a}+x_{b})} \; = \; 
 \det\Biggl[\frac{1}{x_{a} + y_{b}} \Biggr]_{(a,b)=1, ..., N}
\end{equation}
As a consequence, for any given set of parameters $\{Q_{1}, Q_{2}, ..., Q_{N}\}$ and for any $x$
\begin{equation}
 \label{D.2}
 \prod_{1<a<b}^{N} \Bigg|\frac{Q_{a} - Q_{b}}{Q_{a} - Q_{b} + ix} \Bigg|^{2} \; = \; 
 x^{N} \, \det\Biggl[\frac{1}{x - iQ_{a} + iQ_{b}}\Biggr]_{(a,b)=1, ..., N}
\end{equation}

\vspace{5mm}

2. For any given set of parameters $\{Q_{1}, Q_{2}, ..., Q_{N}\}$ and for any $x$
\begin{equation}
 \label{D.3}
\sum_{P\in S_{N}} \Biggl\{ 
\prod_{1<a<b}^{N} \Biggl[\frac{Q_{P_a} - Q_{P_b} - x}{Q_{P_a} - Q_{P_b}} \Biggr] 
\, \frac{1}{Q_{P_1} \, \bigl(Q_{P_1} + Q_{P_2}\bigr) \, ... \, \bigl(Q_{P_1} + ... + Q_{P_N}\bigr)} 
\Biggr\} 
\; = \; 
\frac{1}{\prod_{a=1}^{N}Q_{a}} \, \prod_{1<a<b}^{N} \Biggl[\frac{Q_{_a} + Q_{_b} + x}{Q_{_a} + Q_{_b}} \Biggr] 
\end{equation}
where the summation is taken over all permutations of $N$ elements $\{Q_{1}, Q_{2}, ..., Q_{N}\}$.

\vspace{5mm}

In particular, in the case $x=0$ one gets
\begin{equation}
 \label{D.4}
\sum_{P\in S_{N}} \,  
\frac{1}{Q_{P_1} \, \bigl(Q_{P_1} + Q_{P_2}\bigr) \, ... \, \bigl(Q_{P_1} + ... + Q_{P_N}\bigr)}  
\; = \; 
\frac{1}{\prod_{a=1}^{N}Q_{a}} 
\end{equation}

\vspace{5mm}

Another version of the identity (\ref{D.3}) is

\begin{equation}
 \label{D.5}
\sum_{P\in S_{N}} \Biggl\{ 
(-1)^{[P]} \frac{\prod_{1<a<b}^{N} \Bigl(Q_{P_a} - Q_{P_b} - x\Bigr)}{ 
              Q_{P_1} \, \bigl(Q_{P_1} + Q_{P_2}\bigr) \, ... \, \bigl(Q_{P_1} + ... + Q_{P_N}\bigr)} 
\Biggr\} 
\; = \; 
\frac{1}{\prod_{a=1}^{N}Q_{a}} \, 
\prod_{1<a<b}^{N} \Biggl[\frac{\bigl(Q_{_a} + Q_{_b} + x\bigr)\bigl(Q_{a} - Q_{b}\bigr)}{Q_{_a} + Q_{_b}} \Biggr] 
\end{equation}
where $(-1)^{[P]}$ denotes the parity of the permutation $P$.


\vspace{10mm}

\begin{center}

\appendix{\large \bf Appendix E: Technical part of Chapter V}

\end{center}

\newcounter{E}
\setcounter{equation}{0}
\renewcommand{\theequation}{E.\arabic{equation}}

\vspace{5mm}

In this  Appendix it will be shown how  the factor $I_{K,L}({\bf q}, {\bf n})$, eq.(\ref{5.15})
can be reduced to the product of the  Gamma functions.
 In terms of the parameters $\{m_{\alpha}\}$ and $\{s_{\alpha}\}$
the product factors in eq.(\ref{5.15}) are expressed as follows:
\begin{eqnarray}
\label{E.1}
\prod_{a=1}^{K} q^{(-)}_{{\cal P}_{a}^{(K)}}
&=&
\prod_{\alpha=1}^{M} \prod_{r=1}^{m_{\alpha}}
{q^{\alpha}_{r}}^{(-)}
\\
\nonumber
\\
\label{E.2}
\prod_{a=1}^{L} q^{(+)}_{{\cal P}_{a}^{(L)}}
&=&
\prod_{\alpha=1}^{M} \prod_{r=1}^{s_{\alpha}}
{{q^{*}}^{\alpha}_{r}}^{(+)}
\end{eqnarray}
\begin{eqnarray}
\label{E.3}
\prod_{a<b}^{K}
\Biggl[
\frac{
q^{(-)}_{{\cal P}_a^{(K)}} + q^{(-)}_{{\cal P}_b^{(K)}}  + i \kappa }{
q^{(-)}_{{\cal P}_a^{(K)}} + q^{(-)}_{{\cal P}_b^{(K)}}}
\Biggr]
&=&
\prod_{\alpha=1}^{M} \prod_{1\leq r< r'}^{m_{\alpha}}
\Biggl[
\frac{
{q^{\alpha}_{r}}^{(-)}+{q^{\alpha}_{r'}}^{(-)}+i\kappa}{
{q^{\alpha}_{r}}^{(-)}+{q^{\alpha}_{r'}}^{(-)}}
\Biggr]
\times
\prod_{1\leq\alpha<\beta}^{M} \prod_{r=1}^{m_{\alpha}}\prod_{r'=1}^{m_{\beta}}
\Biggl[
\frac{
{q^{\alpha}_{r}}^{(-)}+{q^{\beta}_{r'}}^{(-)}+i\kappa}{
{q^{\alpha}_{r}}^{(-)}+{q^{\beta}_{r'}}^{(-)}}
\Biggr]
\\
\nonumber
\\
\nonumber
\\
\label{E.4}
\prod_{c<d}^{L}
\Biggl[
\frac{
q^{(+)}_{{\cal P}_c^{(L)}} + q^{(+)}_{{\cal P}_d^{(L)}}-i \kappa }{
q^{(+)}_{{\cal P}_c^{(L)}} + q^{(+)}_{{\cal P}_d^{(L)}}}\Biggr]
&=&
\prod_{\alpha=1}^{M} \prod_{1\leq r< r'}^{s_{\alpha}}
\Biggl[
\frac{
{{q^{*}}^{\alpha}_{r}}^{(+)}+{{q^{*}}^{\alpha}_{r'}}^{(+)}-i\kappa}{
{{q^{*}}^{\alpha}_{r}}^{(+)}+{{q^{*}}^{\alpha}_{r'}}^{(+)}}
\Biggr]
\times
\prod_{1\leq\alpha<\beta}^{M} \prod_{r=1}^{s_{\alpha}}\prod_{r'=1}^{s_{\beta}}
\Biggl[
\frac{
{{q^{*}}^{\alpha}_{r}}^{(+)}+{{q^{*}}^{\beta}_{r'}}^{(+)}-i\kappa}{
{{q^{*}}^{\alpha}_{r}}^{(+)}+{{q^{*}}^{\beta}_{r'}}^{(+)}}
\Biggr]
\\
\nonumber
\\
\nonumber
\\
\nonumber
\prod_{a=1}^{K} \prod_{c=1}^{L}
\Biggl[
\frac{
q_{{\cal P}_a^{(K)}} - q_{{\cal P}_c^{(L)}} - i\kappa}{
q_{{\cal P}_a^{(K)}} - q_{{\cal P}_c^{(L)}}}
\Biggr]
&=&
\prod_{1\leq\alpha<\beta}^{M}
\Biggl\{
\prod_{r=1}^{m_{\alpha}}\prod_{r'=1}^{s_{\beta}}
\Biggl[
\frac{
q^{\alpha}_{r} - {q^{*}}^{\beta}_{r'} - i\kappa}{
q^{\alpha}_{r} + {q^{*}}^{\beta}_{r'}  }
\Biggr]
\times
\prod_{r=1}^{s_{\alpha}}\prod_{r'=1}^{m_{\beta}}
\Biggl[
\frac{
{q^{*}}^{\alpha}_{r} - q^{\beta}_{r'} - i\kappa}{
{q^{*}}^{\alpha}_{r} - q^{\beta}_{r'}  }
\Biggr]
\Biggr\}
\times
\\
\nonumber
\\
\nonumber
\\
\label{E.5}
&\times&
\prod_{\alpha=1}^{M} \prod_{r=1}^{m_{\alpha}}\prod_{r'=1}^{s_{\alpha}}
\Biggl[
\frac{
q^{\alpha}_{r} - {q^{*}}^{\alpha}_{r'} - i\kappa}{
q^{\alpha}_{r} - {q^{*}}^{\alpha}_{r'}  }
\Biggr]
\end{eqnarray}

Substituting eqs.(\ref{E.1})-(\ref{E.5}) into eq.(\ref{5.15}),
and then substituting the resulting expression into eq.(\ref{5.22})
we obtain
\begin{eqnarray}
 \nonumber
W(f) &=& \lim_{t\to\infty} \sum_{M=0}^{\infty} \frac{(-1)^{M}}{M!}
\prod_{\alpha=1}^{M} \Biggl[
\sum_{m_{\alpha}+s_{\alpha}\geq 1}^{\infty} \, (-1)^{m_{\alpha}+s_{\alpha}-1} \, \kappa^{m_{\alpha}+s_{\alpha}}
\int_{-\infty}^{+\infty} \frac{dq_{\alpha} }{2\pi} \,
{\cal G}\bigl(q_{\alpha}, m_{\alpha}, s_{\alpha}\bigr)
\\
\nonumber
\\
\nonumber
&\times&
\exp\Bigl\{
\lambda(t) (m_{\alpha}+s_{\alpha}) \, f 
-\frac{t}{2\beta} \, (m_{\alpha}+s_{\alpha}) q_{\alpha}^{2} \, + \, 
\frac{\kappa^{2}\, t}{24\beta} (m_{\alpha}+s_{\alpha})^{3}
\Bigr\}
\Biggr] 
\\
    \label{E.6}
\\
\nonumber
&\times&
\det\Biggl[
  \frac{1}{\frac{1}{2}\kappa (m_{\alpha}+s_{\alpha}) - i q_{\alpha}
          + \frac{1}{2}\kappa (m_{\beta}+s_{\beta}) + iq_{\beta}}\Biggr]_{\alpha,\beta=1,...M} 
\times {\cal G}_{\alpha\beta} \bigl({\bf q}, {\bf m}, {\bf s}\bigr)
\end{eqnarray}
where
\begin{equation}
 \label{E.7}
{\cal G} =
\frac{(-1)^{s_{\alpha}} (-i\kappa)^{(m_{\alpha}+s_{\alpha})}}{
\prod_{r=1}^{m_{\alpha}}{q^{\alpha}_{r}}^{(-)}
\prod_{r=1}^{s_{\alpha}}{{q^{*}}^{\alpha}_{r}}^{(+)}}
\prod_{r< r'}^{m_{\alpha}}
\Biggl[
\frac{
{q^{\alpha}_{r}}^{(-)}+{q^{\alpha}_{r'}}^{(-)}+i\kappa}{
{q^{\alpha}_{r}}^{(-)}+{q^{\alpha}_{r'}}^{(-)}}
\Biggr]
\prod_{r< r'}^{s_{\alpha}}
\Biggl[
\frac{
{{q^{*}}^{\alpha}_{r}}^{(+)}+{{q^{*}}^{\alpha}_{r'}}^{(+)}-i\kappa}{
{{q^{*}}^{\alpha}_{r}}^{(+)}+{{q^{*}}^{\alpha}_{r'}}^{(+)}}
\Biggr]
\prod_{r=1}^{m_{\alpha}}\prod_{r'=1}^{s_{\alpha}}
\Biggl[
\frac{
q^{\alpha}_{r} - {q^{*}}^{\alpha}_{r'} - i\kappa}{
q^{\alpha}_{r} - {q^{*}}^{\alpha}_{r'}  }
\Biggr]
\end{equation}
and
\begin{equation}
 \label{E.8}
{\cal G}_{\alpha\beta} =
\prod_{r=1}^{m_{\alpha}}\prod_{r'=1}^{m_{\beta}}
\Biggl[
\frac{
{q^{\alpha}_{r}}^{(-)}+{q^{\beta}_{r'}}^{(-)}+i\kappa}{
{q^{\alpha}_{r}}^{(-)}+{q^{\beta}_{r'}}^{(-)}}
\Biggr]
\prod_{r=1}^{s_{\alpha}}\prod_{r'=1}^{s_{\beta}}
\Biggl[
\frac{
{{q^{*}}^{\alpha}_{r}}^{(+)}+{{q^{*}}^{\beta}_{r'}}^{(+)}-i\kappa}{
{{q^{*}}^{\alpha}_{r}}^{(+)}+{{q^{*}}^{\beta}_{r'}}^{(+)}}
\Biggr]
\prod_{r=1}^{m_{\alpha}}\prod_{r'=1}^{s_{\beta}}
\Biggl[
\frac{
q^{\alpha}_{r} - {q^{*}}^{\beta}_{r'} - i\kappa}{
q^{\alpha}_{r} + {q^{*}}^{\beta}_{r'}  }
\Biggr]
\times
\prod_{r=1}^{s_{\alpha}}\prod_{r'=1}^{m_{\beta}}
\Biggl[
\frac{
{q^{*}}^{\alpha}_{r} - q^{\beta}_{r'} - i\kappa}{
{q^{*}}^{\alpha}_{r} - q^{\beta}_{r'}  }
\Biggr]
\end{equation}
The product factors in eq.(\ref{E.7}) can be easily expressed it terms of the
Gamma functions:
\begin{eqnarray}
\label{E.9}
\prod_{r=1}^{m_{\alpha}}{q^{\alpha}_{r}}^{(-)} &=&
\prod_{r=1}^{m_{\alpha}}
\Bigl[
{q_{\alpha}}^{(-)} - \frac{i\kappa}{2}(m_{\alpha}+s_{\alpha}+1) + i\kappa r
\Bigr]
\; = \;
(i\kappa)^{m_{\alpha}}
\frac{
\Gamma\Bigl(
\frac{1}{2}-\frac{s_{\alpha}-m_{\alpha}}{2} - \frac{i{q_{\alpha}}^{(-)}}{\kappa}
\Bigr)}{
\Gamma\Bigl(
\frac{1}{2}-\frac{s_{\alpha}+m_{\alpha}}{2} - \frac{i{q_{\alpha}}^{(-)}}{\kappa}
\Bigr)}
\\
\nonumber
\\
\nonumber
\\
\prod_{r=1}^{s_{\alpha}}{{q^{*}}^{\alpha}_{r}}^{(+)} &=&
\prod_{r=1}^{s_{\alpha}}
\Bigl[
{q_{\alpha}}^{(+)} + \frac{i\kappa}{2}(m_{\alpha}+s_{\alpha}+1) - i\kappa r
\Bigr]
\; = \;
(-i\kappa)^{s_{\alpha}}
\frac{
\Gamma\Bigl(
\frac{1}{2}-\frac{m_{\alpha}-s_{\alpha}}{2} + \frac{i{q_{\alpha}}^{(+)}}{\kappa}
\Bigr)}{
\Gamma\Bigl(
\frac{1}{2}-\frac{m_{\alpha}+s_{\alpha}}{2} + \frac{i{q_{\alpha}}^{(+)}}{\kappa}
\Bigr)}
\label{E.10}
\end{eqnarray}
\begin{eqnarray}
\label{E.11}
\prod_{r< r'}^{m_{\alpha}}
\Biggl[
\frac{
{q^{\alpha}_{r}}^{(-)}+{q^{\alpha}_{r'}}^{(-)}+i\kappa}{
{q^{\alpha}_{r}}^{(-)}+{q^{\alpha}_{r'}}^{(-)}}
\Biggr]
&=&
2^{-(m_{\alpha}-1)}
\frac{
\Gamma\Bigl(
m_{\alpha}-s_{\alpha} - \frac{2i{q_{\alpha}}^{(-)}}{\kappa}
\Bigr)
\Gamma\Bigl(
1-\frac{m_{\alpha}+s_{\alpha}}{2} - \frac{i{q_{\alpha}}^{(-)}}{\kappa}
\Bigr)}{
\Gamma\Bigl(
\frac{m_{\alpha}-s_{\alpha}}{2} - \frac{i{q_{\alpha}}^{(-)}}{\kappa}
\Bigr)
\Gamma\Bigl(
1 - s_{\alpha} - \frac{2i{q_{\alpha}}^{(-)}}{\kappa}
\Bigr)}
\\
\nonumber
\\
\nonumber
\\
\label{E.12}
\prod_{r< r'}^{s_{\alpha}}
\Biggl[
\frac{
{{q^{*}}^{\alpha}_{r}}^{(+)}+{{q^{*}}^{\alpha}_{r'}}^{(+)}-i\kappa}{
{{q^{*}}^{\alpha}_{r}}^{(+)}+{{q^{*}}^{\alpha}_{r'}}^{(+)}}
\Biggr]
&=&
2^{-(s_{\alpha}-1)}
\frac{
\Gamma\Bigl(
s_{\alpha}-m_{\alpha} + \frac{2i{q_{\alpha}}^{(+)}}{\kappa}
\Bigr)
\Gamma\Bigl(
1-\frac{m_{\alpha}+s_{\alpha}}{2} + \frac{i{q_{\alpha}}^{(+)}}{\kappa}
\Bigr)}{
\Gamma\Bigl(
\frac{s_{\alpha}-m_{\alpha}}{2} + \frac{i{q_{\alpha}}^{(+)}}{\kappa}
\Bigr)
\Gamma\Bigl(
1 - m_{\alpha} + \frac{2i{q_{\alpha}}^{(+)}}{\kappa}
\Bigr)}
\\
\nonumber
\\
\nonumber
\\
\label{E.13}
\prod_{r=1}^{m_{\alpha}}\prod_{r'=1}^{s_{\alpha}}
\Biggl[
\frac{
q^{\alpha}_{r} - {q^{*}}^{\alpha}_{r'} - i\kappa}{
q^{\alpha}_{r} - {q^{*}}^{\alpha}_{r'}  }
\Biggr]
&=&
\frac{
\Gamma\bigl(1 + m_{\alpha} + s_{\alpha}\bigr)}{
\Gamma\bigl(1 + m_{\alpha}\bigr) \Gamma\bigl(1 + s_{\alpha}\bigr)}
\end{eqnarray}
Substituting the above expressions into eq.(\ref{E.7}) and using the
standard relations for the Gamma functions,
\begin{eqnarray}
\label{E.14}
\Gamma(z) \,\Gamma(1-z) &=& \frac{\pi}{\sin(\pi z)}
\\
\nonumber
\\
\label{E.15}
\Gamma(1+z) &=& z \, \Gamma(z)
\\
\nonumber
\\
\label{E.16}
\Gamma\Bigl(\frac{1}{2} + z\Bigr) &=&
\frac{\sqrt{\pi} \, \Gamma\bigl(1 + 2z\bigr)}{
2^{2z} \, \Gamma\bigl(1 + z\bigr)}
\end{eqnarray}
 we get
\begin{equation}
\label{E.17}
{\cal G}\bigl(q_{\alpha}, m_{\alpha}, s_{\alpha}\bigr) \; = \;
\frac{
\Gamma\Bigl(
s_{\alpha} + \frac{2i}{\kappa} {q_{\alpha}}^{(-)}
\Bigr) \,
\Gamma\Bigl(
m_{\alpha} - \frac{2i}{\kappa} {q_{\alpha}}^{(+)}
\Bigr) \,
\Gamma\bigl(1 + m_{\alpha} + s_{\alpha}\bigr)}{
2^{(m_{\alpha} + s_{\alpha})}
\Gamma\Bigl(
m_{\alpha} + s_{\alpha} + \frac{2i}{\kappa} {q_{\alpha}}^{(-)}
\Bigr) \,
\Gamma\Bigl(
m_{\alpha} + s_{\alpha} - \frac{2i}{\kappa} {q_{\alpha}}^{(+)}
\Bigr) \,
\Gamma\bigl(1 + m_{\alpha}\bigr) \Gamma\bigl(1 + s_{\alpha}\bigr)}
\end{equation}
Similar calculations for the factor ${\cal G}_{\alpha\beta}$, eq.(\ref{E.8}),
yield the following expression
\begin{eqnarray}
\nonumber
{\cal G}_{\alpha\beta} \bigl({\bf q}, {\bf m}, {\bf s}\bigr) &=&
\frac{
\Gamma
\Bigl[
1 + \frac{m_{\alpha} + m_{\beta} - s_{\alpha} - s_{\beta}}{2}
-\frac{i}{\kappa}\bigl({q_{\alpha}}^{(-)} + {q_{\beta}}^{(-)}\bigr)
\Bigr] \,
\Gamma
\Bigl[
1 - \frac{m_{\alpha} + m_{\beta} + s_{\alpha} + s_{\beta}}{2}
-\frac{i}{\kappa}\bigl({q_{\alpha}}^{(-)} + {q_{\beta}}^{(-)}\bigr)
\Bigr]}{
\Gamma
\Bigl[
1 - \frac{m_{\alpha} - m_{\beta} + s_{\alpha} + s_{\beta}}{2}
-\frac{i}{\kappa}\bigl({q_{\alpha}}^{(-)} + {q_{\beta}}^{(-)}\bigr)
\Bigr] \,
\Gamma
\Bigl[
1 + \frac{m_{\alpha} - m_{\beta} - s_{\alpha} - s_{\beta}}{2}
-\frac{i}{\kappa}\bigl({q_{\alpha}}^{(-)} + {q_{\beta}}^{(-)}\bigr)
\Bigr]}
\times
\\
\nonumber
\\
\nonumber
\\
\nonumber
&\times&
\frac{
\Gamma
\Bigl[
1 - \frac{m_{\alpha} + m_{\beta} - s_{\alpha} - s_{\beta}}{2}
+\frac{i}{\kappa}\bigl({q_{\alpha}}^{(+)} + {q_{\beta}}^{(+)}\bigr)
\Bigr] \,
\Gamma
\Bigl[
1 - \frac{m_{\alpha} + m_{\beta} + s_{\alpha} + s_{\beta}}{2}
+\frac{i}{\kappa}\bigl({q_{\alpha}}^{(+)} + {q_{\beta}}^{(+)}\bigr)
\Bigr]}{
\Gamma
\Bigl[
1 - \frac{m_{\alpha} + m_{\beta} + s_{\alpha} - s_{\beta}}{2}
+\frac{i}{\kappa}\bigl({q_{\alpha}}^{(+)} + {q_{\beta}}^{(+)}\bigr)
\Bigr] \,
\Gamma
\Bigl[
1 - \frac{m_{\alpha} + m_{\beta} - s_{\alpha} + s_{\beta}}{2}
+\frac{i}{\kappa}\bigl({q_{\alpha}}^{(+)} + {q_{\beta}}^{(+)}\bigr)
\Bigr]}
\times
\\
\nonumber
\\
\nonumber
\\
\nonumber
&\times&
\frac{
\Gamma
\Bigl[
1 + \frac{m_{\alpha} + m_{\beta} + s_{\alpha} + s_{\beta}}{2}
+\frac{i}{\kappa}\bigl(q_{\alpha} - q_{\beta}\bigr)
\Bigr] \,
\Gamma
\Bigl[
1 + \frac{-m_{\alpha} + m_{\beta} + s_{\alpha} - s_{\beta}}{2}
+\frac{i}{\kappa}\bigl(q_{\alpha} - q_{\beta}\bigr)
\Bigr]}{
\Gamma
\Bigl[
1 + \frac{-m_{\alpha} + m_{\beta} + s_{\alpha} + s_{\beta}}{2}
+\frac{i}{\kappa}\bigl(q_{\alpha} - q_{\beta}\bigr)
\Bigr] \,
\Gamma
\Bigl[
1 + \frac{m_{\alpha} + m_{\beta} + s_{\alpha} - s_{\beta}}{2}
+\frac{i}{\kappa}\bigl(q_{\alpha} - q_{\beta}\bigr)
\Bigr]}
\times
\\
\nonumber
\\
\nonumber
\\
\label{E.18}
&\times&
\frac{
\Gamma
\Bigl[
1 + \frac{m_{\alpha} + m_{\beta} + s_{\alpha} + s_{\beta}}{2}
-\frac{i}{\kappa}\bigl(q_{\alpha} - q_{\beta}\bigr)
\Bigr] \,
\Gamma
\Bigl[
1 + \frac{m_{\alpha} - m_{\beta} - s_{\alpha} + s_{\beta}}{2}
-\frac{i}{\kappa}\bigl(q_{\alpha} - q_{\beta}\bigr)
\Bigr]}{
\Gamma
\Bigl[
1 + \frac{m_{\alpha} + m_{\beta} - s_{\alpha} + s_{\beta}}{2}
-\frac{i}{\kappa}\bigl(q_{\alpha} - q_{\beta}\bigr)
\Bigr] \,
\Gamma
\Bigl[
1 + \frac{m_{\alpha} - m_{\beta} + s_{\alpha} + s_{\beta}}{2}
-\frac{i}{\kappa}\bigl(q_{\alpha} - q_{\beta}\bigr)
\Bigr]}
\end{eqnarray}


\end{document}